\documentclass[10pt,fleqn]{report}

\usepackage[usenames,dvipsnames]{xcolor}

\definecolor{hp}{RGB}{228,26,28}         
\definecolor{cp}{RGB}{55,126,184}       
\definecolor{ci}{RGB}{77,175,74}           
\definecolor{cd}{RGB}{152,78,163}       
\definecolor{pt}{RGB}{255,127,0}          
\definecolor{ut}{RGB}{166,86,40}          

\usepackage{sectsty}
\colorlet{LightMidnightBlueCustom}{MidnightBlue!80!white}
\chapterfont{\color{MidnightBlue}}  
\sectionfont{\color{MidnightBlue}}  

\usepackage{hyperref}
\hypersetup{
  colorlinks = true,
  linkbordercolor = {white},
  urlcolor    = LightMidnightBlueCustom, 
  linkcolor   = LightMidnightBlueCustom, 
  citecolor   = black 
}

\usepackage{graphicx}

\usepackage{float}

\usepackage{epstopdf}

\usepackage{tikz}
\usetikzlibrary{arrows,positioning,decorations.markings}
\usetikzlibrary{patterns} 

\tikzstyle{myarrow} = [thick, decoration={markings,mark=at position
   0.999 with {\arrow[semithick]{open triangle 60}}},
   double distance=1.4pt, shorten >= 5.5pt,
   preaction = {decorate},
   postaction = {draw,line width=1.4pt, white,shorten >= 4.5pt}
   ]

\usepackage{amsmath,bm}
\usepackage{amssymb}

\usepackage{natbib}
\bibpunct{(}{)}{;}{a}{}{,}

\usepackage{bibentry}
\nobibliography*

\newcommand{\ptlder}[2]{\frac{\partial #1}{\partial #2}}
\newcommand{\totder}[2]{\frac{d #1}{d #2}}
\newcommand{\inverse}[1]{\frac{1}{#1}}
\newcommand{\Grad}{\vec{\nabla}}

\newcommand{\ptlsqd}[2]{\frac{\partial^{2} #1}{\partial #2^{2}}}

\newcommand{\PathToDirMicroSrcVarCovar}{.}  

\usepackage{mdwlist}

\usepackage{listings}
\newif\ifGridDiagrams
\GridDiagramsfalse

\begin{document}



\title{\color{MidnightBlue} \textbf{CLUBB-SILHS: A parameterization of subgrid variability in the atmosphere}}
\date{\Large \color{LightMidnightBlueCustom} \today}
\author{\Large \color{LightMidnightBlueCustom} Vincent E.~Larson}
\maketitle

\section{What are CLUBB and SILHS?}  

CLUBB (Cloud Layers Unified By Binormals) is a parameterization of subgrid-scale 
(i.e.~unresolved) variability in atmospheric models
\citep{golaz_et_al_02a,larson_golaz_05a,bogenschutz_et_al_2013a,larson_2019_momentum}.  
CLUBB parameterizes clouds and turbulence, but it is more general than that.  It also parameterizes
the subgrid variability in hydrometeors, and in principle it could be extended to parameterize variability 
in aerosol-related processes or radiative transfer.  


SILHS (Subgrid Importance Latin Hypercube Sampler) is a Monte Carlo method of integrating over subgrid variability
\citep{larson_et_al_05a,larson_schanen_2013a,raut_larson_2016_flex_sampling,thayer-calder_et_al_2015_cam_clubb_silhs}.
It can be used in conjunction with CLUBB in order to estimate the effects of subgrid variability on microphysical
process rates.

CLUBB-SILHS is targeted at host models with horizontal grid spacings of 2 km and coarser.  
This includes convection permitting models, regional weather forecast models, and global 
climate models.  

This document focuses on a unified configuration of CLUBB-SILHS that does not include
the use of a separate deep convective parameterization.  This configuration differs from the implementations of CLUBB in
version 6 of the Community Atmosphere Model (CAM6) and version 2 of the Energy Exascale Earth System Model 
(E3SMv2) model.  In those models, the Zhang-McFarlane deep convective parameterization \citep{zhang_mcfarlane_1995a} 
is used simultaneously
with CLUBB.  Despite the difference in configuration, this document contains much relevant information
for users of CAM6 and E3SMv2.

The document describes git revision \href{https://github.com/larson-group/clubb_release/commit/da4fc00e153ee358203253ad4afde70d7ed206a5}{da4fc00} of CLUBB-SILHS, committed on 25 Mar 2022.

\section{What's in this document?}

This document is intended for several different audiences.  It describes
the rationale and philosophy behind one general approach to cloud and turbulence parameterization, 
the ``PDF method," and it also 
contains detailed information about a particular parameterization, CLUBB-SILHS.  Many readers will 
opt to pick and choose those chapters that are most relevant to their interests.

Curious newcomers may wish to review the cloud parameterization problem (Chapter \ref{chapt:cloud_param_problem}), 
read an outline of the CLUBB-SILHS method (Chapter \ref{chapt:outline_clubb-silhs}), 
see a comparison of CLUBB-SILHS with other parameterization methodologies 
(Chapter \ref{chapt:clubb_comparison}), and peruse 
the Frequently Asked Questions (Chapter \ref{chapt:faq}).

Those readers who wish to learn more technical details about CLUBB-SILHS' methodology 
may wish to examine a technical description of CLUBB (Chapter \ref{chapt:clubb_tech_descr}),
a summary list of closed equations in CLUBB (Chapter \ref{chapt:clubb_closed_eqns}), or 
a technical description of SILHS (Chapter \ref{chapt:silhs_tech_descr}).  

Large-scale modelers who are considering implementing or tuning CLUBB-SILHS in a host model may want
to consult the comparison of CLUBB-SILHS with other parameterizations (Chapter \ref{chapt:clubb_comparison}),
citations to papers on simulations that include CLUBB or SILHS (Chapter \ref{chapt:results}), 
or source code documentation (Chapter \ref{chapt:code_doc}).  

Those who wish to modify or extend CLUBB-SILHS' source code may want to read the 
technical descriptions (Chapters \ref{chapt:clubb_tech_descr}, 
\ref{chapt:clubb_closed_eqns}, \ref{chapt:silhs_tech_descr}), or the code documentation 
(Chapter \ref{chapt:code_doc}).

For those readers who seek further information, the final chapter provides an annotated bibliography
of articles about CLUBB and SILHS (Chapter \ref{chapt:annot_bib}).

If you have comments, questions, or suggestions, please do not hesitate to email them to Vincent Larson 
at vlarson@uwm.edu.  Both CLUBB-SILHS and this document are undergoing continual revision.


\section{Acknowledgments}

Numerous people have contributed to CLUBB-SILHS' documentation over the years,
especially Chris Golaz, Eric Raut, Brian Griffin, and David Schanen.  I thank
Wuyin Lin, Hugh Morrison, Ben Stephens, and Jan Gr\"{u}nenwald for extensive comments 
on a draft of this document.
V.~Larson's contributions to this document were supported by a Climate Process Team grant, 
Award Number 0968640, from the National Science Foundation.

\tableofcontents

\newpage

\chapter{What is the cloud parameterization problem?}

\label{chapt:cloud_param_problem}


Global numerical weather predictions and climate simulations are both created by large-scale numerical models.
These models solve, in an approximate way, the equations of fluid flow in the earth's atmosphere.

The models cover the globe with a mesh of grid points, arranged in contiguous grid columns.  
Each grid column has dozens of grid levels in the vertical,
from the ground surface up to the model top at high altitudes in the atmosphere.  At each grid point and time step,
only one value of, e.g., temperature is computed.  Values of temperature between grid points are not directly computed.  

Even with today's supercomputers,
the spacing between grid points is coarse.  For instance, a global numerical weather model 
might have a horizontal grid spacing of about 10 km,
and the number of vertical grid levels might number 100.  Some clouds are much smaller than this,
with a horizontal width, e.g., of 1 km.  These small clouds are important because they 
reflect sunlight and transport heat, moisture, and momentum
in the vertical.  To approximate their effects, large-scale models employ ``cloud parameterizations."

A cloud parameterization estimates the effects of small-scale cloud elements and turbulence on
the larger resolved scales in a host (fluid-flow) model.  When asked about cloud parameterization, 
our minds instinctively flit to images
of convective plumes (i.e.~buoyant coherent structures) 
or entrainment processes (i.e.~turbulent mixing).  But it is instructive to ask, 
What is the minimum set of information that a coarse-resolution 
host model requires from a cloud parameterization?  That is, what terms are missing from a host model's
equation set?  

The equation set in a host model must be spatially filtered over grid-box-sized volumes,
and the filtering process introduces unclosed (i.e.~unknown) terms.  
Ultimately, these terms are what must be parameterized, and
if they are parameterized accurately, no other terms need to be parameterized.  

Ignoring horizontal advection terms, the primary grid-mean equations predicted by a host model are: 
\begin{equation}
\label{eq_rtm_unclosed}
\ptlder{\color{hp}\overline{r_t}}{t}
= \underbrace{ - {\color{hp} \overline{w}}\ptlder{\color{hp}\overline{r_t}}{z} }_\mathrm{mean\;adv} \
  \underbrace{ - \inverse{{\color{hp}\rho_s}}\ptlder{{\color{hp}\rho_s} {\color{cp}\overline{w'r'_t}}}{z} }_\mathrm{turb\;adv} \
  \underbrace{\color{ci} {\color{black}+} \overline{\color{hp} \left.\ptlder{ r_t }{ t }\right|_{\mathrm{mc}}}}_\mathrm{microphys}  
\end{equation}
\begin{equation}
\label{eq_thlm_unclosed}
\ptlder{\color{hp}\overline{\theta_l}}{t} 
= \underbrace{- {\color{hp} \overline{w}}\ptlder{\color{hp} \overline{\theta_l}}{z} }_\mathrm{mean\;adv} \
  \underbrace{ - \inverse{{\color{hp}\rho_s}}\ptlder{{\color{hp}\rho_s}{\color{cp} \overline{w'\theta'_l}}}{z} }_\mathrm{turb\;adv} \ 
  \underbrace{+{\color{ci} \overline{\color{hp} \left.\ptlder{ \theta_{l} }{ t }\right|_{\mathrm{RT}} }}}_\mathrm{radiation}  \
  \underbrace{\color{ci} {\color{black}+} \overline{\color{hp} \left.\ptlder{ \theta_l }{ t }\right|_{\mathrm{mc}}  }}_\mathrm{microphys} 
\end{equation}
\begin{equation}
\label{eq_um_unclosed}
\ptlder{\color{hp}\overline{u}}{t} 
= \underbrace{ - {\color{hp} \overline{w}}\ptlder{\color{hp} \overline{u}}{z} }_\mathrm{mean\;adv} \
  \underbrace{\color{hp} - f (v_g - \overline{v}) }_\mathrm{Coriolis/press} \
  \underbrace{ - \inverse{{\color{hp}\rho_s}}\ptlder{{\color{hp}\rho_s}{\color{cp} \overline{u'w'}}}{z} }_\mathrm{turb\;adv} 
\end{equation}
\begin{equation}
\label{eq_vm_unclosed}
\ptlder{\color{hp}\overline{v}}{t} 
= \underbrace{- {\color{hp} \overline{w} }\ptlder{\color{hp}\overline{v}}{z} }_\mathrm{mean\;adv} \
  \underbrace{\color{hp} + f (u_g - \overline{u}) }_\mathrm{Coriolis/press}  \
  \underbrace{ - \inverse{{\color{hp}\rho_s}}\ptlder{{\color{hp}\rho_s}{\color{cp} \overline{v'w'}}}{z} }_\mathrm{turb\;adv} ,
\end{equation}
%
%
where $r_t$ is the total water mixing ratio (vapor + liquid cloud water), 
$\theta_l$ is the liquid water potential temperature, $w$ is the upward wind, 
$u$ is the eastward wind, $v$ is the northward wind, 
$\rho_s$ is the basic state air density, $z$ is altitude, and $t$ is time.
Also, ${\color{ci} {\color{hp} \mathrm{RT}}}$ is the radiative heating rate, $\mathrm{mc}$ denotes a 
microphysical tendency, $f$ the Coriolis (Earth rotational) parameter, 
and $u_g$ and $v_g$ the geostrophic winds.  
Overbars denote grid box spatial averages.  Color coding of the equations
is described in Table \ref{tab:color_coding}.  These equations, or similar ones that predict vapor and liquid separately, 
are solved on the grid mesh by the host model.

We use an overbar to denote a grid cell spatial mean (or, more briefly, ``grid-mean").  A prime denotes a deviation from
the grid cell mean.

Microphysical variables, such as precipitation mixing ratios, are predicted by a microphysics scheme,
not a cloud parameterization.  Nevertheless, averaging up 
to the grid scale requires input about subgrid variability from the cloud parameterization.  
We list one example, 
a prognostic equation for rain water mixing ratio, $r_r$,
where ice processes have been omitted for simplicity:

\begin{equation}
\label{eq_rrm_unclosed}
\ptlder{\color{hp}\overline{r_r}}{t}
= \underbrace{ - {\color{hp} \overline{w}}\ptlder{\color{hp}\overline{r_r}}{z} }_\mathrm{mean\;adv} \
  \underbrace{ - \inverse{{\color{hp}\rho_s}}\ptlder{{\color{hp}\rho_s} {\color{cp}\overline{w'r'_r}}}{z} }_\mathrm{turb\;adv} \
  \underbrace{{\color{ci} {\color{black}+} \overline{\color{hp} \left.\ptlder{ r_{r} }{ t }\right|_{\mathrm{Autoconv}} }}  \
  {\color{ci} {\color{black}+} \overline{\color{hp} \left.\ptlder{ r_{r} }{ t }\right|_{\mathrm{Accretion}} }}  \
  {\color{ci} {\color{black}+} \overline{\color{hp} \left.\ptlder{ r_{r} }{ t }\right|_{\mathrm{Evap}} }}}_\mathrm{microphys} \  . \  . \  .
\end{equation}

\begin{table}
\caption{Color coding in equations.}
\begin{tabular}{ll}
\hline
\textbf{Color}     &  \textbf{Meaning}            \\
\hline
{\color{hp} Red}   & {\color{hp} Variables predicted by the host model, microphysics scheme, or radiative transfer scheme}         \\
{\color{cp} Blue}   & {\color{cp} Variables predicted by CLUBB and returned to the host model}         \\
{\color{cd} Purple}   & {\color{cd} Variables predicted by CLUBB and used internally within CLUBB}         \\
{\color{ci} Green}   & {\color{ci} Terms or grid-averaging operators calculated using CLUBB's subgrid PDF}         \\
{\color{ut} Brown}   & {\color{ut} Terms within CLUBB that are not closed by use of CLUBB's subgrid PDF}         \\
{\color{olive} Olive} & {\color{olive} A lovely color.  Someday we must invent a new class of CLUBB terms for it}         \\
\hline
\end{tabular}
\label{tab:color_coding}
\end{table}

Inspection of Eqns.~(\ref{eq_rtm_unclosed})-(\ref{eq_rrm_unclosed}) tells us what a parameterization
needs to supply to a host model: 
the turbulent fluxes of scalars and momentum (the ``blue fluxes") and the subgrid
distributions needed to average quantities to the grid scale (a key part of the ``green bars").  
These quantities are what a parameterization ought to focus on.  If they can be parameterized accurately,
no other quantities are needed for closure.

The green bars are important
because microphysical processes are non-linear, and for non-linear functions, feeding in grid means
yields the wrong answer, because $\overline{f(x)} \ne f(\overline{x})$.  

Conspicuously absent from the equation set are
obvious signs of convective plumes and entrainment.  These quantities are of known importance, and 
they must be buried somewhere within the governing equations.  Hence their physics ought to be included 
at least implicitly in the parameterized equations, 
but they do not pop out of Eqns.~(\ref{eq_rtm_unclosed})-(\ref{eq_rrm_unclosed}).

\chapter{ Brief outline of the CLUBB-SILHS method }

\label{chapt:outline_clubb-silhs}

\section{CLUBB-SILHS is a LES emulator}

\label{sec:LES_emulator}


Consider the following thought experiment.  Imagine that a high-resolution, 3-dimensional large-eddy simulation (LES)
model is used to simulate a field of shallow cumuli that evolve and develop into deep cumuli. 
LESs are accurate enough to serve as a useful surrogate for nature.
Although LES models are formulated in terms of local derivatives rather than vertical integrals, 
LES models are nonetheless capable of simulating penetrative convection.
Suppose that the LES is phrased in terms of variables such as vertical velocity $w$, 
total water mixing ratio (vapor + liquid) $r_t$,
liquid water potential temperature $\theta_l$, and microphysical quantities such as rain water mixing ratio $r_r$.  
Suppose further that every 5 minutes during the simulation, profiles of selected horizontally averaged (i.e. averaged over the horizontal extent of the LES domain)
moments --- such as $\overline{w'r_t'}$, $\overline{\theta_l'^2}$, $\overline{r_r}$, etc.  --- are output to disk.

Now imagine, in this shallow-to-deep transitional simulation, how these output moments would appear.  
The LES would simulate parcels (i.e.~``point" volumes of air) that ascend quickly in convective plumes and transport moisture upward.  
Nevertheless, although the updraft speeds might be large, the turbulent cloud layer would 
deepen only gradually as successive clouds moisten higher altitudes.  
Typically, a cloud layer deepens less than, usually much less than, 1 km per every 5 min interval.  
As a consequence of the evolution, the layer would exhibit slow deepening 
and gradual strengthening of fluxes such as $\overline{w'r_t'}$ 
and variances such as $\overline{\theta_l'^2}$.  

The moments that are output every 5 minutes are what CLUBB-SILHS seeks to parameterize.  
The moments contain what is needed to update a host model (blue fluxes plus information to perform 
the green-bar integrals), plus a wealth of further detail.
If CLUBB-SILHS can approximate these moments adequately, then it can approximate deep convective flows.
The horizontal domain of the LES corresponds to the grid-box width in CLUBB.  
CLUBB-SILHS' formulation mimics aspects of the 
LES formulation, but CLUBB-SILHS uses horizontally averaged equations, rather than fine-scale, 3D equations.

CLUBB-SILHS does not attempt to parameterize individual convective plumes directly.  
Instead, CLUBB-SILHS parameterizes horizontally averaged turbulent fields, 
which evolve more slowly.  

\section{Overview of CLUBB}

\hypertarget{url:overview_clubb}{}

CLUBB's goal, as for all cloud/turbulence parameterization suites, is to estimate the green bars 
and the four blue fluxes.  

CLUBB's main inputs and outputs are as follows:
\begin{enumerate}

\item \textit{Inputs:}

\begin{enumerate}

\item Grid mean fields.

\item Higher-order moments, including the four blue fluxes.

\end{enumerate}

\item \textit{Outputs:}

\begin{enumerate}

\item Updated values of the higher-order moments, including the four blue fluxes.

\item Parameters that govern the shape of the PDF and are needed to perform microphysical green-bar integrals.

\end{enumerate}

\end{enumerate}

In order to estimate the green-bar integrals, CLUBB models the subgrid probability density function (PDF)
in each grid box and time step, and then integrates the quantities that need to be closed, 
such as the relevant microphysical process rates, over the subgrid PDF.  
This is the first reason that modeling the subgrid PDF is a central preoccupation of CLUBB.

CLUBB prognoses the four blue fluxes.  It turns out that some (but not all) of the terms in the equations 
for the fluxes can be closed using CLUBB's subgrid PDF.  This is a second reason that the subgrid PDF is central to CLUBB.

In order to constrain the PDF, CLUBB prognoses several subgrid moments: means shift the position of the PDF
to higher or lower values, variances determine the width of the PDF, covariances determine the 
covariability among variates, and third-order moments determine whether the PDF is skewed to the left or the right.
Predicting an infinite number of moments would define the PDF shape exactly \citep{shohat_tamarkin_43a}, 
but a parameterization can afford to predict only a few moments.  
Hence CLUBB resorts to the assumption that the shape of the PDF is a normal/lognormal mixture.   

In summary, then, CLUBB predicts various moments, some of which are required
directly by the host model (the blue fluxes) and some of which are useful for constraining
the PDF (e.g., scalar variances).  The PDF, in turn, is needed to close certain terms by 
estimating various grid averages (the green bars).

Phrased more conventionally, CLUBB uses a higher-order closure approach, with several terms closed by
integration over a PDF.  In order to advance the solution one time step in one grid box,
CLUBB carries out the following 4-step procedure:

\begin{enumerate}

\item \textit{Advance higher-order moments one time step.}  CLUBB prognoses several subgrid turbulent fluxes, 
plus several variances and covariances, and finally one third-order moment, $\overline{w'^3}$.  
The prognostic equations for these moments are spatially filtered (similar to Reynolds-averaged) versions of the 
governing equations of fluid flow, in particular, the Navier-Stokes equations for momentum
and the advection-diffusion equations for thermodynamic and microphysical scalars.  
Much of the physics contained in the governing Navier-Stokes and advection-diffusion equations
is inherited by the filtered versions of those equations in CLUBB.

\item \textit{Given the updated higher-order moments, diagnose the subgrid PDF.}
The subgrid variability in each grid box and time step are represented by a single, multivariate PDF
whose shape (i.e., functional form) is assumed.
CLUBB assumes that the marginals of $w$, $r_t$, and $\theta_l$ are distributed according to a mixture of normals, 
a.k.a.~a ``double Gaussian" PDF \citep{larson_et_al_02a}.
This is the sum of two normals.  For cloud droplet number concentration, CLUBB assumes a single lognormal marginal.
For other hydrometeors, CLUBB assumes a mixture of two lognormals \citep{griffin_larson_2016_hydrometeor_PDF}.  
However, all these distributions are marginals of a single
subgrid multivariate PDF with co-varying variates.  Given the assumption of a double normal/lognormal shape,
the PDF for a particular grid box and time step is defined by ``PDF parameters," which describe the mean,
variance, covariance, and mixture fraction of each of the two normal/lognormal \textit{components}.

\item \textit{Given the subgrid PDF, close some terms via integration over the PDF.}  
This is a spatial grid-box average, where the PDF weights the area covered by particular values of the integrand.
The averaging integral is a continuous form of a weighted sum (average).  These unclosed, integrable 
terms fall into two classes:

\begin{enumerate}

\item  \textit{Turbulent advection terms and buoyancy-related terms.}  
Every prognostic equation for a moment, such as $\overline{r_t'^2}$, contains a turbulent advection
term, such as $\overline{w'r_t'^2}$.  This term represents the vertical transport of $r_t'^2$ by
the turbulent vertical velocity, $w'$.  Every prognostic equation for a moment that includes $w'$
also contains a buoyancy-related term, such as $\overline{w'\theta_v'}$. In CLUBB, these terms are always integrated 
over the PDF \textit{analytically}.  

\item  \textit{Microphysical terms.}  
Microphysical processes may be a source or sink of either grid means (see, e.g., Eqn.~\ref{eq_rrm_unclosed}) 
or higher-order moments.  Microphysical terms are complex and are typically integrated using 
a Monte Carlo method, namely SILHS.  
However, for simple, warm-rain microphysics, CLUBB includes an option to perform the integrals 
analytically \citep{larson_griffin_2013a,griffin_larson_2016_microphys_covar}.  

\end{enumerate}

\item \textit{Close pressure and dissipation terms using classical closures.}
Prognostic equations for moments that include velocity include a pressure term, 
such as $\overline{w'\partial p'/\partial z}$.
Every moment equation also contains a turbulent dissipation term that 
represents the homogenization arising from turbulent mixing.  For instance, the $\overline{r_t'^2}$
equation contains a dissipation term proportional to $-\overline{\Grad r_t' \cdot \Grad r_t'}$.
The pressure and dissipation terms cannot be integrated because CLUBB's PDF 
does not include pressure perturbations or gradients.  Instead,
these terms are closed using classical methods from the turbulence literature.   


\end{enumerate}

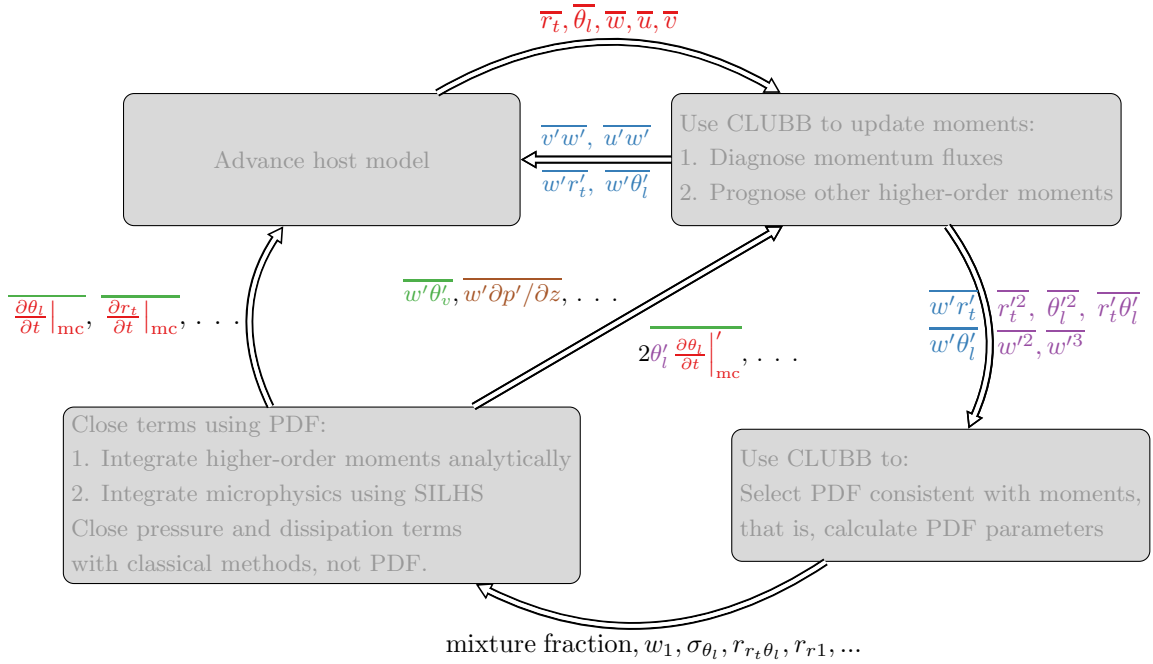
\begin{figure}
\centering
\begin{tikzpicture}[
      mycircle/.style={
         rectangle,
         rounded corners,
         align=left,
         draw=black,
         fill=gray,
         fill opacity = 0.3,
         text opacity=1,
         inner sep=3pt,
         minimum width=150pt,
         minimum height=50pt,
         font=\small},
      node distance=2.4cm and 2cm
      ]
      \node[mycircle] (Host) {Advance host model}
            ;
      \node[mycircle,below=of Host] (Integrate) 
           {\color{black} Close terms using PDF: \\
            1. Integrate higher-order moments analytically \\
            2. Integrate microphysics using SILHS \\
            Close pressure and dissipation terms \\
            with classical methods, not PDF. 
           }
            ;
      \node[mycircle,right=of Integrate] (PDF) 
            {Use CLUBB to:  \\
             Select PDF consistent with moments, \\
             that is, calculate PDF parameters
            }
            ;
      \node[mycircle,right=of Host] (Update) 
            {Use CLUBB to update moments: \\
             1. Diagnose momentum fluxes \\
             2. Prognose other higher-order moments
            }
            ;

   \foreach \i/\j/\txt/\p in {
      Update/Host/${ \color{cp} \overline{w'r_t'}, \; \overline{w'\theta_l'} }$/below,
      Update/Host/${ \color{cp} \overline{v'w'}, \; \overline{u'w'} }$/above,
      Integrate/Update/{${{\color{ci} \overline{w'\theta_v'}}, 
                         {\color{ut} \overline{w'\partial p'/\partial z}} }$, . . .}/above left,
      Integrate/Update/{$ 2 {\color{ci} \overline{ {\color{cd} \theta_l'} 
                            {\color{hp} \left.\ptlder{\theta_l}{ t }\right|_{\mathrm{mc}}'} }}$, . . .}/below right}
       \draw [myarrow] (\i) -- node[font=\small,\p] {\txt} (\j);

     \path[]
     (Host) edge[myarrow, bend left] node [above] 
     {${\color{hp} \overline{r_t}, \overline{\theta_l}, \overline{w}, \overline{u}, \overline{v}}$} 
     (Update);

     \path[]
     (Update) edge[myarrow, bend left] node [right, align=left] 
     { ${\color{cd}  \overline{r_t'^2}, \; \overline{\theta_l'^2}, \; \overline{r_t'\theta_l'} }$  \\
        ${\color{cd} \overline{w'^2}, \overline{w'^3} }$
     } 
     (PDF);

     \path[]
     (Update) edge[myarrow, bend left] node [left, align=left] 
     {${\color{cp}  \overline{w'r_t'}}$ \\
      ${\color{cp} \overline{w'\theta_l'}}$   } 
     (PDF);

     \path[]
     (PDF) edge[myarrow, bend left] node [below] 
     {${ \textrm{mixture\;fraction}, w_1, \sigma_{\theta_l}, r_{r_t \theta_l}, r_{r1},  . . .   }$} 
     (Integrate);

     \path[]
     (Integrate) edge[myarrow, bend left] node [left] 
     {${\color{ci} \overline{\color{hp} \left.\ptlder{ \theta_l }{ t }\right|_{\mathrm{mc}} }{\color{black},} \
        \overline{\color{hp} \left.\ptlder{ r_t }{ t }\right|_{\mathrm{mc}} }
     }$, . . .} 
     (Host);

     \node[align=center,font=\bfseries, yshift=2em] (title) 
         at (current bounding box.north)
         {A CLUBB-SILHS time step, \\ illustrating the main calculations and flow of information};

\end{tikzpicture}
\caption{The main calculations performed in a CLUBB-SILHS time step.  
Arrows labelled with variables depict inputs and outputs.  This schematic depicts
the flow of information within CLUBB-SILHS, not the strict ordering of the sequence of calculations.  
Although it is not depicted in the schematic, CLUBB, SILHS, and microphysics are often substepped 
together in a loop several times during a single host model time step.}
\label{fig:clubb-silhs_flowchart}
\end{figure}

Once the prognostic equations are closed, they can be advanced another time step, and the procedure is repeated.
The procedure is illustrated schematically in Fig.~\ref{fig:clubb-silhs_flowchart}. 



\section{Overview of SILHS}

\hypertarget{url:overview_silhs}{}
\label{chapt:overview_silhs}

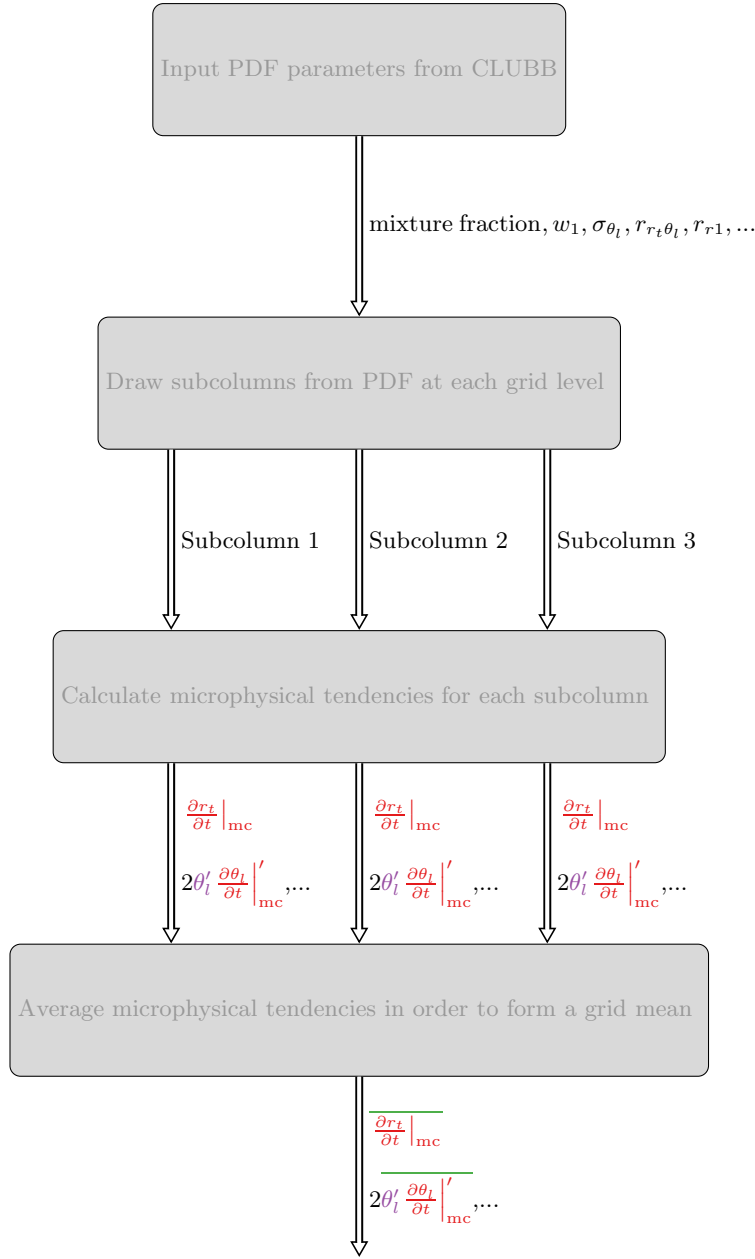
\begin{figure}
\centering
\begin{tikzpicture}[
      mycircle/.style={
         rectangle,
         rounded corners,
         align=left,
         draw=black,
         fill=gray,
         fill opacity = 0.3,
         text opacity=1,
         inner sep=3pt,
         minimum width=150pt,
         minimum height=50pt,
         font=\small},
      node distance=2.4cm and 2cm
      ]
      \node[mycircle] (PDF_parms) {Input PDF parameters from CLUBB}
            ;
      \node[mycircle,below=of PDF_parms] (Draw_subcols) 
           {Draw subcolumns from PDF at each grid level 
           }
            ;
      \node[mycircle,below=of Draw_subcols] (Calc_microphys) 
            {Calculate microphysical tendencies for each subcolumn
            }
            ;
      \node[mycircle,below=of Calc_microphys] (Avg_microphys) 
            {Average microphysical tendencies in order to form a grid mean
            }
            ;
      \coordinate[below=of Avg_microphys] (Output);

      \draw [myarrow] 
        (PDF_parms) 
        -- node[font=\small,right,align=left] 
        {${ \textrm{mixture\;fraction}, w_1, \sigma_{\theta_l}, r_{r_t \theta_l}, r_{r1},  . . .   }$} 
        (Draw_subcols);

      \draw [myarrow] 
        (Draw_subcols)
        -- node[font=\small,right,align=left] 
        {Subcolumn 2} 
        (Calc_microphys);

      \draw [myarrow] 
        ([xshift=-2.5cm]Draw_subcols.south)
        -- node[font=\small,right,align=left] 
        {Subcolumn 1} 
        ([xshift=-2.5cm]Calc_microphys.north);

      \draw [myarrow] 
        ([xshift=2.5cm]Draw_subcols.south)
        -- node[font=\small,right,align=left] 
        {Subcolumn 3} 
        ([xshift=2.5cm]Calc_microphys.north);

      \draw [myarrow] 
        (Calc_microphys)
        -- node[font=\small,right,align=left] 
        {${\color{hp} \left.\ptlder{ r_t }{ t }\right|_{\mathrm{mc}}  }$ \\ \\
           $2 { {\color{cd} \theta_l'} 
                            {\color{hp} \left.\ptlder{\theta_l}{ t }\right|_{\mathrm{mc}}'} }$,...}
        (Avg_microphys);

      \draw [myarrow] 
        ([xshift=-2.5cm]Calc_microphys.south)
        -- node[font=\small,right,align=left] 
        {${\color{hp} \left.\ptlder{ r_t }{ t }\right|_{\mathrm{mc}} }$ \\ \\
           $2 { {\color{cd} \theta_l'} 
                            {\color{hp} \left.\ptlder{\theta_l}{ t }\right|_{\mathrm{mc}}'} }$,...}
        ([xshift=-2.5cm]Avg_microphys.north);

      \draw [myarrow] 
        ([xshift=2.5cm]Calc_microphys.south)
        -- node[font=\small,right,align=left] 
        {${\color{hp} \left.\ptlder{ r_t }{ t }\right|_{\mathrm{mc}} }$ \\ \\
           $2 { {\color{cd} \theta_l'} 
                            {\color{hp} \left.\ptlder{\theta_l}{ t }\right|_{\mathrm{mc}}'} }$,...}
        ([xshift=2.5cm]Avg_microphys.north);

      \draw [myarrow] 
        (Avg_microphys.south)
        -- node[font=\small,right,align=left] 
        {${\color{ci} \overline{\color{hp} \left.\ptlder{ r_t }{ t }\right|_{\mathrm{mc}} } }$ \\ \\
           $2 {\color{ci} \overline{ {\color{cd} \theta_l'} 
                            {\color{hp} \left.\ptlder{\theta_l}{ t }\right|_{\mathrm{mc}}'} }}$,...}
        (Output);

     \node[align=center,font=\bfseries, yshift=2em] (title) 
         at (current bounding box.north)
         {Main SILHS calculations};

\end{tikzpicture}
\caption{The main calculations performed by SILHS.  
Arrows labelled with variables depict inputs and outputs.  }
\label{fig:silhs_flowchart}
\end{figure}

The purpose of SILHS is to calculate grid-box averages of physical process rates 
\citep{larson_et_al_05a,larson_schanen_2013a,raut_larson_2016_flex_sampling}.  The chief input
and output of SILHS are:

\begin{enumerate}

\item \textit{Input:} PDF parameters.

\item \textit{Output:} Grid-averaged microphysical tendencies of means and higher-order moments.

\end{enumerate}

Currently, SILHS is used 
only to average microphysical process rates, but in the future it could be applied to radiative transfer or aerosols.  
Such averages 
are integrals of the process rate over the subgrid PDF.  In some cases, the processes might
be represented by complex numerical subroutines.  In such cases, analytic integration is infeasible.  Instead,
SILHS uses Monte Carlo integration.  It chooses sample points from the PDF at each grid level, 
forms them into a set of profiles, i.e., ``subcolumns," and feeds each subcolumn 
into the microphysics parameterization, one by one.  
The resulting profiles of microphysical tendencies are then averaged to form a grid mean.

What does a SILHS subcolumn represent?  Think back to the LES thought experiment of Section \ref{sec:LES_emulator}.  
Imagine if a soda straw were stuck vertically into the LES domain and then pulled out, 
drawing with it the profiles of moisture, temperature, and so forth.  That multivariate profile 
is what is represented by a subcolumn.  If the soda straw is wider, 
the subcolumn has greater weight and represents more of the domain area.  If the soda straw is narrower, 
it has less weight and represents less area.  In SILHS, the soda straws are assumed to be oriented 
vertically and to have equal width at all altitudes.  These assumptions will facilitate future 
usage in radiative transfer schemes.

Because SILHS uses a sampling approach, it can be thought of as an interface between the 
cloud parameterization and the microphysics parameterizations.  It ``sits" between CLUBB and a microphysics parameterization.  
Thereby, SILHS modularizes the two: it allows CLUBB to focus on subgrid variability
and the microphysics to focus on local process rates.

SILHS involves the following steps \citep{larson_schanen_2013a}:

\begin{enumerate}

\item  \textit{Draw a set of sample points from CLUBB's subgrid PDF at a starting mid-altitude grid level.}   
Each sample point is multivariate, with one random value for each variate: $w$, $r_t$, $\theta_l$, $r_r$, etc.  The starting level
is chosen to be at an important altitude, such as the altitude at which cloud water mixing ratio maximizes.

\item  \textit{Form a set of subcolumns by choosing sample points above and below the starting level.}
Based on a sample value at the starting level, a correlated sample value is drawn for the grid level above.  
That sample value is used to choose the value for the next higher grid level, and we work our way up the profile.  
The vertical correlations are assumed to diminish 
with the vertical distance between grid levels.  In a similar manner, 
we choose the sample values from the starting level down to the ground.

\item  \textit{Feed each subcolumn, one by one, into a microphysics parameterization.} 
The microphysics code does not know that the profile is a sample of the grid column, 
rather than the grid-mean (population mean) profile, 
and it does not need to know.  It computes the microphysical tendencies for each subcolumn
on the assumption that the profile is horizontally uniform.  

\item  \textit{Average the resulting set of microphysics tendencies in order to form an average for each grid level.}
A grid average is estimated as a sample average over subcolumns.  If desired, the average may be weighted, 
where each weight is associated with a subcolumn.  The averaging produces grid mean microphysical tendencies, 
which appear on the right-hand side of, e.g., prognostic equations for grid means of microphysical variables.
Separate averages produce various covariances of microphysical rates and microphysical variables, which
appear in the equations for scalar variances, e.g., $\overline{\theta_l'^2}$.

\end{enumerate}

\chapter{Technical description of CLUBB}

\label{chapt:clubb_tech_descr}

This chapter elaborates on the brief outline provided in Chapter \ref{chapt:outline_clubb-silhs}.

\section{Filtering the equations of motion.}

No numerical model of the atmosphere --- except a direct numerical simulation model, which uses a grid spacing
on the order of 1 mm --- simulates the Navier-Stokes equations directly.  Attempting to do so
at coarse resolution with a non-dissipative numerical scheme would lead to an accumulation of kinetic energy 
at the grid scale and ultimately a model crash.  At best, atmospheric models solve a modified version 
of the Navier-Stokes equations that contains extra smoothing at small spatial scales.  A common way to view 
what is done in practice is the following.
First, the Navier-Stokes equation is spatially filtered (i.e.~averaged over a horizontal area) 
to remove the scales smaller than the grid scale.  Second,
the filtered equations are solved numerically \citep{leonard_1974_turb_filter,germano1992a,colucci_et_al_98a}.  
We adopt this viewpoint.

Filtering is similar in spirit to Reynolds averaging, in which fields are broken into mean, $\overline{()}$,
and perturbation, $()'$, parts.  The equations are then averaged using Reynolds rules of averaging, which assume,
for instance, that $\overline{()'}\approx0$.  (For a clear and detailed introduction to Reynolds averaging 
and higher-order closure, see \citet{stull1988a}.)
In contrast, filtering involves applying a running-mean spatial average to all prognostic equations.  
In general, this procedure violates Reynolds rules of averaging, and hence awkward extra terms appear 
\citep{leonard_1974_turb_filter}. 
However, if each averaged moment, e.g.~$\overline{x' y'}$, is replaced with an appropriate analog, 
e.g.~$\overline{x\,y} - \overline{x} \; \overline{y}$, then the extra terms are absorbed and the 
simple Reynolds averaged form of the equations is recovered \citep{germano1992a}.

This document will use the familiar Reynolds averaging notation (overbars and primes),
but we intend these to denote filtering operations.  In other words,
$\overline{x'y'}$, for instance, should be interpreted as $\overline{x\,y} - \overline{x} \; \overline{y}$.

\section{Unclosed, higher-order prognostic equations in CLUBB}

Picture a drop of ink that is dropped into a jar of water.  Then imagine that the drop 
is carried around by swirling eddies in the jar.  
The concentration of ink is governed by the advection-diffusion equation.  
Analogously, one may think of a prognosed moment, such as total water variance $\overline{r_t'^2}$, 
as a scalar or ``ink" that is transported by a turbulent flow field.  However, the evolution 
of $\overline{r_t'^2}$ is influenced by several perhaps unfamiliar source and sink terms that arise from the influence
of turbulence on moments.  These prognostic moment equations are reviewed in this section.

First, we will list CLUBB's equations; then we will discuss the physics of various terms within those equations. 

\subsection{List of unclosed, prognostic equations}

\label{sec:list_unclosed_eqns}


The following equations are prognosed in CLUBB:

\begin{equation}
\label{eq_wprtp_unclosed}
\ptlder{\color{cp}\overline{w'r'_t}}{t} 
=  \underbrace{ - {\color{hp} \overline{w}}\ptlder{\color{cp} \overline{w'r'_t}}{z} }_\mathrm{mean\;adv} \
    \underbrace{ - \inverse{{\color{hp}\rho_s}}\ptlder{{\color{hp}\rho_s} {\color{ci}\overline{w^{'2}r'_t}}}{z} }_\mathrm{turb\;adv} \
    \underbrace{ - {\color{cd} \overline{w^{'2}} }\ptlder{\color{hp}\overline{r_t}}{z} }_\mathrm{turb\;prod} \
    \underbrace{ - {\color{cp} \overline{w'r'_t} } \ptlder{\color{hp}\overline{w}}{z} }_\mathrm{accum} \
    \underbrace{ + \frac{g}{\theta_{vs}} {\color{ci} \overline{r'_t\theta'_v}} }_\mathrm{buoy\;prod}  \
    \underbrace{ + {\color{ci} \overline{ {\color{cd} w'} {\color{hp} \left.\ptlder{ r_{t} }{ t }\right|_{\mathrm{mc}}'} }} }_\mathrm{microphys} \
    \underbrace{  - \frac{1}{{\color{hp}\rho_s}} {\color{ut} \overline{r'_t\ptlder{p'}{z}} } }_\mathrm{pressure}  \
    \underbrace{ - {\color{ut} \epsilon_{w r_t} } }_\mathrm{dissip} 
\end{equation}
\begin{equation}
\label{eq_wpthlp_unclosed}
\ptlder{\color{cp}\overline{w'\theta'_l}}{t}
=  \underbrace{ - {\color{hp} \overline{w} }\ptlder{\color{cp} \overline{w'\theta'_l}}{z} }_\mathrm{mean\;adv}	\
    \underbrace{ - \inverse{{\color{hp}\rho_s}}\ptlder{{\color{hp}\rho_s}{\color{ci} \overline{w^{'2}\theta'_l}}}{z} }_\mathrm{turb\;adv} \
    \underbrace{ - {\color{cd}\overline{w^{'2}}}\ptlder{\color{hp}\overline{\theta_l}}{z} }_\mathrm{turb\;prod}  \
    \underbrace{ - {\color{cp}\overline{w'\theta'_l}}\ptlder{\color{hp}\overline{w}}{z} }_\mathrm{accum}  \
    \underbrace{ + \frac{g}{\theta_{vs}} {\color{ci}\overline{\theta'_l\theta'_v}} }_\mathrm{buoy\;prod} \
    \underbrace{ +{\color{ci} \overline{{\color{cd} w'} {\color{hp} \left.\ptlder{\theta_l}{t}\right|_{\mathrm{mc}}'} }} }_\mathrm{microphys} \
     \underbrace{  - \frac{1}{{\color{hp}\rho_s}} {\color{ut} \overline{\theta'_l\ptlder{p'}{z}} } }_\mathrm{pressure} \ 
      \underbrace{ - {\color{ut} \epsilon_{w \theta_l} }  }_\mathrm{dissip} 
\end{equation}
\begin{equation}
\label{eq_upwp_unclosed}
\ptlder{\color{cp}\overline{u_h'w'}}{t} 
=  \underbrace{ - {\color{hp} \overline{w}}\ptlder{\color{cp} \overline{u_h'w'}}{z} }_\mathrm{mean\;adv} \
    \underbrace{ - \inverse{{\color{hp}\rho_s}}\ptlder{{\color{hp}\rho_s} {\color{ci}\overline{w^{'2}u_h'}}}{z} }_\mathrm{turb\;adv} \
    \underbrace{ - {\color{cd} \overline{w^{'2}} }\ptlder{\color{hp}\overline{u_h}}{z} }_\mathrm{turb\;prod} \
    \underbrace{ - {\color{cp} \overline{u_h'w'} } \ptlder{\color{hp}\overline{w}}{z} }_\mathrm{accum} \
    \underbrace{ + \frac{g}{\theta_{vs}} {\color{ci} \overline{u_h'\theta'_v}} }_\mathrm{buoy\;prod}  \
    \underbrace{  - \frac{1}{{\color{hp}\rho_s}} 
        \left( 
            {\color{ut} \overline{u_h'\ptlder{p'}{z}} 
            + \overline{w'\ptlder{p'}{x_h}}
            }
        \right) }_\mathrm{pressure}  \
    \underbrace{ - {\color{ut} \epsilon_{u_h w} } }_\mathrm{dissip} 
\end{equation}
\begin{equation}
\label{eq_rtp2_unclosed}
\ptlder{\color{cd}\overline{r_t^{'2}}}{t}
= \underbrace{ - {\color{hp} \overline{w}}\ptlder{{\color{cd}\overline{r^{'2}_t}}}{z} }_\mathrm{mean\;adv} \
   \underbrace{ - \inverse{{\color{hp}\rho_s}}\ptlder{{\color{hp}\rho_s}{\color{ci}\overline{w'r_t^{'2}}}}{z} }_\mathrm{turb\;adv} \
   \underbrace{ - 2{\color{cp} \overline{w'r'_t}} \ptlder{{\color{hp}\overline{r_t}}}{z} }_\mathrm{turb\;prod} \ 
    \underbrace{ +2 {\color{ci} \overline{ {\color{cd} r_t'} {\color{hp} \left.\ptlder{r_t}{ t }\right|_{\mathrm{mc}}'} }} }_\mathrm{microphys} \
   \underbrace{ - {\color{ut} \epsilon_{r_t r_t}}  }_\mathrm{dissip} 
\end{equation}
\begin{equation}
\label{eq_thlp2_unclosed}
\ptlder{\color{cd} \overline{\theta_l^{'2}}}{t}
= \underbrace{ - {\color{hp} \overline{w}}\ptlder{{\color{cd}\overline{\theta^{'2}_l}}}{z} }_\mathrm{mean\;adv} \
   \underbrace{ - \inverse{{\color{hp}\rho_s}}\ptlder{{\color{hp}\rho_s}{\color{ci}\overline{w'\theta_l^{'2}}}}{z} }_\mathrm{turb\;adv} \
   \underbrace{ - 2{\color{cp} \overline{w'\theta'_l}} \ptlder{{\color{hp}\overline{\theta_l}}}{z} }_\mathrm{turb\;prod} \
    \underbrace{ +2 {\color{ci} \overline{ {\color{cd} \theta_l'} {\color{hp} \left.\ptlder{\theta_l}{ t }\right|_{\mathrm{mc}}'} }} }_\mathrm{microphys} \
   \underbrace{ - {\color{ut} \epsilon_{\theta_l \theta_l}}  }_\mathrm{dissip} 
\end{equation}
\begin{equation}
\label{eq_rtpthlp_unclosed}
\ptlder{\color{cd}\overline{r'_t\theta'_l}}{t}
= \underbrace{ - {\color{hp}\overline{w}}\ptlder{\color{cd}\overline{r'_t\theta'_l}}{z} }_\mathrm{mean\;adv} \
    \underbrace{ - \inverse{{\color{hp}\rho_s}}\ptlder{{\color{hp}\rho_s}{\color{ci}\overline{w'r'_t\theta'_l}}}{z} }_\mathrm{turb\;adv} \
    \underbrace{ - {\color{cp} \overline{w'r'_t}}\ptlder{\color{hp}\overline{\theta_l}}{z} }_\mathrm{turb\;prod\;1} \
    \underbrace{ - {\color{cp} \overline{w'\theta'_l}}\ptlder{\color{hp}\overline{r_t}}{z} }_\mathrm{turb\;prod\;2} \
    \underbrace{ + {\color{ci} \overline{ {\color{cd} r_t'} {\color{hp} \left.\ptlder{\theta_l}{ t }\right|_{\mathrm{mc}}'} }} }_\mathrm{microphys\;1} \
    \underbrace{ + {\color{ci} \overline{ {\color{cd} \theta_l'} {\color{hp} \left.\ptlder{r_t}{ t }\right|_{\mathrm{mc}}'} }} }_\mathrm{microphys\;2} \
    \underbrace{  - {\color{ut} \epsilon_{r_t \theta_l}} }_\mathrm{dissip} 
\end{equation}
\begin{equation}
\label{eq_wp2_unclosed}
\ptlder{\color{cd}\overline{w^{'2}}}{t}  \
= \underbrace{ -{\color{hp} \overline{w}}\ptlder{\color{cd}\overline{w^{'2}}}{z} }_\mathrm{mean\;adv} \
   \underbrace{ - \inverse{{\color{hp}\rho_s}}\ptlder{{\color{hp}\rho_s}{\color{cd}\overline{w^{'3}}}}{z} }_\mathrm{turb\;adv}  \
   \underbrace{ - 2{\color{cd}\overline{w^{'2}}}\ptlder{\color{hp}\overline{w}}{z} }_\mathrm{accum} \
   \underbrace{ + \frac{2g}{\theta_{vs}} {\color{ci} \overline{w'\theta'_v}} }_\mathrm{buoy\;prod} \
   \underbrace{ - \frac{2}{{\color{hp}\rho_s}} {\color{ut} \overline{w'\ptlder{p'}{z}} } }_\mathrm{pressure}  \
   \underbrace{- {\color{ut} \epsilon_{ww} }}_\mathrm{dissip}
\end{equation}
\begin{equation}
\label{eq_wp3_unclosed}
\ptlder{\color{cd} \overline{w^{'3}}}{t}
= \underbrace{ - {\color{hp}\overline{w}}\ptlder{\color{cd}\overline{w^{'3}}}{z} }_\mathrm{mean\;adv} \
    \underbrace{ - \inverse{{\color{hp}\rho_s}}\ptlder{{\color{hp}\rho_s} {\color{ci}\overline{w^{'4}}}}{z} }_\mathrm{turb\;adv} \
    \underbrace{ + 3\frac{\color{cd}\overline{w^{'2}}}{{\color{hp}\rho_s}}\ptlder{{\color{hp}\rho_s}\overline{\color{cd}w^{'2}}}{z} }_\mathrm{turb\;prod} \
    \underbrace{ - 3{\color{cd} \overline{w^{'3}}}\ptlder{\color{hp}\overline{w}}{z} }_\mathrm{accum} \
    \underbrace{ + \frac{3g}{\theta_{vs}} {\color{ci}\overline{w^{'2}\theta'_v}} }_\mathrm{buoy\;prod} \
   \underbrace{ - \frac{3}{{\color{hp}\rho_s}}{\color{ut} \overline{ w^{'2} \ptlder{p'}{z} }  } }_\mathrm{pressure}  \
   \underbrace{ - { \color{ut} \epsilon_{www} }  }_\mathrm{dissip} 
\end{equation}
%
%
\begin{equation}
\label{eq_up2_unclosed}
\ptlder{\color{cd}\overline{u^{'2}}}{t}
= \underbrace{ - {\color{hp}\overline{w}}\ptlder{\color{cd}\overline{u^{'2}}}{z} }_\mathrm{mean\;adv} \
    \underbrace{ - \inverse{{\color{hp}\rho_s}}\ptlder{{\color{hp}\rho_s}{\color{ci}\overline{w'u^{'2}}}}{z} }_\mathrm{turb\;adv} \
    \underbrace{ - 2{\color{cp}\overline{u'w'}}\ptlder{\color{hp}\overline{u}}{z} }_\mathrm{turb\;prod} \
    \underbrace{   - \frac{2}{{\color{hp}\rho_s}} {\color{ut} \overline{u'\ptlder{p'}{x}} }  }_\mathrm{pressure}  \
    \underbrace{ - {\color{ut} \epsilon_{uu} }  }_\mathrm{dissip}  
\end{equation}
\begin{equation}
\label{eq_vp2_unclosed}
\ptlder{\color{cd}\overline{v^{'2}}}{t}
= \underbrace{ - {\color{hp}\overline{w}}\ptlder{\color{cd}\overline{v^{'2}}}{z} }_\mathrm{mean\;adv} \
    \underbrace{ - \inverse{{\color{hp}\rho_s}}\ptlder{{\color{hp}\rho_s}{\color{ci}\overline{w'v^{'2}}}}{z} }_\mathrm{turb\;adv} \
    \underbrace{ - 2{\color{cp}\overline{v'w'}}\ptlder{\color{hp}\overline{v}}{z} }_\mathrm{turb\;prod} \
    \underbrace{   - \frac{2}{{\color{hp}\rho_s}} {\color{ut} \overline{v'\ptlder{p'}{y}} }  }_\mathrm{pressure}  \
   \underbrace{ - {\color{ut} \epsilon_{vv} }  }_\mathrm{dissip}  
\end{equation}
%
%
%


The notation here is mostly described in Chapter \ref{chapt:cloud_param_problem}, but in addition,
$\theta_v$ denotes the virtual potential temperature (which is a measure of buoyancy), 
$\overline{\theta_{vs}}$ is a dry, base-state value of virtual potential temperature,
$p$ denotes pressure, and $\epsilon$ denotes turbulent dissipation. $u_h$ denotes either 
horizontal component of velocity (i.e.~$u$ or $v$). 

CLUBB's equations satisfy the anelastic approximation.  That is,
basic state density $\rho_s=\rho_s(z)$, and $\rho'$ only appears 
in terms that contain
the acceleration due to gravity, $g$.  CLUBB's vertical coordinate
is height (altitude), not pressure.  CLUBB's height coordinate must be
converted to pressure by the hydrostatic relationship when CLUBB
is implemented in a hydrostatic global model.

\subsection{The unclosed equations obey a reflectional symmetry}

\label{sec:parity_symm}

All the equations listed in Section \ref{sec:list_unclosed_eqns} satisfy a reflectional, 
i.e.~parity, symmetry.  That is, if the equation set has one solution, then there also exists
another, ``mirror-image" solution about the mid-point of the domain.  
For example, if the equations admit a solution with 
a warm thermal ($\theta'_v > 0$) that is rising ($w'>0$) from below ($z<0$), then they also admit
mirror-image solution with a cool thermal ($\theta'_v < 0$) that is sinking ($w'<0$) from above ($z<0$).
We mention this symmetry because we would like our parameterized equation set in Section \ref{sec:closed_eqns} 
to satisfy the same symmetry.

To pass from one solution to its mirror-image solution, make the following transformation in every equation:

\begin{equation}
   \begin{split}
      t & \rightarrow t \\
      \vec{x} & \rightarrow -\vec{x} \\
      \vec{u}' & \rightarrow -\vec{u}' \\
      \vec{\overline{u}} & \rightarrow -\vec{\overline{u}} \\
      p' & \rightarrow p' \\
      \rho_s & \rightarrow \rho_s \\
      \theta_v' & \rightarrow -\theta_v' \\
      \theta_l' & \rightarrow -\theta_l' \\
      \totder{\overline{\theta_l}}{z} & \rightarrow \totder{\overline{\theta_l}}{z} \\
      r_t' & \rightarrow -r_t' \\
      \totder{\overline{r_t}}{z} & \rightarrow \totder{\overline{r_t}}{z} \\
      \left.\ptlder{ u_h }{ t }\right|'_{\mathrm{mc}} & \rightarrow -\left.\ptlder{ u_h }{ t }\right|'_{\mathrm{mc}} \\
      \left.\ptlder{ \theta_l }{ t }\right|'_{\mathrm{mc}} & \rightarrow -\left.\ptlder{ \theta_l }{ t }\right|'_{\mathrm{mc}} \\
      \left.\ptlder{ r_t }{ t }\right|'_{\mathrm{mc}} & \rightarrow -\left.\ptlder{ r_t }{ t }\right|'_{\mathrm{mc}} \\
      \epsilon_{xx} & \rightarrow \epsilon_{xx} \\
      \epsilon_{xxx} & \rightarrow -\epsilon_{xxx}
   \end{split}
\nonumber
\end{equation}
\noindent
where $\epsilon_{xx}$ is a dissipation term in an even-order equation, such as a flux, and 
$\epsilon_{xxx}$ is a dissipation term in an odd-order equation, such as that for $\overline{w'^3}$.

The Reynolds-averaged equations listed in Section \ref{sec:list_unclosed_eqns} inherit this 
symmetry from the Navier-Stokes equations \citep[p. 17][]{frisch_1995_turb_book}.

\subsection{Main classes of terms in the prognostic equations}

\label{sec:classes_of_terms}

Let's speak first in broad-brush terms about the basic balance of processes in a turbulent layer 
that has reached a steady equilibrium.  Consider first a moment 
that does not contain $w'$, such as a scalar variance.  The moment has sources and sinks 
that may or may not balance each other.  
At the altitudes with an imbalance, 
turbulent advection transports the moment from the altitudes of excess source to the altitudes of 
excess sink.  In cumulus layers, this often means transporting excess variability from near cloud base
to near cloud top, a process that tends to deepen the cloud layer.  

Consider now the turbulent fluxes, which do include $w'$.  The fluxes obey a similar balance, 
except that they include a pressure term.  
This pressure term is usually a sink, because pressure perturbations tend to oppose the vertical motion of parcels.
In the equations for the velocity variances, the return-to-isotropy pressure term tends to take 
the ``excess" vertical motion $\overline{w'^2}$ generated by buoyant drafts and convert it to horizontal motions 
$\overline{u'^2}$ and $\overline{v'^2}$.  

The prognostic equation for a 2nd-order moment, $\overline{x'y'}$, is derived by adding two moment equations
\begin{equation}
    \ptlder{x'y'}{t} = x' \ptlder{y'}{t} + y' \ptlder{x'}{t}
\end{equation}
and applying a horizontal average.  The time tendencies of $x'$ and $y'$ are replaced by the corresponding
terms on the right-hand sides of these equations.  Therefore, each right-hand side term 
in a 2nd-order moment equation is the product of a perturbation field and a perturbation tendency.
Noting this helps us see how various tendencies, such as the pressure gradient tendency, 
are sprinkled throughout the equation set.  

The main categories of terms may be interpreted as follows:
\begin{itemize}

\hypertarget{url:categories_of_terms}{}

\item \textit{Mean vertical transport (``mean adv" or ``ma").}  Each of the prognosed moments is vertically advected by the mean vertical velocity, $\overline{w}$.  The mean advection is handled by a host model.  

\item \textit{Turbulent vertical transport (``turb adv" or ``ta").}  Turbulent advection represents vertical transport 
by updrafts and downdrafts, $w'$.  
The density-weighted vertical integral of turbulent advection is zero, 
aside from contributions from the top and bottom boundaries.  This means that turbulent advection 
is neither a net source nor a net sink: it merely tranports the prognosed variable from one altitude to another.  
It tends to move excess variability at one altitude to altitudes with less variability.


\item \textit{Turbulent production (``turb prod" or ``tp").}  In an equation for a 2nd-order moment $\overline{x'y'}$, 
the turbulent production term is a
turbulent flux multiplied by the vertical derivative of a mean field, 
$\overline{w'x'}\partial\overline{y}/\partial z + \overline{w'y'}\partial\overline{x}/\partial z$.  
This term tends to generate variability strongly when 
the vertical gradient is large, as in, e.g., cumulus (Cu) fields.  When the vertical gradient is large, a parcel
that is lifted finds itself surrounded by air with different properties.  This generates horizontal variability
at the new vertical level.

\item \textit{Buoyancy (``buoy prod" or ``bp").}  The buoyancy terms appear in the equation for any moment that contains $w'$.    
These terms are recognizable because they are prefixed by the acceleration due to gravity, $g$, 
and because they contain the perturbation virtual potential temperature, $\theta_v'$, 
which is an approximate indicator of buoyancy.  The buoyancy terms arise from the 
$g \theta_v'/\theta_{vs}$ term in the vertical velocity equation.  
Buoyancy causes parcels to rise or sink, which itself generates turbulence.

\item \textit{Microphysics (``microphys" or ``mc").}  Microphysical terms appear in the equation for any prognosed moment 
that contains $r_t'$ or $\theta_l'$.  The variables $r_t'$ and $\theta_l'$ do not include 
precipitation mixing ratio in their definitions, and
 hence they are not conserved with respect to the formation or evaporation of precipitation.  
Usually, the microphysical terms act as sinks of variability, but evaporation of rain near the ground 
can generate cold pools and hence act as a source of variability in temperature \citep{griffin_larson_2016_microphys_covar}.

\item \textit{Pressure (``pr").}  These terms appear in the equation for any moment that contains a velocity fluctuation, that is, $w'$, $u'$, or $v'$.  Pressure fluctuations act to equalize the intensity of all three components of velocity and also to damp
the motion of buoyant parcels.

\item \textit{Dissipation (``dissip" or ``dp").}  These terms represent the smoothing of variability by molecular diffusivity or viscosity.  
For instance, the dissipation of $r_t$ equals

\begin{equation}
      -\epsilon_{r_t r_t} \equiv -\kappa \overline{\Grad r_t' \cdot \Grad r_t'} ,
\end{equation}

where $\kappa$ is the molecular, kinematic diffusivity of total water.  Dissipation is always a sink and never a source.  
In strong turbulence, the off-diagonal dissipation terms (e.g., $\epsilon_{wr_t}$) are thought to be minor 
relative to pressure terms.  

\item \textit{Accumulation  (``accum" or ``ac").}  These terms are proportional to the vertical convergence of $\overline{w}$, 
$-\partial w/\partial z$.  They are small because the grid-scale vertical motion $\overline{w}$ is usually weak.  

\end{itemize}

By scanning the equations and noting the green terms, we notice that 3 kinds of terms 
involve integration over the subgrid PDF: turbulent advection, buoyancy terms, and 
microphysical terms.

\section{Example budgets from single-column simulations.}  

\begin{figure}[H]
\centering
\includegraphics[width=0.45\textwidth]
{\PathToDirMicroSrcVarCovar/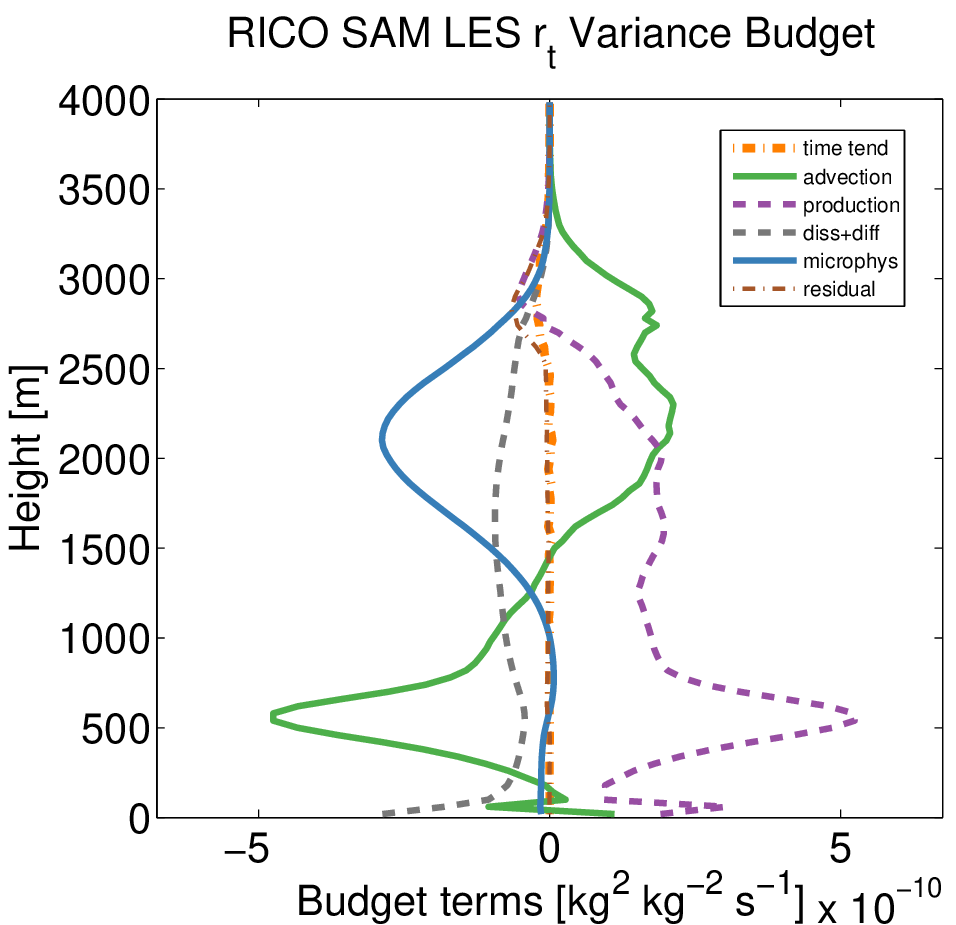}
\includegraphics[width=0.45\textwidth]
{\PathToDirMicroSrcVarCovar/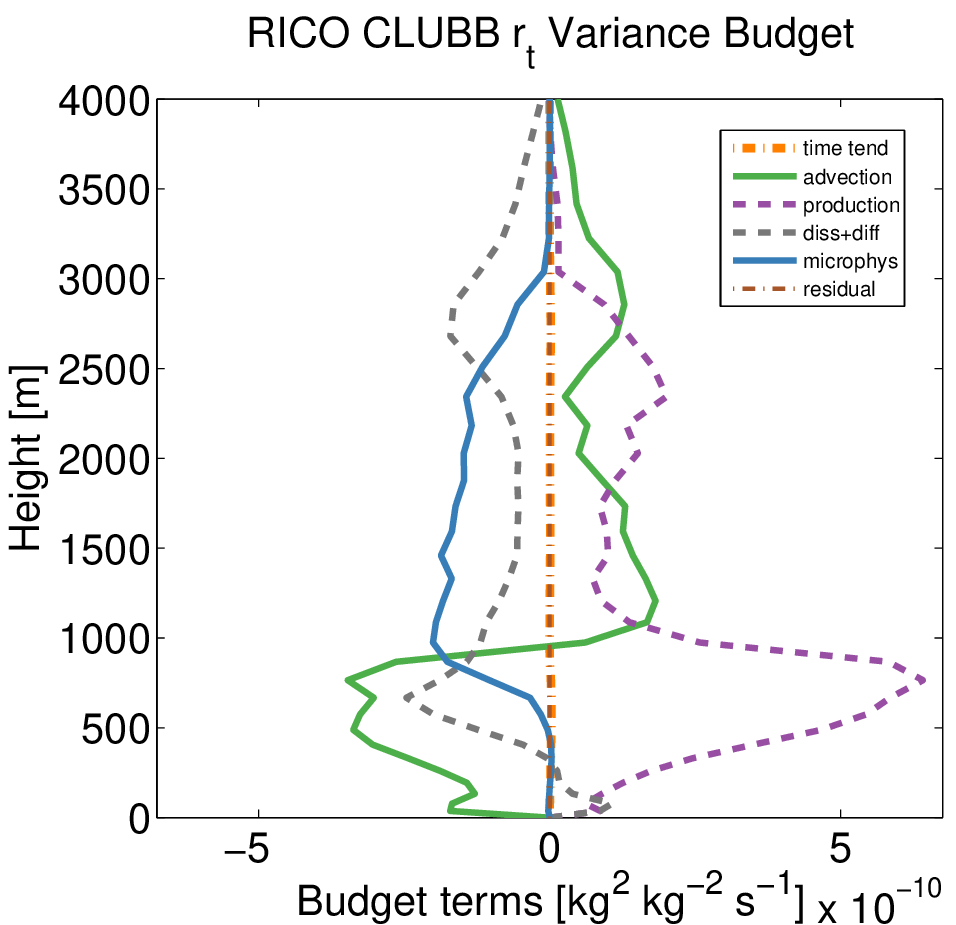}
\caption{   
Budget of $\overline{r_t'^2}$ as simulated by LES (left panel) and CLUBB-SILHS (right panel).  
A typical diagnostic closure for total water variance ($\overline{r_t'^2}$) balances
the source due to {\color{violet} turbulent production} with the sink due to {\color{gray} turbulent dissipation}
\citep[e.g.,][]{bogenschutz_krueger_2013_shoc}.
In fact, however, {\color{gray} turbulent dissipation} is weaker than {\color{ForestGreen} turbulent advection}
and {\color{Cerulean} microphysics} (see left panel).  
{\color{ForestGreen} Turbulent advection} transports the ``excess" $\overline{r_t'^2}$
produced near cloud base up to cloud top, thereby helping to deepen the cumulus layer.  {\color{Cerulean} Microphysics}
reduces variability in $r_t$ by removing cloud droplets where cloud water is large.
However, {\color{Cerulean} microphysical damping} is not an interchangeable substitute for {\color{gray} turbulent damping}.  
Rather, {\color{Cerulean} microphysical damping} kicks in only when rain forms, thereby producing a different feedback 
than {\color{gray} turbulent damping}.   
Figure reproduced from \citet{griffin_larson_2016_microphys_covar} (Creative Commons Attribution 3.0 License).
}
\label{fig:RICO_budgets_rtp2}
\end{figure}

\begin{figure}[H]
\centering
\includegraphics[width=0.45\textwidth]
{\PathToDirMicroSrcVarCovar/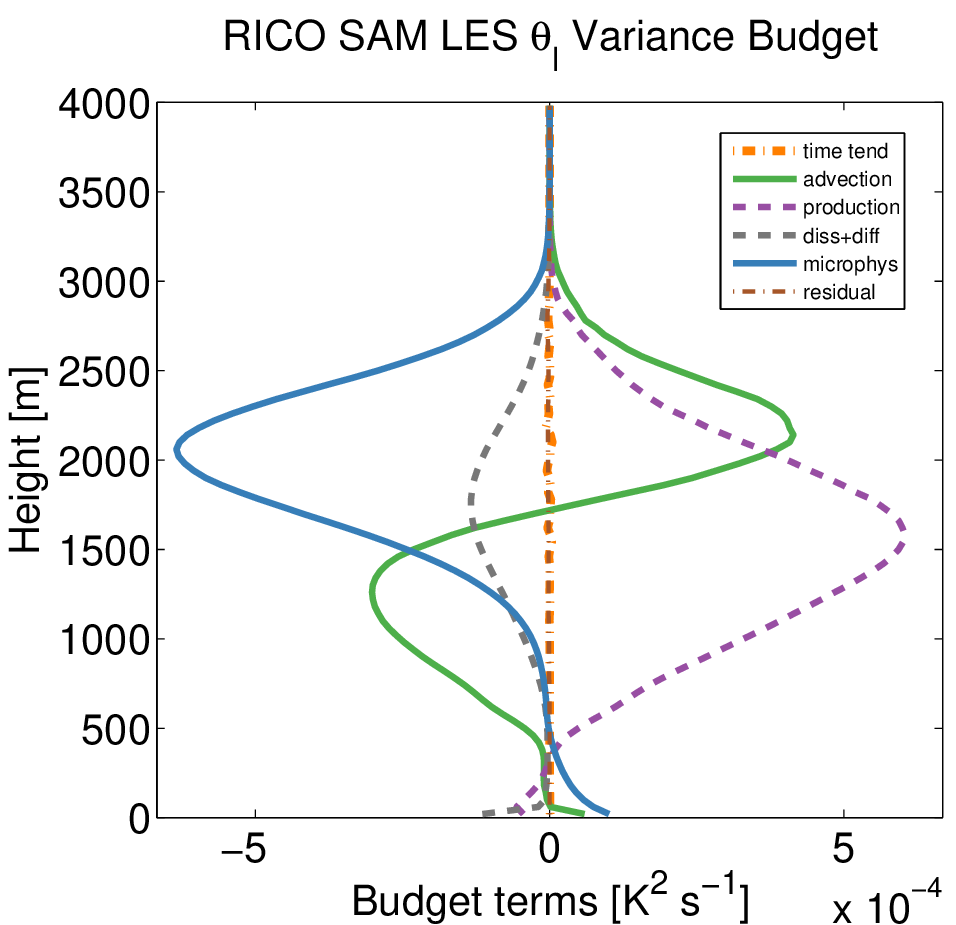}
\includegraphics[width=0.45\textwidth]
{\PathToDirMicroSrcVarCovar/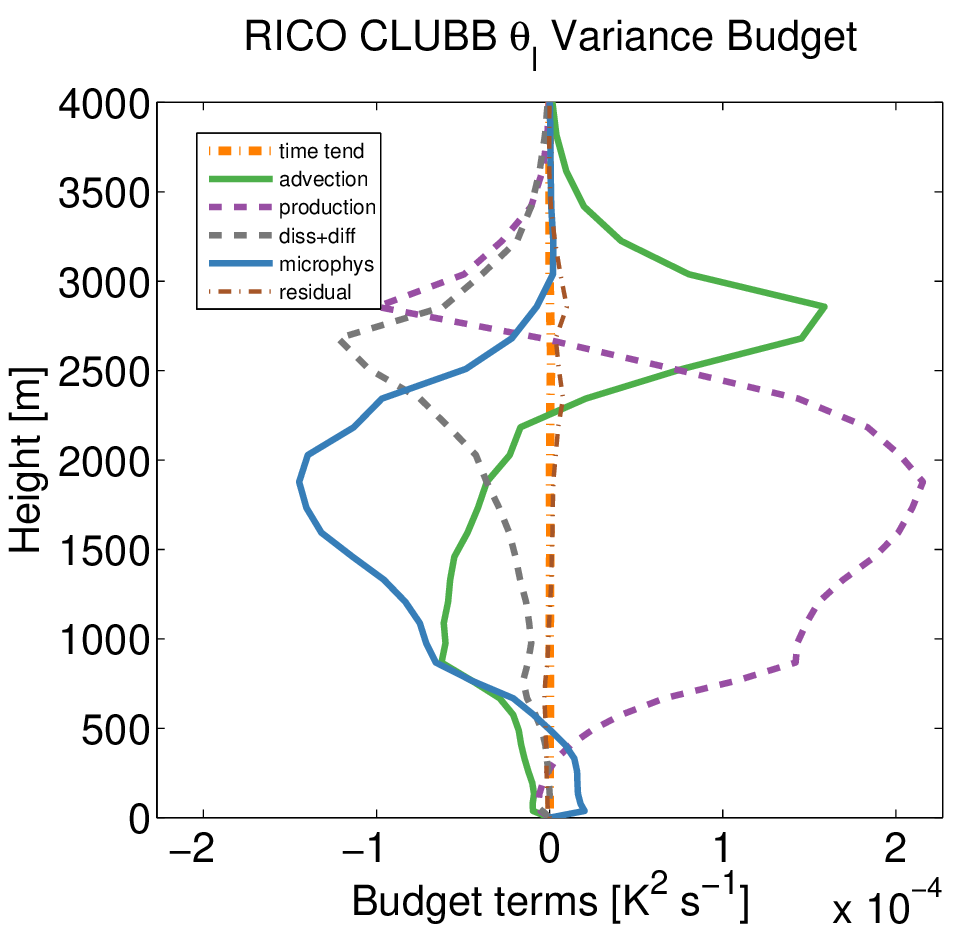}
\caption{
Budget of $\overline{\theta_l'^2}$ as simulated by LES (left panel) and CLUBB-SILHS (right panel).  
The budget of $\overline{\theta_l'^2}$ is similar in most respects to the budget of $\overline{r_t'^2}$
shown in Fig.~\ref{fig:RICO_budgets_rtp2}:  {\color{violet} turbulent production} is balanced 
not primarily by {\color{gray} turbulent dissipation}, but rather by a combination of
{\color{ForestGreen} turbulent advection} and {\color{Cerulean} microphysics} (see left panel).
One key difference is that the {\color{Cerulean} microphysical term} in the $\overline{\theta_l'^2}$
budget is \textit{positive} near the ground.  That is, microphysics \textit{creates} variability in temperature.
This indicates that cold pools are forming in the simulation.  The budget provides useful guidance on
the strength of the cold pool formation.
Figure reproduced from \citet{griffin_larson_2016_microphys_covar} (Creative Commons Attribution 3.0 License).
}
\label{fig:RICO_budgets_thlp2}
\end{figure}

\begin{figure}[H]
\centering
\includegraphics[width=0.45\textwidth]
{\PathToDirMicroSrcVarCovar/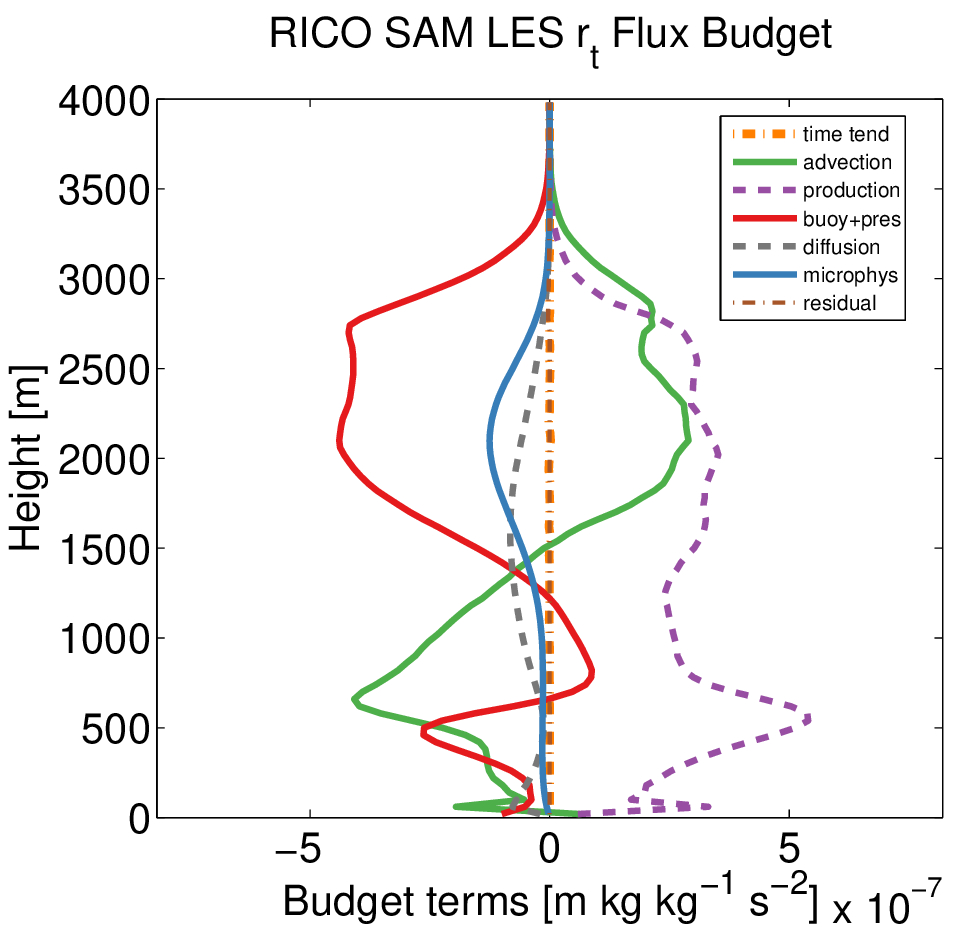}
\includegraphics[width=0.45\textwidth]
{\PathToDirMicroSrcVarCovar/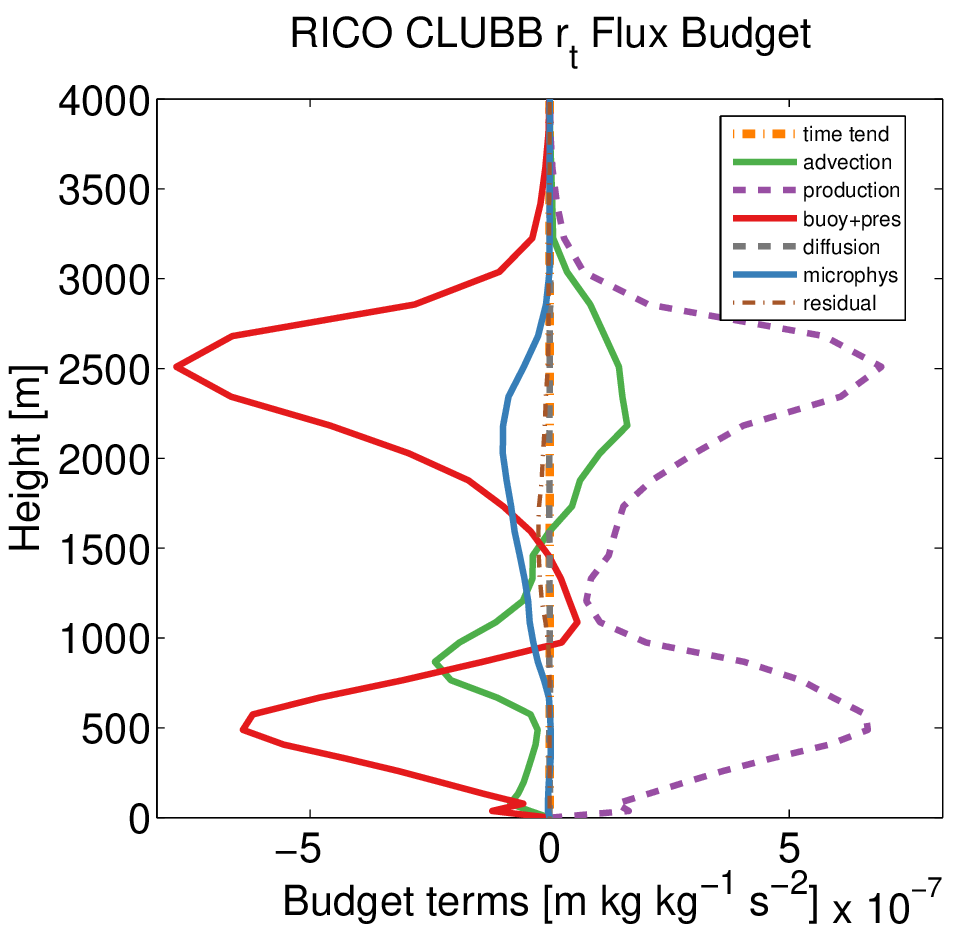}
\caption{
Budget of $\overline{w'r_t'}$ as simulated by LES (left panel) and CLUBB-SILHS (right panel).  
In the budget for $\overline{w'r_t'}$, the source due to {\color{violet} turbulent production} is not balanced
by {\color{gray} turbulent dissipation} or {\color{Cerulean} microphysical damping}, but rather 
{\color{ForestGreen} turbulent advection} and the sum of {\color{red} buoyancy and pressure}. 
The buoyancy term, which is proportional to $\overline{r_t'\theta_v'}$, is negative within cloud. 
Figure reproduced from \citet{griffin_larson_2016_microphys_covar} (Creative Commons Attribution 3.0 License).
}
\label{fig:RICO_budgets_wprtp}
\end{figure}

\begin{figure}[H]
\centering
\includegraphics[width=0.45\textwidth]
{\PathToDirMicroSrcVarCovar/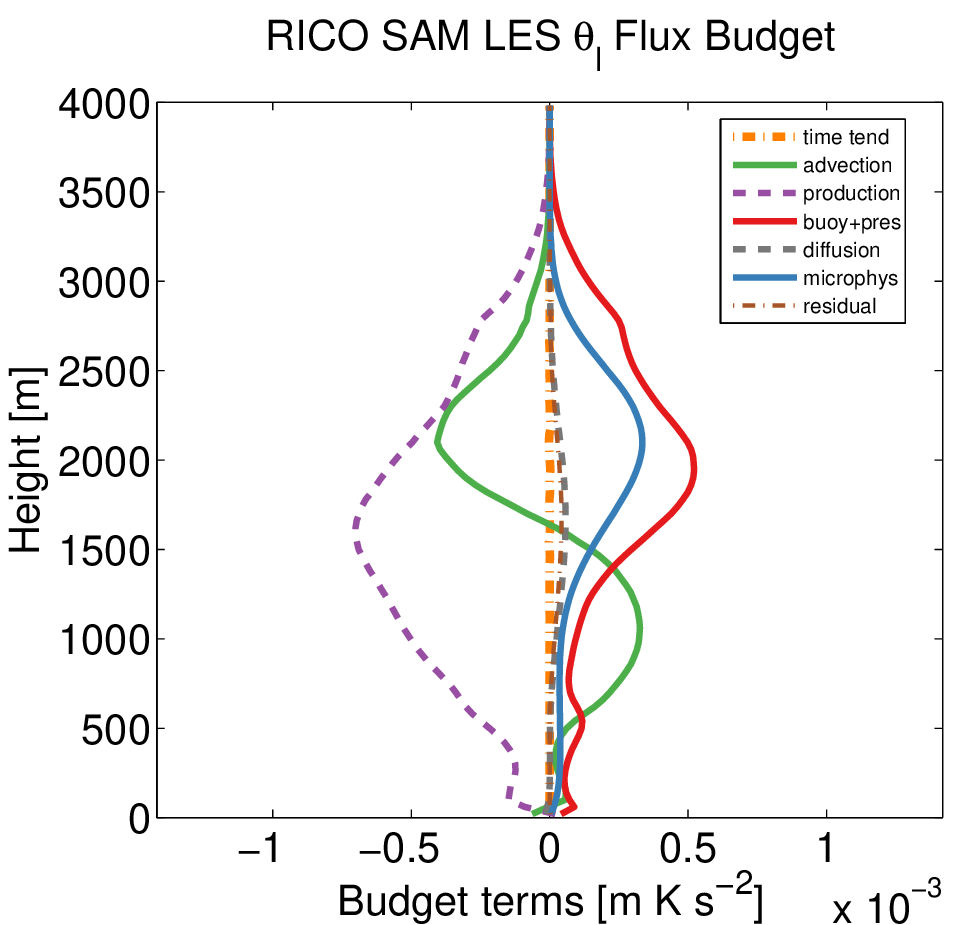}
\includegraphics[width=0.45\textwidth]
{\PathToDirMicroSrcVarCovar/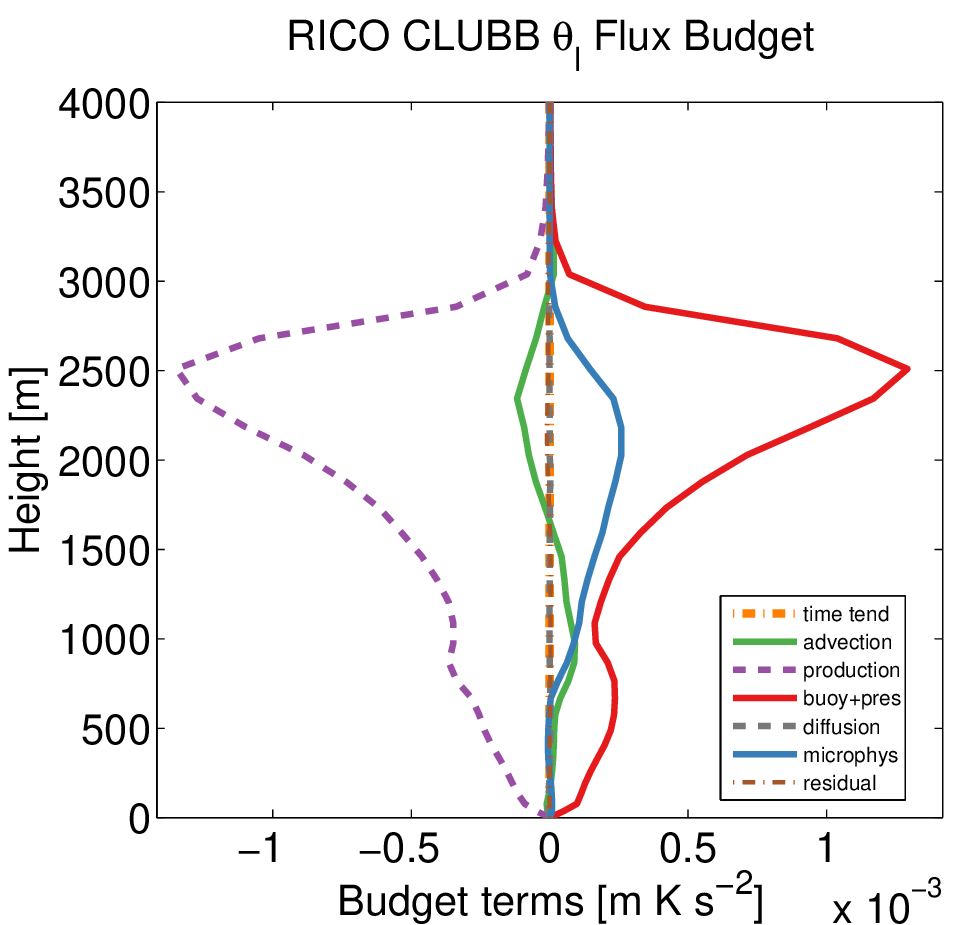}
\caption{
Budget of $\overline{w'\theta_l'}$ as simulated by LES (left panel) and CLUBB-SILHS (right panel).  
The budget for $\overline{w'\theta_l'}$ is a pseudo mirror image of the budget of $\overline{w'r_t'}$.
This is because $\theta_l$ and $r_t$ are negatively correlated.
Figure reproduced from \citet{griffin_larson_2016_microphys_covar} (Creative Commons Attribution 3.0 License).
}
\label{fig:RICO_budgets_wpthlp}
\end{figure}

In order to give more insight into how the higher-order budget terms behave, 
Figures~(\ref{fig:RICO_budgets_rtp2})-(\ref{fig:RICO_budgets_wpthlp}) display
some example budgets.  For example, the budget of $\overline{r_t'^2}$ is
the equation that governs the time rate of change of $\overline{r_t'^2}$, and it
includes all processes that change the value of $\overline{r_t'^2}$.  
Simulated in these figures is the RICO case of drizzling, shallow cumulus over the Atlantic
\citep{vanzanten_et_al_2011a}.  The left-hand panel of each budget shows a LES performed using the SAM model 
\citep{khairoutdinov_randall_03a}.  The right-hand panel shows output from CLUBB.  See the figure
captions for interpretation of the plots.

The budgets shown in Figures~(\ref{fig:RICO_budgets_rtp2})-(\ref{fig:RICO_budgets_wpthlp}) show both
interesting differences and interesting commonalities.  For instance, the major sink terms differ between budgets.  
In the turbulent flux budgets, the major sink is the sum of pressure and buoyancy.
In contrast, in the scalar variance budgets, the major sink term is the microphysical loss 
of cloud water through autoconversion and accretion.  
However, the budgets also have interesting similarities.  In all four budgets, the major source term is turbulent production.
And in all four budgets, turbulent advection is a major term: it acts to deepen the layer and hence is
vital for cumulus-layer formation, growth, and maintenance.

\section{Closures that do \textit{not} use CLUBB's PDF} 

\label{sec:closures_no_pdf}

CLUBB's PDF is \textit{not} used to derive terms involving pressure ($p'$), turbulent dissipation 
(e.g., $-\Grad x \cdot \Grad x$),
or the momentum fluxes, $\overline{u'w'}$ and $\overline{v'w'}$.  By closing these terms without consulting the PDF,
CLUBB is not inconsistently assuming one PDF for some processes 
and another PDF for other processes.
Rather, CLUBB merely makes no assumption at all about the PDF in the closures for $p'$, gradients, and $u'$ or $v'$.
CLUBB's PDF does not extend to those variables.

This section reviews CLUBB's closures that do not make use of a PDF: pressure, dissipation, and momentum.
For these terms, CLUBB employs standard parameterizations from the turbulence literature.

\subsection{Background on closure of pressure terms}


The parameterization of pressure perturbation terms can be illuminated by writing a Poisson equation for pressure. 
To do so, we write the equation for the $i$th component of perturbation velocity.  
This is obtained by subtracting the equation for the mean wind from the equation for the total wind 
\citep[][Eqn.~4.1.1]{stull1988a}:
\begin{equation}
   \ptlder{u_i'}{t} 
   + \overline{u_j} \ptlder{u_i'}{x_j} 
   + u_j' \ptlder{\overline{u_i}}{x_j}
   + u_j' \ptlder{u_i'}{x_j}
   - \ptlder{}{x_j} \overline{u_i' u_j'}
  = -\frac{1}{\rho} \ptlder{p'}{x_i}
    + g \frac{\theta_v'}{\theta_{ref}} \delta_{i3}  .
\end{equation}
Assume that the fluid is Boussinesq.  That is, we assume that buoyancy appears nowhere
in the fluid equations except alongside the acceleration due to gravity, $g$.  
Now exploit this assumption in order to  
pull $u_j'$ from the $u_j' \partial u_i'/\partial x_j$ term inside the derivative.  
Then, in order to eliminate the time derivative, 
take the divergence of this equation.  That is, apply $\partial/\partial x_i$ to all terms in the above equation. 
The result is \citep[e.g.,][]{gibson_launder_1978_press} or \citep[Eqn.~11.9,][]{pope_00a}:
\begin{equation}
-\frac{1}{\rho} \ptlsqd{p'}{x_i} = 
       \underbrace{
       \frac{\partial^2 
               \left[ u'_i u'_j - \overline{ u'_i u'_j } \right]
            }
            { \partial x_j \partial x_i }
                  }_{\;\; turbulence \;\;}
 + \underbrace{
   2 \ptlder{ \overline{u}_i }{ x_j } \ptlder{ u'_j }{ x_i }
              }_{\;\; mean \; strain \;\;}
 + \underbrace{ 
   \delta_{j3}\frac{ g }{ \theta_{ref} } \ptlder{ \theta_v' }{ x_j }
              }_{\;\; buoyancy \;\;} ,
                                        \label{eq:press_Poisson}
\end{equation}
\noindent
where $p'$ is a pressure perturbation and $\theta_v$ is the virtual potential temperature.  
To derive the mean strain term, we have assumed that the fluid is Boussinesq.  

Eqn.~\ref{eq:press_Poisson} may be interpreted by means of an analogy to electrostatics,
which obeys a similar equation \citep[Section 7.2,][]{houze_93a}.  
Each source term in Eqn.~\ref{eq:press_Poisson} is 
analogous to a distribution of electrical charge.  The pressure perturbation is analogous
to the electrostatic potential induced by that charge distribution.  

Eqn.~\ref{eq:press_Poisson} shows that the pressure term has three components that add linearly to form $p'$:
\begin{equation}
   p' = p'^{(s)} + p'^{(r)} + p'^{(b)} , 
\end{equation}
\noindent
where $p'^{(s)}$ is the ``slow" term, so-called because it responds only indirectly to changes in the mean wind; 
$p'^{(r)}$ is the ``rapid" term, so-called because it responds directly to changes in the mean wind, $\overline{u_i}$; 
and $p'^{(b)}$ is the buoyancy term.  One may solve the Poisson equation separately for each source term 
on the right-hand side, obtaining $p'^{(s)}$, $p'^{(r)}$, and $p'^{(b)}$ separately and directly. 
The full pressure perturbation is then formed by adding the contributions.  However, each of the three components 
of the pressure fluctuation may be conceptualized separately, which is illuminating.  

Unfortunately, the problem is not as simple as parameterizing the pressure fluctuation itself.
Instead, the equations we must close (see Section \ref{sec:list_unclosed_eqns}) 
contain terms that include the covariance of pressure fluctuations and other fields, such as wind.  
They have the form
\begin{equation}
            \overline{  
                       \frac{p'^{(x)}}{\rho} 
       \left( \ptlder{u_i'}{x_j} + \ptlder{u_j'}{x_i} \right) 
                     } ,
\end{equation}
\noindent
where $x$ denotes either the slow, rapid, or buoyancy component of pressure.

Consider first the slow term (the first term on the right-hand side), which contains only turbulent fluctuations.  
It is reasonable to assume that the effect of pressure perturbations on turbulence fluctuations 
is to make them more isotropic.  That is, pressure acts to make all components of turbulence 
--- $\overline{u'^2}$, $\overline{v'^2}$, and $\overline{w'^2}$ --- more similar to each other in magnitude.  
Based on this idea, \citet{rotta_1951_return_iso} proposed \citep[see also Eqn.~(11.24) of][]{pope_00a}:
\begin{equation}
            \overline{  
                       \frac{p'^{(s)}}{\rho} 
       \left( \ptlder{u_i'}{x_j} + \ptlder{u_j'}{x_i} \right) 
                     }
         = - \frac{C_s}{\tau}
    \left( 
        \overline{u_i' u_j'} - \frac{2}{3} \overline{e} \delta_{ij}
    \right)
\end{equation}
\noindent
Since $\overline{e} = 0.5 ( \overline{u'^2} + \overline{v'^2} + \overline{w'^2} )$, we can see that 
if the turbulence is isotropic, then $\overline{u'^2} = \overline{v'^2} = \overline{w'^2} = (2/3) \overline{e}$.  
In this case, the return-to-isotropy term is zero. This is expected, because in this case the turbulence 
is already isotropic. Also, we see that return to isotropy drives non-diagonal terms ($i \ne j$) toward zero.


The rapid term can be parameterized as \citep{gibson_launder_1978_press}:
\begin{equation}
                  \overline{  
                       \frac{p'^{(r)}}{\rho} 
       \left( \ptlder{u_i'}{x_j} + \ptlder{u_j'}{x_i} \right) 
                     }
      = -C_r \left( 
              P_{ij} - \frac{2}{3} \delta_{ij} P
             \right)
\end{equation}
\noindent
where 
\begin{equation}
    P \equiv -\overline{u_i' u_k'} \ptlder{\overline{u_i}}{x_k} 
\end{equation}
\noindent
and
\begin{equation}
    P_{ij} \equiv 
      -\overline{u_i' u_k'} \ptlder{\overline{u_j}}{x_k} 
      -\overline{u_j' u_k'} \ptlder{\overline{u_i}}{x_k}. 
\end{equation}
Analogously, the buoyancy term can be parameterized as
\begin{equation}
                  \overline{  
                       \frac{p'^{(b)}}{\rho} 
       \left( \ptlder{u_i'}{x_j} + \ptlder{u_j'}{x_i} \right) 
                     }
      = -C_b \left( 
              G_{ij} - \frac{2}{3} \delta_{ij} G
             \right)
\end{equation}
\noindent
where
\begin{equation}
    G \equiv - \frac{g_i}{\theta_{ref}}
       \overline{u_i' \theta_v'} 
\end{equation}
\noindent
and
\begin{equation}
    G_{ij} \equiv - \frac{1}{\theta_{ref}}
      \left(
       g_j \overline{u_i' \theta_v'} 
     + g_i \overline{u_j' \theta_v'} 
      \right)
\end{equation}



\subsection{Background on closure of dissipation terms}

The higher-order moment equations contain dissipation terms that act as sinks of variability. 
Consider, for example, the scalar dissipation of $\overline{r_t'^2}$ \eqref{eq_rtp2_unclosed} \citep{pope_00a}:
\begin{equation}
   -\epsilon_{r_t r_t}  
   = - \kappa_{r_t} \Grad r_t \cdot \Grad r_t  =  - \kappa_{r_t} \left| \Grad r_t \right|^2, 
\end{equation}
where $\kappa_{r_t}$ is a ``molecular" diffusivity of total water.   By inspection, we see that the scalar dissipation
is always either negative or zero.  Dissipation occurs when thin filaments of fluid are homogenized by molecular diffusivity.
The filaments are created by turbulent mixing and stretching, which forms ever finer length scales 
in a downward cascade from the turbulence kinetic energy-producing (TKE-producing) scales.  Hence the scalar dissipation can be parameterized in terms
of properties of the large eddies, independently of the value of the molecular diffusivity.  
CLUBB uses a standard parameterization \citep[e.g.,][]{pope_00a}:
\begin{equation}
   -\epsilon_{r_t r_t}  
   = - \kappa_{r_t} \Grad r_t \cdot \Grad r_t  
   \approx  - \dfrac{C}{\tau} \overline{r_t^{'2}} , 
\end{equation}
where $\tau$ is an eddy turnover time scale.
The dissipation rate is stronger when the variability in $r_t$, $\overline{r_t^{'2}}$, is greater 
or when the turbulent time scale, $\tau$, is shorter.  The time scale $\tau$ is \citep{golaz_et_al_02a}:
\begin{equation}
\label{eq_tau}
\tau = \left\{
\begin{array}{ll}
\displaystyle \frac{L}{\sqrt{\, \overline{e} \,}}; 
& L / \sqrt{\, \overline{e} \,} \leqslant \tau_{\rm max} \\
\displaystyle \tau_{\rm max};
&  L / \sqrt{\, \overline{e} \,} > \tau_{\rm max}
\end{array}
\right. \hbox{ .}
\end{equation}
$L$ is CLUBB's turbulent mixing length scale, which is, roughly speaking, the distance between a buoyant parcel 
and its level of neutral buoyancy.  Hence, $L$ becomes small in stably stratified layers (even highly sheared ones),
and $\tau$ inherits this property, damping turbulent fields in stable stratification.  
$L$ is described in more detail in the next section.  
The turbulence kinetic energy, $\overline{e}$, is a function of the vertical 
velocity variance $\overline{w^{'2}}$, 
north-south wind variance $\overline{v^{'2}}$, and 
east-west wind variance $\overline{u^{'2}}$:
\begin{equation}
\label{eq_aniso_tke}
\overline{e} \equiv \frac{1}{2} 
  \left( 
    \overline{w^{'2}}+\overline{u^{'2}}+\overline{v^{'2}}
  \right).
\end{equation}
%
Hence $\tau$ may be interpreted as the time required for a parcel traveling with a velocity 
$\sqrt{\overline{e}}$ to reach its level of neutral buoyancy.  When the eddies are small and/or turbulence
is strong, $\tau$ is short and dissipation is strong.

\subsection{How CLUBB's turbulent mixing length scale is calculated in CAM6 and E3SMv2} 

\label{sec:mixing_length}

CLUBB contains a single turbulent length scale, $L$, that appears throughout CLUBB.  $L$ is used to compute CLUBB's 
turbulent time scale, $\tau$ (see Eqn.~\eqref{eq_tau}), which appears in several dissipation and pressure terms.  
Through $\tau$, large values of $L$ allow turbulence to grow easily; small values of $L$ suppress turbulence.
$L$ also appears in CLUBB's eddy diffusivity, $K$, which is used to supply background
numerical smoothing.  $K$ can also be used optionally in CLUBB to diagnose momentum fluxes.

$L$ is a buoyant length scale whose formulation is inspired by \citet{bougeault_andre_86a}.
Roughly speaking, in CLUBB, at a given grid level, a parcel starts with the TKE at that level 
and moves upward under the action of buoyancy 
until the initial TKE is exhausted.  During transit, the parcel undergoes entrainment and may release latent heat.  
Using a similar calculation,
a parcel is moved downward.  The upward and downward lengths are averaged by a suitable formula.  
For details, see \citet{golaz_et_al_02a}.

Because $L$ is based on buoyancy, it has small values in stable stratification, 
and in that case it suppresses vertical
motions.  In contrast, in shallow cumulus layers, $L$ has values of several hundred meters.  

At horizontal grid spacings less than a kilometer or so, the length scale is truncated 
to one-quarter the horizontal grid spacing in order to damp subgrid turbulence 
and permit resolved motions to dominate \citep{larson_2012_2km_16km}.

\subsection{A new option to calculate CLUBB's turbulent mixing length scale}

CLUBB's old length scale is somewhat expensive and inflexible.  CLUBB's new approach is
to diagnose the turbulent time scale, $\tau$, directly.  CLUBB's new $\tau$ includes
the effects of not only buoyancy, but also shear.  In fact, different time scales
are calculated for fluxes and variances.  Use of multiple time scales 
is more appropriate than the use of a single time scale in 
stably stratified inversions above cloud layers. 
For details, see \citet{guo_et_al_2021_taus}.

\subsection{Summary list of pressure and dissipation closures}

This section lists specific pressure and dissipation closures used in CLUBB, with
labeling of the terms described \hyperlink{url:categories_of_terms}{here}, and some of the 
constants listed in Table \ref{tab:tuning_parameters_in_clubb}.  The pressure closures in
CLUBB include slow (i.e.~return to isotropy), fast (i.e.~mean field), and buoyancy terms.  The dissipation
closures in CLUBB are zero for off-diagonal terms in the dissipation tensor.

\hypertarget{url:wpxp_pr}{}

The pressure and dissipation terms in the turbulent vertical flux equations
for ${\color{cp} \overline{w'r_t'}}$ (Eqn.~\ref{eq_wprtp_unclosed})
and for ${\color{cp} \overline{w'\theta_l'}}$ (Eqn.~\ref{eq_wpthlp_unclosed})
are closed as follows \citep[][Eqn.~20]{golaz_et_al_02a}:
\begin{equation}
    \underbrace{  - \frac{1}{{\color{hp}\rho_s}} {\color{ut} \overline{r'_t\ptlder{p'}{z}} } }_\mathrm{pressure}  
    \approx      \underbrace{ - \frac{C_6}{\color{cd}\tau}{\color{cp} \overline{w'r'_t} } }_{pr1} \
    \underbrace{ + C_7 {\color{cp} \overline{w'r'_t}} \ptlder{\color{hp}\overline{w}}{z} }_{pr2}  \
    \underbrace{ - C_7 \frac{g}{\theta_{vs}} {\color{ci}\overline{r'_t\theta'_v}} }_{pr3}  
\end{equation}
\begin{equation}
    \underbrace{ - {\color{ut} \epsilon_{w r_t} } }_\mathrm{dissip} 
   \approx
   0 .
\end{equation}
The pressure and dissipation terms in the equation
for ${\color{cp} \overline{u_h' w'}}$ (Eqn.~\ref{eq_upwp_unclosed})
is closed as:
\begin{equation}
    \underbrace{  - \frac{1}{{\color{hp}\rho_s}} {\color{ut} \overline{u'_h\ptlder{p'}{z}} } }_\mathrm{pressure}  
    \approx      \underbrace{ - \frac{C_6}{\color{cd}\tau}{\color{cp} \overline{u'_h w'} } }_{pr1} \
    \underbrace{ + C_7 {\color{cp} \overline{u'_h w'}} \ptlder{\color{hp}\overline{w}}{z} }_{pr2}  \
    \underbrace{ - C_7 \frac{g}{\theta_{vs}} {\color{ci}\overline{u'_h \theta'_v}} }_{pr3} \
    \underbrace{ + C_7 {\color{cp} \overline{w'^2}} \ptlder{\color{hp}\overline{u_h}}{z} }_{pr4}  \
\end{equation}
\begin{equation}
    \underbrace{ - {\color{ut} \epsilon_{u_h w} } }_\mathrm{dissip} 
   \approx
   0 .
\end{equation}
\hypertarget{url:xp2_dp}{}
The dissipation terms in the scalar variance equations
for ${\color{cd} \overline{r_t'^2}}$ (Eqn.~\ref{eq_rtp2_unclosed}),
${\color{cd} \overline{\theta_l'^2}}$ (Eqn.~\ref{eq_thlp2_unclosed}), and
${\color{cd} \overline{r_t' \theta_l'}}$ (Eqn.~\ref{eq_rtpthlp_unclosed})
are closed as follows \citep[][Eqn.~24b]{golaz_et_al_02a}:
\begin{equation}
   \underbrace{ - {\color{ut} \epsilon_{r_t r_t}}  }_\mathrm{dissip} 
   \approx  \underbrace{ - \dfrac{C_2}{\color{cd}\tau} \left(   {\color{cd} \overline{r_t^{'2}} }  
                               - r_{t,\rm{tol}}^{2} \right) }_{dp1} . 
\end{equation}
\hypertarget{url:wp2_pr}{}
The pressure and dissipation terms in the variance equation for ${\color{cd} \overline{w'^2}}$ (Eqn.~\ref{eq_wp2_unclosed})
are closed as follows \citep[][Eqn.~19]{golaz_et_al_02a}:
\begin{equation}
   \underbrace{ - \frac{2}{{\color{hp}\rho_s}} {\color{ut} \overline{w'\ptlder{p'}{z}} } }_\mathrm{pressure}  
   \approx
   \underbrace{ - \frac{C_4}{\color{cd} \tau} \left( {\color{cd} \overline{w^{'2}}} -\frac{2}{3}\overline{e} \right) }_{pr1} \
   \underbrace{ 
       +C^\mathrm{uu}_\mathrm{shr} 2{\color{cd}       \overline{w^{'2}}}\ptlder{\color{hp}\overline{w}}{z}
       - C^\mathrm{uu}_\mathrm{buoy} \frac{2g}{\theta_{vs}} {\color{ci} \overline{w'\theta'_v}}
      }_{pr2}  \
   \underbrace{ 
     + \frac{2}{3} C^\mathrm{uu}_\mathrm{buoy}
                \frac{g}{\theta_{vs}} {\color{ci} \overline{w'\theta'_v} } 
     + \frac{2}{3} C^\mathrm{uu}_\mathrm{shr}
     \left(
       - {\color{cp} \overline{u'w'}}\ptlder{\color{hp}\overline{u}}{z} 
       - {\color{cp} \overline{v'w'}}\ptlder{\color{hp}\overline{v}}{z} 
     \right) }_{pr3} .
                                                \label{eq:wp2_pr_closure}
\end{equation}
and \citep[][Eqn.~24a]{golaz_et_al_02a}:
\begin{equation}
   \underbrace{- {\color{ut} \epsilon_{ww} }}_\mathrm{dissip}
   \approx
   \underbrace{ - \dfrac{C_1}{\color{cd} \tau} \left( {\color{cd}  \overline{w^{'2}}} 
                              - w_{\rm{tol}}^{2} \right) }_{dp1} .
\end{equation}

\hypertarget{url:up2_pr}{}
The pressure and dissipation terms in the variance equations
            for ${\color{cd} \overline{u'^2}}$ (Eqn.~\ref{eq_up2_unclosed}) and
            for ${\color{cd} \overline{v'^2}}$ (Eqn.~\ref{eq_vp2_unclosed})
are closed as follows \citep[][Eqn.~15]{andre_et_al_78a}:
\begin{equation}
     \underbrace{   - \frac{2}{{\color{hp}\rho_s}} {\color{ut} \overline{u'\ptlder{p'}{x}} }  }_\mathrm{pressure}  
     \approx
    C^\mathrm{uu}_\mathrm{shr} 2{\color{cp}\overline{u'w'}}\ptlder{\color{hp}\overline{u}}{z} 
    \underbrace{ 
      + \frac{2}{3} C^\mathrm{uu}_\mathrm{buoy}
        \frac{g}{\theta_{vs}}  {\color{ci}\overline{w'\theta'_v}} 
      + \frac{2}{3} C^\mathrm{uu}_\mathrm{shr}
      \left(
        - {\color{cp}\overline{u'w'}}\ptlder{\color{hp}{\overline{u}}}{z} 
        -  {\color{cp} \overline{v'w'}}\ptlder{\color{hp}\overline{v}}{z} 
      \right) }_{pr2} \
    \underbrace{ - \dfrac{C_4}{\color{cd}\tau} \left( {\color{cd}\overline{u^{'2}}} - \frac{2}{3} {\color{cd}\overline{e}} \right) }_{pr1} 
\end{equation}
%
%
\begin{equation}
    \underbrace{ - {\color{ut} \epsilon_{uu} }  }_\mathrm{dissip}  
    \approx
    \underbrace{ - \frac{2}{3} 
                  C_{14} \frac{\color{cd}\overline{e}}{\color{cd}\tau}
               }_{dp1}  .
\end{equation}
%

%
\hypertarget{url:wp3_pr}{}
The pressure and dissipation terms in the third-order moment equation
            for ${\color{cd} \overline{w'^3}}$ (Eqn.~\ref{eq_wp3_unclosed})
are closed as follows \citep[][Eqn.~23]{golaz_et_al_02a}:
\begin{equation}
       \underbrace{ - \frac{3}{{\color{hp}\rho_s}}{\color{ut} \overline{ w^{'2} \ptlder{p'}{z} }  } }_\mathrm{pressure}  
       \approx
   \underbrace{ - \frac{C_8}{\color{cd}\tau} {\color{cd}\overline{w^{'3}}} }_{pr1}
    \underbrace{ - C_{11} \left(
                 - 3 {\color{cd} \overline{w^{'3}}}\ptlder{\color{hp} \overline{w}}{z}
                 + \frac{3g}{\theta_{vs}} {\color{ci} \overline{w^{'2}\theta'_v} }
             \right) }_{pr2} 
    \underbrace{ - C_{15} K_m \ptlder{}{z} \left( 
                           \frac{g}{\theta_{vs}} {\color{ci} \overline{w'\theta'_v}}  
                        -  {\color{cp}\overline{u'w'}} \ptlder{\color{hp}\overline{u}}{z}
                        -  {\color{cp}\overline{v'w'}} \ptlder{\color{hp}\overline{v}}{z}
                        \right)
               }_{pr\_turb} 
                                                  \label{eq:wp3_pr_closure}
\end{equation}
%
%
\begin{equation}
   \underbrace{ - { \color{ut} \epsilon_{www} }  }_\mathrm{dissip} 
   \approx
    0  .
\end{equation}

\subsection{A heuristic rationalization of the $pr\_turb$ term}

Equation \eqref{eq:wp3_pr_closure} contains a term whose formulation warrants extra discussion:
\begin{equation}
  \left. - \frac{3}{{\color{hp}\rho_s}}{\color{ut} \overline{ w^{'2} \ptlder{p'}{z} }  } \right|_{pr\_turb}
      = - C_{15} K_m \ptlder{}{z} \left( 
                            \frac{g}{\theta_{vs}} {\color{ci} \overline{w'\theta'_v}}  
                         -  {\color{cp}\overline{u'w'}} \ptlder{\color{hp}\overline{u}}{z}
                         -  {\color{cp}\overline{v'w'}} \ptlder{\color{hp}\overline{v}}{z}
                                 \right) .
                                                  \label{eq:bp2_closure}
\end{equation}
This term appears in the $\overline{w'^3}$ equation \eqref{eq_wp3}, and it parameterizes a pressure
effect.    
The term contains an eddy diffusivity, $K_m$, and is
of the form typically used to model a down-gradient turbulent flux:
\begin{equation}
   \overline{w'x'} = - K_m \ptlder{\overline{x}}{z}.
                          \label{eq:proto_K_diff}
\end{equation}
To obtain the form \eqref{eq:bp2_closure}, we let
\begin{equation}
    \overline{x} \equiv \left. \overline{w' \ptlder{p'}{z} } \right|_{pr\_turb}
                 \propto -\left( \frac{g}{\theta_{vs}} {\color{ci} \overline{w'\theta'_v}}  
                         -  {\color{cp}\overline{u'w'}} \ptlder{\color{hp}\overline{u}}{z}
                         -  {\color{cp}\overline{v'w'}} \ptlder{\color{hp}\overline{v}}{z} \right)
                                 \label{eq:xbar_def}
\end{equation}
from Eqn. \eqref{eq:wp2_pr_closure}.  We also let
\begin{equation}
    x' \equiv \left. w' \ptlder{p'}{z} \right|_{pr\_turb}
\end{equation}
so that
\begin{equation}
   \overline{w'x'} \equiv \overline{ w'\left(\left. w'\ptlder{p'}{z} \right|_{pr\_turb} \right) } 
              = \left. \overline{w'^2 \ptlder{p'}{z} } \right|_{pr\_turb}.
                            \label{eq:wpxp_def}
\end{equation}
Combining equations \eqref{eq:proto_K_diff}, \eqref{eq:xbar_def}, and \eqref{eq:wpxp_def} yields \eqref{eq:bp2_closure}, 
up to a constant.  Clearly this is merely a plausibility argument, but it conveys the physical reasoning 
behind the formulation.

This term turns out to play an important role in generating positive $\overline{w'^3}$ in subcloud layers.
Without this term, there is little preference for positive or negative $\overline{w'^3}$
in layers with no cloud.  This is a consequence of the reflectional symmetry 
of the $\overline{w'^3}$ equation (see Section \ref{sec:parity_symm})
and the fact that the lower boundary condition for the $\overline{w'^3}$ equation in CLUBB is 
$\left. \overline{w'^3} \right|_{z=0} = 0$.  In particular, in a dry layer, CLUBB's parameterization of 
the buoyancy term, $\overline{w'^2\theta_v'}$,
does not determine the sign of $\overline{w'^3}$ but instead has whatever sign $\overline{w'^3}$ has 
(see Eqn. \ref{eq_wp2thlp}).  However, typically both 
$\overline{w'\theta_v'}$ and $-\overline{u'w'}\partial \overline{u}/\partial z$ 
decrease with increasing altitude in clear, convective layers.  That will lead to positive generation 
of $\overline{w'^3}$, if $C15 > 0$. 

We note parenthetically that \citet{moeng_rotunno_1990_skewness} show that skewness is \textit{negative} near the lower
boundary in Rayleigh-Benard convection.  However, in this case, $\overline{w'\theta'}$ \textit{increases} with increasing
height in the viscous sublayer at the lower boundary, which is still consistent with CLUBB's parameterization.

\section{Closures that \textit{do} use CLUBB's PDF}

The use of a subgrid PDF to diagnose cloud fraction, cloud liquid water mixing ratio, 
and various moments has a long history \citep[e.g.,][]{sommeria_deardorff_77a, mellor_77a, 
smith_90a, lewellen_yoh_93a, tompkins_02a}.  A subgrid PDF has also been used
to close moments that include $w$ \citep[e.g.,][]{pope_1985_vel_scalar_pdf,randall_et_al_92a,lappen_randall_01a,
lappen_randall_01b,lappen_randall_01c}.  Subgrid PDFs also have a long history in combustion research
\citep[e.g.,][]{obrien_80a,frankel_et_al_93a,bray_libby_94a,cook_riley_94a}.

CLUBB's PDF is used to close a variety of terms.  These include buoyancy ($\theta_v$) terms, 
such as $\overline{w'\theta_v'}$; third-order turbulent terms, such as $\overline{w'r_t'^2}$;
one fourth-order term, $\overline{w'^4}$; various mean microphysical terms, such as 
$\overline{ \left.\ptlder{ r_t }{ t }\right|_{\mathrm{mc}}}$; and finally, various second-order
microphysical terms, such as $\overline{ w' \left.\ptlder{ r_{t} }{ t }\right|_{\mathrm{mc}}' }$. 
CLUBB's PDF is also used to calculate cloud fraction and cloud liquid water mixing ratio,
but these quantities are not needed for closure.

\subsection{Background on univariate normal and lognormal PDF shapes}

CLUBB's PDF is both multi-variate (i.e., it includes multiple variables such as $w$, $r_t$, $\theta_l$, $r_r$, etc.) 
and multi-component (i.e., it is the sum of \textit{two} normal/lognormal components).  
Before introducing CLUBB's PDF, we pause to review its building blocks, 
the univariate normal and lognormal distributions.

First, a definition: a marginal distribution is the distribution that remains when a subset of variables
is integrated out.  For instance, if we start with a bivariate distribution $P(w,r_t)$, and we integrate
over all values of $r_t$, a univariate marginal results, $P(w)$.

A univariate normal has the form, for an arbitrary variable $x$:
\begin{equation}
P_n\left( x \right) 
= \frac{1}{\left( 2\pi \sigma \right)^{1/2}}
  \exp\left[ -\frac{1}{2}\left( \frac{x - \mu}{\sigma} \right)^{2} 
      \right] \mathrm{,}
\end{equation}
\noindent
where the mean is $\mu$ and the standard deviation is $\sigma$.  A normal distribution is useful
for a variable that can take on either positive or negative values (e.g., $w$), or a variable that
never approaches values near zero in the atmosphere (e.g., $\theta_l$).

A lognormal is the distribution of a random variable whose logarithm has a normal distribution.
A univariate lognormal has the form for an arbitrary variable $x$ \citep{larson_griffin_2013a}:
\begin{equation}
P_L\left( x \right) 
= \frac{1}{ \left( 2\pi \right)^{1/2} \sigma x }
  \exp\left[ \frac{ -\left( \ln x - \mu \right)^{2} }
                   { 2 \sigma^{2} }
      \right] .
\end{equation}
\noindent
Here $\mu$ and $\sigma$ are the mean and standard deviation of $\ln x$, not $x$.  
The lognormal distribution can be transformed to a normal distribution with mean 
$\mu$ and standard deviation $\sigma$ by making the substitution $\tilde{x} = \ln x$
and using the fundamental transformation law of probabilities \citep{ press_et_al_07a}.

A lognormal distribution is useful for representing variables that are non-negative (e.g., hydrometeor mixing ratios).  
Lognormal distributions are positively skewed and have long tails to the right.


\subsection{Assumed shape of CLUBB's PDF}

\hypertarget{url:clubb_pdf}{}
\label{sec:clubb_pdf}

CLUBB's subgrid PDF has the following functional form \citep{raut_larson_2016_flex_sampling}:
\begin{gather}
\begin{split}
P(\mathbf{x}) = \sum_{m=1}^{N_\mathrm{comp}} \xi_{(m)}\ [\ & f_{p(m)}
 P_{(m)}(\chi,\eta,w,N_{cn},\mathbf{hm})\ + \\
& (1-f_{p(m)})\ \delta(\mathbf{hm})\ P_{(m)}(\chi,\eta,w,N_{cn})\ ]  .
\end{split}
\label{eq:CLUBB_PDF}
\end{gather}
Here $\delta$ denotes the Dirac delta function.  
The PDF has $N_\mathrm{comp}$ components; currently, in CLUBB, $N_\mathrm{comp}
= 2$. Each component $m$ has a weight $\xi_{(m)}$, where
$\sum_{m=1}^{N_\mathrm{comp}} \xi_{(m)} = 1$. In each component, a fraction
$f_{p(m)}$ of the component is allowed to contain precipitation, where $0 \le
f_{p(m)} \le 1$. The vector $\mathbf{hm}$ contains hydrometeor species (e.g.,
rain, snow, cloud ice, etc.). The type and number of hydrometeors depends on the
microphysics scheme used.

In the portions of the PDF that contain precipitation,
$P_{(m)}(\chi,\eta,w,N_{cn},\mathbf{hm})$ is a multivariate normal-lognormal
distribution, where $\chi$, $\eta$, and $w$ are normally distributed, and
$N_{cn}$ and all the variables in $\mathbf{hm}$ are lognormally distributed. 
($\chi$ and $\eta$, denoted $s$ and $t$ in \citet{mellor_77a}, 
are a convenient linearized combination of $r_t$ and $\theta_l$.
Increases in $\chi$ correspond to increases in supersaturation.  $\eta$ is ``orthogonal" to $\chi$,
and increases in $\eta$ do not alter the degree of supersaturation.
See Section \ref{sec:anl_int_cloud} for a mathematical definition of $\chi$.)
In the parts of the PDF that don't contain precipitation,
$P_{(m)}(\chi,\eta,w,N_{cn})$ is a multivariate normal-lognormal distribution, as in the 
precipitating part, but here all the hydrometeors are zero, rather than lognormally
distributed.  ($\delta(\mathbf{hm})$ is short for
$\delta(\mathrm{hm}_1)\delta(\mathrm{hm}_2)\cdots\delta(\mathrm{hm}_n)$.)

A general expression for normal/lognormal PDFs such as $P_{(m)}(\chi,\eta,w,N_{cn},\mathbf{hm})$
may be written as follows \citep{griffin_larson_2016_hydrometeor_PDF}.  Suppose $P$ has $d$ variates: 
the first $j$ variables are normally distributed, 
and the remaining $k$ variables are lognormally distributed, so that $d=j+k$.  
Now transform the lognormal variates to normal space.  
Then rename the variates $(\chi,\eta,w,N_{cn},\mathbf{hm})$ to $\left( x_{1}, x_{2}, \ldots, x_{d} \right)$. 
This yields \citep{fletcher_zupanski_06a,griffin_2016}:
\begin{equation}
\label{eq:multivar_normal_lognormal_PDF}
\begin{split}
P_{\left(m\right)} \left( x_{1}, x_{2}, \ldots, x_{d} \right) =\
& \inverse{ \left( 2 \pi \right)^{\frac{d}{2}}
            \left| \mathbf{\Sigma}_{\left(m\right)} \right|^{\frac{1}{2}} }
  \left( \displaystyle\prod_{k=j+1}^{d} \inverse{ x_k } \right)  \\
& \times
  \exp\left\{ -\dfrac{1}{2}
               \left( \vec{\boldsymbol{x}}
                      - \vec{\boldsymbol{\mu}}_{\left(m\right)}
               \right)^{\mathrm{T}}
               \mathbf{\Sigma}_{\left(m\right)}^{-1} 
               \left( \vec{\boldsymbol{x}}
                      - \vec{\boldsymbol{\mu}}_{\left(m\right)} \right)
      \right\}
\mathrm{.}
\end{split}
\end{equation}
Both $\vec{\boldsymbol{x}}$ and $\vec{\boldsymbol{\mu}}_{\left(m\right)}$ are
$d\times1$ vectors and are given by
\begin{equation}
\vec{\boldsymbol{x}} =
\left[
\begin{array}{c}
x_{1}        \\
\vdots       \\
x_{j}        \\
\ln x_{j+1}  \\
\vdots       \\
\ln x_{d}    \\
\end{array}
\right]
\qquad
\mathrm{and}
\qquad
\vec{\boldsymbol{\mu}}_{\left(m\right)} =
\left[
\begin{array}{c}
\mu_{x_{1}\left(m\right)}            \\
\vdots                               \\
\mu_{x_{j}\left(m\right)}            \\
\mu_{x_{j+1}\left(m\right)}  \\
\vdots                               \\
\mu_{x_{d}\left(m\right)}    \\
\end{array}
\right]
\mathrm{,}
\end{equation}
where $\mu_{x\left(m\right)}$ is the $m$th component mean of $x$ for the first $j$ variates and where
$\mu_{x\left(m\right)}$ is the $m$th component mean of $\ln x$ for the last $k$ variates.  
$\mathbf{\Sigma}_{\left(m\right)}$ is the covariance matrix with dimension $d \times d$, and its determinant 
is $\left| \mathbf{\Sigma}_{\left(m\right)} \right|$.

In order to define the PDF given in \eqref{eq:CLUBB_PDF}, numerous PDF parameters must be provided, 
such as the mean total water for each component, $r_{t1}$ and $r_{t2}$.  These
component means are related to, but distinct from, the grid mean, $\overline{r_t}$.  
Because the number of PDF parameters is large and the number of prognosed moments
is limited, assumptions about the relationships between PDF parameters are required.
The default version of CLUBB calculates 
many of the PDF parameters from the moments using the ``ADG1" assumptions
of \citet{larson_et_al_02a}.  The PDF parameters that must be provided are:

\begin{enumerate}

\item The weight of each PDF component, $\xi_{(m)}$ (given by ADG1, \citet{larson_et_al_02a}).  

\item The precipitation fraction in each component, $f_{p(m)}$ (given by Eqns.~(12)-(14) of \citet{griffin_larson_2016_hydrometeor_PDF}).  

\item $\bm{\mu}_{(m)}$: a vector of $d$ means corresponding to each variate in
the PDF, for the $m$th PDF component. These means are in ``normal space," which means
that for the lognormal variates, the corresponding mean that appears in
$\bm{\mu}_{(m)}$ is the mean of the natural logarithm of the variate. Also,
these are in-precipitation means. Therefore, in the within-precipitation portion 
of the PDF, $P_{(m)}(\chi,\eta,w,N_{cn},\mathbf{hm})$ is distributed
such that each (normal-space) variate has a mean given by $\bm{\mu}_{(m)}$ \citep{griffin_larson_2016_hydrometeor_PDF}.
Outside of precipitation, the four variates of $P_{(m)}(\chi,\eta,w,N_{cn})$
have the same means ($\bm{\mu}_{(m)}$), but the hydrometeors are zero.

\item $\bm{\sigma}_{(m)}$: a vector of $d$ in-precipitation normal-space standard
deviations for the $m$th PDF component (given by Section 3.1 of \citet{griffin_larson_2016_hydrometeor_PDF} 
for hydrometeors and by ADG1 for non-hydrometeors).  

\item $\bm{\Sigma}_{(m)}$: a $d\times d$ covariance matrix, where
$\bm{\Sigma}_{(m)i,j}$ is the in-precipitation, normal-space covariance between the
$i$\textsuperscript{th} and $j$\textsuperscript{th} variates in PDF component
$m$.  $\bm{\Sigma}_{(m)}$ is constructed by pre-multiplying and post-multiplying
a correlation matrix by a diagonal matrix with the standard deviations $\bm{\sigma}_{(m)}$
placed along the diagonal.    
At present, CLUBB's correlation matrix has values prescribed as in \citet{storer_et_al_2015a}.  


\end{enumerate}

\begin{figure}[ht]
\centering
\includegraphics[width=0.45\textwidth]{\PathToDirMicroSrcVarCovar/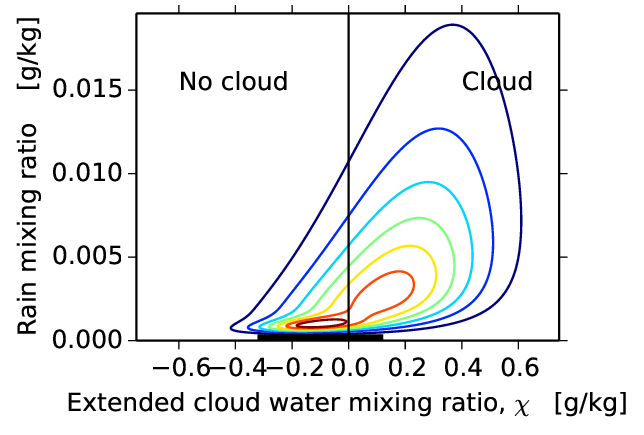}
\caption{   
An example bivariate marginal PDF of CLUBB's extended cloud liquid water variable, $\chi$, versus rain mixing ratio.  
The PDF is assumed to be a two-component mixture.  Each component is a normal in the $\chi$ variate and a 
lognormal in the rain variate.  In addition, the drier component has a delta function at zero rain 
(denoted by the thick horizontal
line on axis) representing rain-free areas.  The left half of the figure indicates that in this example, 
there is rain within clear air; the right half shows rain within cloud. 
}
\label{fig:chi_rain_PDF_schematic}
\end{figure}

To give a sense of what CLUBB's PDF looks like, the bivariate marginal PDF of $\chi$ and rain 
are shown in Fig.~\ref{fig:chi_rain_PDF_schematic}.  It shows an example in which
rain is falling primarily, but not entirely, through cloud.  This might occur in tilted cumulus (Cu), 
where some rain falls through Cu and grows by accretion but other rain falls alongside Cu
and starts to evaporate.  

\subsection{Analytic integration: 3rd-order turbulent transport moments}

\label{sec:3rd_order_moment_closures}

The integrals for the third-order moments are simple enough that they can be performed analytically.
An example of such an integral is
\begin{equation}
          \overline{w' r_t' \theta_l'} = 
                  \int_{\theta_l'=-\infty}^{\theta_l'=\infty} 
                  \int_{r_t'=-\infty}^{r_t'=\infty}
                  \int_{w'=-\infty}^{w'=\infty}  
                  w' r_t' \theta_l' \, P(w',r_t',\theta_l') \, dw' \, dr_t' \, d\theta_l'  .
\label{eq:wprtpthlp_integral}
\end{equation}
Here, $P(w',r_t',\theta_l')$ is the marginal PDF of $w$, $r_t$, and $\theta_l$, that is, the integral
that remains when all other variates are integrated over.  In addition, $P(w',r_t',\theta_l')$
has been written in terms of the perturbation variables, such as $w' = w - \overline{w}$. 

Furthermore, the results of the third-order integrations of the ADG1 parameterization 
are simple enough that they can be re-arranged and cast
in terms of lower-order moments (e.g., $\overline{w' r_t'}$) rather than PDF parameters (e.g., mixture fraction, $a$).
This facilitates the implicit discretization of the transport terms, which in turn improves their numerical stability.

Below we list the formulas that close the turbulent transport terms appearing 
in Eqns.~(\ref{eq_wprtp_unclosed})-(\ref{eq_vp2_unclosed}).
Details of the derivation can be found in \citet{larson_golaz_05a}. 

First, we define some notation.  We define $c_{w\theta_l}$ and $c_{wr_t}$
as in Eqns.~(15) and (16) of \citet{larson_golaz_05a}:
\begin{equation}
\label{eq_cwthl}
c_{w\theta_l} 
= \frac{ \overline{w'\theta_l'} }
       { \sqrt{\overline{w'^2}}\sqrt{\overline{\theta_l'^2}} }
\end{equation}
\begin{equation}
\label{eq_cwrt}
c_{wr_t} 
= \frac{ \overline{w'r_t'} }
       { \sqrt{\overline{w'^2}}\sqrt{\overline{r_t'^2}} } .
\end{equation}
The width of the individual Gaussians in $w$ is given by Eqn.~(37) of \citet{larson_golaz_05a}:
\begin{equation}
\label{eq_sc}
\tilde{\sigma}^2_w 
\equiv \sigma^2_w/\,\overline{w^{'2}}
= \gamma \left[ 1 - \max\left( c^2_{w\theta_l}, c^2_{wr_t} \right) \right]  ,
\end{equation}
where $\gamma$ (``gamma\_coef") is a tunable parameter.
We define the following quantities in order to simplify the notation:
\begin{equation}
\label{eq_a1}
a_1 \equiv \frac{1}{ (1-\tilde{\sigma}_w^2) }
\end{equation}
\begin{equation}
\label{eq_a2}
a_2 \equiv \frac{1}{ (1-\tilde{\sigma}_w^2)^2 }
\end{equation}
\begin{equation}
\label{eq_a3}
a_3 \equiv 3 \tilde{\sigma}_w^4 
      + 6 ( 1 - \tilde{\sigma}_w^2 ) \tilde{\sigma}_w^2
      + ( 1 - \tilde{\sigma}_w^2 )^2 
\end{equation}
The scalar skewnesses are parameterized in terms of the vertical velocity skewness.  For instance,
the skewness of $r_t$ is diagnosed as \citep[Eqn.~38,][]{larson_golaz_05a}:
\begin{equation}
\label{eq:scalar_Sk}
   \mathrm{Sk}_{r_t} = \frac{\mathrm{Sk}_w}{ ( 1 - \tilde{\sigma}_w^2 )^2} c_{wr_t} 
                       \left[  \beta + ( 1 - \beta )  \frac{c_{wr_t}^2}{ ( 1 - \tilde{\sigma}_w^2 )}  \right] ,
\end{equation}
where $\beta$ is a tunable parameter whose realizable values lie in the range $0 < \beta < 3$.
With these preliminaries in hand, we proceed to write down the closures.  
The turbulence moment $\overline{w'^4}$ is given by Eqn.~(40) of \citet{larson_golaz_05a}:
\hypertarget{url:wp4_diagnosis}{}
\begin{equation}
\label{eq_wp4}
\overline{w^{'4}}
= a_3\, \overline{w^{'2}}^2
+ a_1 \frac{ \overline{w^{'3}}^2 }{ \overline{w^{'2}} } .
\end{equation}
The flux transport terms are given by Eqn.~(42) of  \citet{larson_golaz_05a}:
\begin{equation}
\label{eq_wp2thlp}
\overline{w^{'2}\theta_l'}
= a_1 \frac{\overline{w^{'3}}}{\overline{w^{'2}}} \,
  \overline{w'\theta_l'}
\end{equation}
\begin{equation}
\label{eq_wp2rtp}
\overline{w^{'2}r_t'}
= a_1 \frac{\overline{w^{'3}}}{\overline{w^{'2}}} \,
  \overline{w'r_t'} .
\end{equation}
The variance transport terms follow Eqn.~(46)  of \citet{larson_golaz_05a}:
\begin{equation}
\label{eq_wpthlp2}
\begin{split}
\overline{w'\theta_l^{'2}}
& = 
    \frac{1}{3} \beta
    a_1 \frac{\overline{w^{'3}}}{\overline{w^{'2}}} \overline{\theta_l^{'2}}
  + \left( 1 - \frac{1}{3}\beta \right)
    a_2 \frac{\overline{w^{'3}}}{\overline{w^{'2}}^2} \overline{w'\theta_l'}^2 \,
\end{split}
\end{equation}
\begin{equation}
\label{eq_wprtp2}
\begin{split}
\overline{w'r_t^{'2}}
& = 
    \frac{1}{3} \beta
    a_1 \frac{\overline{w^{'3}}}{\overline{w^{'2}}} \overline{r_t^{'2}}
  + \left( 1 - \frac{1}{3}\beta \right)
    a_2 \frac{\overline{w^{'3}}}{\overline{w^{'2}}^2} \overline{w'r_t'}^2 .
\end{split}
\end{equation}
The covariance transport term is obtained substituting Eqn.~(56) of \citet{larson_golaz_05a} into Eqn.~(48) 
of \citet{larson_golaz_05a}:
\begin{equation}
\label{eq_wprtpthlp}
\begin{split}
\overline{w'r_t'\theta'_l}
& =
    \frac{1}{3} \beta
    a_1 \frac{\overline{w^{'3}}}{\overline{w^{'2}}} \,
    \overline{r_t'\theta'_l}
  + \left( 1 - \frac{1}{3}\beta \right)
    a_2 \frac{\overline{w^{'3}}}{\overline{w^{'2}}^2}
    \overline{w'r_t'} \, \overline{w'\theta_l'} .
\end{split}
\end{equation}
The horizontal wind variance terms satisfy:
\begin{equation}
\label{eq_wpup2}
\begin{split}
\overline{w'u^{'2}}
& = 
    \frac{1}{3} \beta
    a_1 \frac{\overline{w^{'3}}}{\overline{w^{'2}}} \overline{u^{'2}}
  + \left( 1 - \frac{1}{3}\beta \right)
    a_2 \frac{\overline{w^{'3}}}{\overline{w^{'2}}^2} \overline{w'u'}^2 \,
\end{split}
\end{equation}
\begin{equation}
\label{eq_wpvp2}
\begin{split}
\overline{w'v^{'2}}
& = 
    \frac{1}{3} \beta
    a_1 \frac{\overline{w^{'3}}}{\overline{w^{'2}}} \overline{v^{'2}}
  + \left( 1 - \frac{1}{3}\beta \right)
    a_2 \frac{\overline{w^{'3}}}{\overline{w^{'2}}^2} \overline{w'v'}^2 .
\end{split}
\end{equation}

\subsection{Analytic integration: Buoyancy terms}

\hypertarget{url:anl_int_buoy_terms}{}

The equations also contain unclosed buoyancy terms.  All buoyancy terms in the equations involve $\theta_v$. 
They are
$\overline{w'\theta'_v}$, $\overline{r'_t\theta'_v}$,
$\overline{\theta'_l\theta'_v}$, and $\overline{w^{'2}\theta'_v}$.  Using a generic co-variate, $\varpi$, 
all these moments can be written as \citep{golaz_et_al_02a}:
\begin{equation}
\overline{\varpi'\theta'_v} 
= \overline{\varpi'\theta'_l} 
+ \underbrace{ \frac{1-\epsilon_0}{\epsilon_0} \theta_0 
             }_{\approx \, 200 K}
  \overline{\varpi'r'_t}
+ \underbrace{
   \left( 
     \frac{L_v}{c_p} \left( \frac{p_0}{p} \right)^{R_d/c_p}      
     - \frac{1}{\epsilon_0}\theta_0
   \right) }_{\approx \, 2000 K}
  \overline{\varpi'r'_c} \hbox{ ,}
\end{equation}
where $\varpi'$ represents $w'$, $r'_t$, $\theta'_l$,  $w^{'2}$, or
a passive scalar.
Here $\epsilon_0 = R_d/R_v$, $R_d$ is the gas constant of dry air, 
$R_v$ is the gas constant of water vapor, $L_v$ is the latent heat 
of vaporization, $c_p$ is the heat capacity of air, 
and $p_0$ is a reference pressure. 

The covariances involving
cloud liquid water ($\overline{\varpi'r'_c}$) can be found by integration over the PDF.
See the next section for specific expressions.

\subsection{Analytic integration: Cloud properties}


\hypertarget{url:anl_int_cloud_terms}{}
\label{sec:anl_int_cloud}

The cloud properties, such as cloud fraction, mean liquid water
and correlations involving liquid water ($\overline{\varpi'r'_c}$)
are obtained by analytic integration over the PDF.  For instance,
the integrals to diagnose cloud fraction and grid-mean liquid water mixing ratio are 
\citep{mellor_77a,sommeria_deardorff_77a,lewellen_yoh_93a,larson_et_al_01b}:
\begin{equation}
  C = \int_{-\infty}^{\infty} H(\chi) P(\chi) d\chi
\label{eq:cloud_frac_integral}
\end{equation}

\begin{equation}
  \overline{r_c} = \int_{-\infty}^{\infty} \chi H(\chi) P(\chi) d\chi ,
\label{eq:rcm_integral}
\end{equation}
where $H$ is the Heaviside step function, and 
\begin{equation}
\label{eq:chi_def}
\chi = r_{t} - r_s(T_l,p) \frac{(1+\beta r_t)}{\left[ 1+\beta r_s(T_l,p) \right]}
\end{equation}
is an ``extended" liquid water mixing ratio.  When $\chi\ge 0$, then $\chi = r_c$,
but $\chi$ can also take on negative values.
To calculate $C$ and $\overline{r_c}$, the following properties
are computed for each Gaussian component ($i=1,2$):
\begin{equation}
T_{li} = \theta_{li} \left( \frac{p}{p_0} \right)^{R_d/c_p}
\end{equation}
\begin{equation}
\label{eq:rs_def}
r_{si} = \frac{R_d}{R_v}\; \frac{e_s(T_{li})}{p-[1-(R_d/R_v)] e_s(T_{li})}
\end{equation}
\begin{equation}
\label{eq:beta_def}
\beta_i =
\frac{R_d}{R_v} \left( \frac{L}{R_d T_{li}} \right) 
\left( \frac{L}{c_p T_{li}} \right)
\end{equation}
\begin{equation}
\label{eq:s_def}
\chi_i = r_{ti} - r_{si}\frac{1+\beta_i r_{ti}}{1+\beta_i r_{si}}
\end{equation}
\begin{equation}
c_{r_{ti}} = \frac{1}{1 + \beta_i r_{si}}
\end{equation}
\begin{equation}
c_{\theta_{li}} 
= \frac{1 + \beta_i r_{ti}}
       {[1 + \beta_i r_{si}]^2}
  \frac{c_p}{L} \beta_i r_{si}
  \left( \frac{p}{p_0} \right)^{R_d/c_p}
\end{equation}
\begin{equation}
\sigma_{\chi i}^2 
= c_{\theta_{li}}^2 \sigma_{\theta_{li}}^2 
+ c_{r_{ti}}^2 \sigma_{r_{ti}}^2
- 2 c_{\theta_{li}} \sigma_{\theta_{li}} 
    c_{r_{ti}} \sigma_{r_{ti}} r_{r_t \theta_l}
\end{equation}
\begin{equation}
\label{eq:C_gauss}
C_i = \frac{1}{2} 
      \left[ 
        1 + \mathrm{erf} \left( \frac{\chi_i}{\sqrt{2}\sigma_{\chi i}} \right) 
      \right]
\end{equation}
\begin{equation}
\label{eq:rl_gauss}
r_{ci} 
= \chi_i C_i
+ \frac{\sigma_{\chi i}}{\sqrt{2\pi}} 
  \exp \left[ 
         -\frac{1}{2}\left( \frac{\chi_i}{\sigma_{\chi i}} \right)^2 
       \right]
\end{equation}
where $C_i$ and $r_{ci}$ are the cloud fraction and cloud liquid water mixing ratio of
the $i$th Gaussian component.

Now, given the expressions $C_i$ \eqref{eq:C_gauss} and $r_{ci}$ \eqref{eq:rl_gauss} for the $i$th component, 
the layer-averaged cloud properties are given by:
\begin{equation}
\overline{C} = \xi_1 C_1 + (1-\xi_1) C_2
\end{equation}
\begin{equation}
\overline{r_c} = \xi_1 r_{c1} + (1-\xi_1) r_{c2}
\end{equation}
\begin{equation}
\overline{w'r_c'} = \xi_1 (w_1-\overline{w}) r_{c1} + (1-\xi_1) (w_2-\overline{w}) r_{c2}
\end{equation}
\begin{equation}
\overline{w^{'2}r_c'} 
= \xi_1 \left( (w_1-\overline{w})^2 + \sigma_{w1}^2 \right) r_{c1} 
+ (1-\xi_1) \left( (w_2-\overline{w})^2 + \sigma_{w2}^2 \right) r_{c2}
- \overline{w^{'2}} \left( \xi_1 r_{c1} + (1-\xi_1) r_{c2} \right)
\end{equation}
\begin{equation}
\begin{split}
\overline{\theta_l'r_c'}
=& \xi_1 \left[ 
       (\theta_{l1} - \overline{\theta_l} ) r_{c1} 
       - C_1 
         \left( 
           c_{\theta_{l1}} \sigma_{\theta_{l1}}^2
           - r_{r_t \theta_l} c_{r_{t1}} \sigma_{r_{t1}} \sigma_{\theta_{l1}}
         \right)
     \right] \\
&+ (1-\xi_1) \left[ 
           (\theta_{l2} - \overline{\theta_l} ) r_{c2} 
           - C_2
             \left( 
               c_{\theta_{l2}} \sigma_{\theta_{l2}}^2
               - r_{r_t \theta_l} c_{r_{t2}} \sigma_{r_{t2}} \sigma_{\theta_{l2}}
             \right)
         \right]
\end{split}
\end{equation}
\begin{equation}
\begin{split}
\overline{r_t'r_c'}
=& \xi_1 
   \left[ 
     (r_{t1} - \overline{r_t} ) r_{c1} 
     + C_1
       \left(
         c_{r_{t1}} \sigma_{r_{t1}}^2
         - r_{r_t \theta_l} c_{\theta_{l1}} \sigma_{r_{t1}} \sigma_{\theta_{l1}}
       \right)
   \right]\\
&+ (1-\xi_1) 
   \left[ 
     (r_{t2} - \overline{r_t} ) r_{c2} 
     + C_2
       \left(
         c_{r_{t2}} \sigma_{r_{t2}}^2
         - r_{r_t \theta_l} c_{\theta_{l2}} \sigma_{r_{t2}} \sigma_{\theta_{l2}}
       \right)
   \right]
\end{split}
\end{equation}

Unfortunately, for quantities involving liquid water mixing ratio ($r_c$), there are no handy expressions in terms of 
grid-box-averaged moments, as in for the 3rd-order turbulent transport moments (see Section \ref{sec:3rd_order_moment_closures}).
Instead, we are left with the more obscure expressions above in terms of the PDF parameters for Gaussian components 1 and 2.  
For more information about moments involving $r_c$, see \citet{ larson_et_al_02a}.

\subsection{Integration over microphysical process rates}

Two types of microphysical terms need to be integrated over subgrid variability:
1) microphysical contributions to the grid-mean equations for $\overline{r_t}$ \eqref{eq_rtm_unclosed} 
and $\overline{\theta_l}$ \eqref{eq_thlm_unclosed}; and 2) microphysical contributions to 
higher-order moments \eqref{eq_wprtp_unclosed}-\eqref{eq_rtpthlp_unclosed}.
CLUBB's closes both types of terms using the same tools, either analytic integration 
or Monte Carlo integration (SILHS).

\subsubsection{Closure of microphysics using analytic integration}

Both the grid-mean and the higher-order microphysical terms can be calculated by analytic integration
for the special case of the ``KK" microphysics scheme of \citet{khairoutdinov_kogan_00a}.  The KK microphysics
includes all major warm-rain processes (autoconversion, accretion, evaporation, and sedimentation)
but not any ice-related processes.  KK microphysics is integrable because it parameterizes 
all process rates in terms of simple power laws.  As an example, consider autoconversion
of small cloud droplets to larger rain drops.  Autoconversion depends strongly on droplet size.  
KK parameterizes this process as a function of cloud droplet mixing ratio, $r_c$, and cloud droplet
number concentration, $N_c$:
\begin{equation}
   \mathrm{Auto} = k \, r_c^{\alpha_a} N_c^{\beta_a} ,
   \label{eq:local_kk_auto}
\end{equation}
where $k$, $\alpha_a>0$, and $\beta_a<0$ are constants.  This expression approximates the \textit{local} rate of autoconversion.
To upscale the autoconversion to the \textit{grid scale}, note that $r_c = \chi H(\chi)$ and perform the integral
\begin{equation}
          \overline{\mathrm{Auto}} = 
                  \int_{\chi=-\infty}^{\chi=\infty} 
                  \int_{N_c=0}^{N_c=\infty}
                  k H(\chi) \chi^{\alpha_a} N_c^{\beta_a} \, P(\chi,N_c) \, dN_c \, d\chi .
\label{eq:mean_auto_integral}
\end{equation}
It turns out that this integral can be solved in terms of special mathematical functions.
In fact, all processes in the KK microphysics have been integrated analytically on the assumption 
that both the mean and variance of droplet number are prescribed, rather than prognosed.
The grid-mean terms are presented in \citet{larson_griffin_2013a} and \citet{griffin_larson_2013a}. 
Higher-order moments are presented in \citet{griffin_larson_2016_microphys_covar}.
The resulting integrals have been implemented in CLUBB and are available using the code options 
\texttt{microphys\_scheme = "khairoutdinov\_kogan"} and \texttt{l\_local\_kk = .false.}.

\subsubsection{Closure of microphysics using SILHS}

Analytic integration is fast and accurate, but it is feasible only for simple microphysical parameterizations.
When CLUBB is used with a more complex microphysical scheme, CLUBB resorts to a more general 
Monte Carlo approach, SILHS.  SILHS samples CLUBB's subgrid PDF and feeds the samples into 
the microphysical scheme.  The microphysical scheme treats the sample point as a uniform parcel 
and needs not know anything about the subgrid variability.  In this way, SILHS acts like an interface 
between CLUBB and the microphysics, allowing CLUBB to focus on subgrid variability and the microphysics scheme
to focus on local microphysical rates.  For this reason, SILHS works with a wide range of bulk microphysical schemes.
The drawbacks of Monte Carlo techniques such as SILHS, as compared to analytic integration, 
are that they introduce statistical noise due to small sample sizes, and the convergence is slow, meaning
that multiple sample points are required for accurate integration.  For more information about SILHS, see 
Chapter \ref{chapt:silhs_tech_descr}.

\section{Numerical limiters}

\subsection{Clipping and hole filling of positive semi-definite quantities}

Recall that CLUBB prognoses subgrid-scale variances of several fields: 
the horizontal components of velocity ($\overline{u'^2}$ and $\overline{v'^2}$), 
the vertical component of velocity ($\overline{w'^2}$), 
and scalar variances  ($\overline{r_t'^2}$ and $\overline{\theta_l'^2}$).  
A variance, by its mathematical definition, cannot be negative.  
In addition, CLUBB computes terms for the equations of 
two grid mean fields ---   total water mixing ratio (liquid + vapor, $\overline{r_t}$) and liquid water potential temperature ($\overline{\theta_l}$)  --- whose values cannot be negative on physical grounds.  
After each of these equations is solved, 
CLUBB takes measures to ensure that each of the resulting fields is non-negative.  

The simplest way to ensure that all values are non-negative is clipping.  
Clipping simply overwrites negative values with small, positive values:
  
\begin{lstlisting}
if ( xp2 < threshold ) then
     xp2 = threshold
endif
\end{lstlisting}

The drawback of clipping is that it is non-conservative.  That is, 
clipping increases the values 
of the negative regions in a profile while leaving the positive regions untouched, 
causing a net increase in the vertical integral.
Such non-conservation can lead to numerical instability in the case of the variances, 
and unphysical results in the case of the grid means.

Instead, CLUBB ensures non-negativity by using a conservative method, namely, 
\href{https://github.com/larson-group/clubb_release/blob/da4fc00e153ee358203253ad4afde70d7ed206a5/src/CLUBB_core/fill_holes.F90#L23-L28}
{hole filling}.  
That is, CLUBB fills the regions of negative values (``holes") using positive ``mass" 
from other altitudes in the profile.  
When CLUBB produces a hole, it is often associated with a positive spike just above and/or 
below the hole.
It appears that the process in CLUBB that spuriously ``digs" the hole often dumps the dug-out mass 
in the layers adjacent to the hole.  
Hence, to undo the pathological digging, CLUBB fills a negative level using mass from the 
\href{https://github.com/larson-group/clubb_release/blob/da4fc00e153ee358203253ad4afde70d7ed206a5/src/CLUBB_core/advance_xp2_xpyp_module.F90#L4964-L4970}{nearest two levels}
on either side
of the negative level.  The mass used to fill the hole is deducted proportionally from the adjacent levels in order to conserve
the vertical mass-weighted integral.  An idealized equation to implement this method for a function $f(z)$ is, 
assuming that holes are filled back to zero,
\begin{equation}
  f_\mathrm{hole-filled} = \frac{ \left< f \right> }{ \left< \max(0,f) \right> }  \max(0,f) ,
                                                                                       \label{eq:hole_filling}
\end{equation}
where $\left< () \right>$ denotes a mass-weighted average over the vertical portion that includes the hole and the $\pm$2 surrounding levels.
Here, $\max(0,f)$ is a clipped version of $f$, and the coefficient $\left< f \right> / \left< \max(0,f) \right>$ reduces $f$ proportionally in the 
layers surrounding the hole. 
If Eqn.~\eqref{eq:hole_filling} is vertically integrated, then we obtain
\begin{equation}
      \left< f_\mathrm{hole-filled} \right> =  \left< f \right> ,
\end{equation}
indicating that the formulation is conservative.

If the mass in this nearby region is non-positive, or is insufficient to fill the hole completely (i.e. $\left< f \right> < 0$), then 
mass is sought from all vertical levels in order to fill the hole.  (In the source code, 
the contributions from the local and global hole filling are combined and stored in budget terms 
of the form \texttt{xm\_cl} for the grid means --- \eqref{eq_rtm} and \eqref{eq_thlm} --- 
and \texttt{xp2\_pd} for the variances --- \eqref{eq_rtp2}, \eqref{eq_thlp2}, \eqref{eq_wp2}, \eqref{eq_up2}, and \eqref{eq_vp2} --- where ``\texttt{pd}" denotes ``positive definite.") 
If even this approach fails to fill the hole completely, 
then, and only then, CLUBB resorts to (non-conservative) 
\href{https://github.com/larson-group/clubb_release/blob/da4fc00e153ee358203253ad4afde70d7ed206a5/src/CLUBB_core/clip_explicit.F90#L897-L902}{clipping}. 
(In the source code, the clipping terms on higher-order moments are
stored in budget terms with the form \texttt{xp2\_cl}.)

\subsection{Monotonic limiter of turbulent fluxes}

In nature, turbulent mixing of a scalar cannot create a new extremum.  However, nothing in 
CLUBB's prognostic equations prevents turbulence-induced 
extrema in the grid means from forming: if a prognosed flux gradient is large enough, 
an extremum in the grid mean will form.  
To prevent the formation of spurious extrema, CLUBB employs a 
\href{https://github.com/larson-group/clubb_release/blob/da4fc00e153ee358203253ad4afde70d7ed206a5/src/CLUBB_core/mono_flux_limiter.F90#L28-L39}{monotonic flux limiter}.

The essential idea behind CLUBB's monotonic flux limiter is to estimate the maximum allowable upper
and lower bounds on the flux, clip the flux where it exceeds the allowable bounds,
and then recalculate the grid mean based on the clipped flux.  A challenge is that CLUBB
``knows" only the horizontally averaged moments of a grid level, 
not the values of a scalar at each horizontal position along the grid level.  Hence CLUBB does not know
the extreme values of the turbulent field at each grid level, and CLUBB cannot precisely
determine the maximum allowable values of the fluxes.  Instead, CLUBB uses a ``soft" limiter
that assumes that the extrema in $x$ can be estimated by $\overline{x} \pm 2 \sqrt{\overline{x'^2}}$.
Then CLUBB predicts the minimum value of flux,  \texttt{wpxp\_min}, and the maximum value of flux, \texttt{wpxp\_max}, 
that would produce these two extrema in $x$.  Finally, CLUBB limits the fluxes where needed and computes a consistent grid mean profile:

\begin{lstlisting}
! Limit turbulent flux
where ( wpxp > wpxp_max ) 
     wpxp_adj = wpxp_max - wpxp
     wpxp = wpxp_max
elsewhere ( wpxp < wpxp_min ) 
     wpxp_adj = wpxp_min - wpxp
     wpxp = wpxp_min    
endwhere

! Adjust mean field in order to be consistent with limited flux
xm = xm - dt*d(wpxp_adj)/dz
\end{lstlisting}
(In the source code, the flux limiter budget tendencies are stored in variables with names like
\texttt{wpxp\_mfl} and \texttt{xm\_mfl}.)

Although CLUBB's monotonic flux limiter can correct spurious extrema,
it does not ensure that the resulting grid means are positive semi-definite.  
In particular, sometimes CLUBB's turbulence
can drive total water mixing ratio, $\overline{r_t}$, negative.  To fix this, CLUBB
uses the 
\href{https://github.com/larson-group/clubb_release/blob/da4fc00e153ee358203253ad4afde70d7ed206a5/src/CLUBB_core/pos_definite_module.F90#L13-L17}{positive definite limiter}
of \citet{smolarkiewicz_1989_pos_def_limiter}.  (In the source code, the budget contributions
of the positive definite limiter are stored in terms of the form 
\texttt{xm\_pd} and \texttt{wpxp\_pd}.)

\subsection{Clipping of covariances}

Mathematically, subgrid correlations between variables cannot exceed the range $[-1,1]$.
Such correlations are implicit in the covariances and variances that CLUBB prognoses. 
If a covariance has grown too large in magnitude, CLUBB 
\href{https://github.com/larson-group/clubb_release/blob/da4fc00e153ee358203253ad4afde70d7ed206a5/src/CLUBB_core/clip_explicit.F90#L360-L367}{clips it}
.  (In the source code, these budget contributions 
have labels of the form \texttt{xpyp\_cl}.)  In addition, if a variance has grown too small,
the variance is increased by 
\href{https://github.com/larson-group/clubb_release/blob/da4fc00e153ee358203253ad4afde70d7ed206a5/src/CLUBB_core/clip_explicit.F90#L897-L901}
{clipping}.

\subsection{Clipping of skewnesses}

In CLUBB, the skewness of vertical velocity, $\mathrm{Sk}_w = \overline{w'^3}/ \left(\overline{w'^2}\right)^{3/2}$, is an 
important ratio because large, positive values of $\mathrm{Sk}_w$ tend to be associated with cumulus clouds,
whereas small or negative values tend to be associated with stratocumulus clouds.  However, when both the numerator $\overline{w'^3} \rightarrow 0$ and the denominator $\overline{w'^2} \rightarrow 0$, then $\mathrm{Sk}_w$ is undefined.  
This occurs commonly, for instance, at high altitudes, above any cloudy and turbulent layers.  In such cases, CLUBB
forces $\mathrm{Sk}_w \rightarrow 0$ by adding a small, positive tolerance to the denominator.  Namely, for a generic variable $x$,
function \texttt{Skw\_func} sets:
\begin{lstlisting}
Skx = xp3 / ( xp2 + Skw_denom_coef * x_tol**2 )**three_halves
\end{lstlisting}
Here \texttt{Skw\_denom\_coef} is a dimensionless tunable parameter, usually set to 4, and \texttt{x\_tol} is a small tolerance
with appropriate units.  

Despite this regularizing feature, sometimes $\mathrm{Sk}_w$ grows too large in magnitude in CLUBB, causing
numerical instability.  To avoid this, $\overline{w'^3}$ is clipped in subroutine \texttt{clip\_skewness\_core}
so that it does not exceed $(\mathrm{Skw\_max\_mag} \times \overline{w'^2}^{3/2})$ in magnitude.

\chapter{List of closed equations in CLUBB}

\label{chapt:clubb_closed_eqns}

For reference, this section lists CLUBB's prognostic equations in closed form.  
Explanation of some of the labels for the budget terms is given \hyperlink{url:categories_of_terms}{here}.
Then, we comment on CLUBB's numerical method of solution of these equations.

\section{List of closed, prognostic equations}

\label{sec:closed_eqns}

First we list the first-order moments.  These are handled by a host model, if CLUBB is embedded 
in a host model.  They are written:

\begin{equation}
\label{eq_rtm}
\ptlder{\color{hp}\overline{r_t}}{t}
= \underbrace{ - {\color{hp} \overline{w}}\ptlder{\color{hp}\overline{r_t}}{z} }_{ma}
  \underbrace{ - \inverse{{\color{hp}\rho_s}}\ptlder{{\color{hp}\rho_s} {\color{cp}\overline{w'r'_t}}}{z} }_{ta}
  + {\color{ci} \overline{\color{hp} \left.\ptlder{ r_t }{ t }\right|_{\mathrm{mc}}}}  
  + {\color{hp} \left. \ptlder{\overline{r_t}}{t} \right|_{\rm{sdmp}} }
  + {\color{ut} \left. \ptlder{\overline{r_t}}{t} \right|_{\rm{cl}} }
  + {\color{ut} \left. \ptlder{\overline{r_t}}{t} \right|_{\rm{mfl}} }
\end{equation}
\begin{equation}
\label{eq_thlm}
\ptlder{\color{hp}\overline{\theta_l}}{t} 
= \underbrace{- {\color{hp} \overline{w}}\ptlder{\color{hp} \overline{\theta_l}}{z} }_{ma}
  \underbrace{ - \inverse{{\color{hp}\rho_s}}\ptlder{{\color{hp}\rho_s}{\color{cp} \overline{w'\theta'_l}}}{z} }_{ta} 
  +{\color{ci} \overline{{\color{hp} \mathrm{RT}}}} 
  + {\color{ci} \overline{\color{hp} \left.\ptlder{ \theta_l }{ t }\right|_{\mathrm{mc}}}}  
  + {\color{hp} \left. \ptlder{\overline{\theta_l}}{t} \right|_{\rm{sdmp}} }
  + {\color{ut} \left. \ptlder{\overline{\theta_l}}{t} \right|_{\rm{cl}} }
  + {\color{ut} \left. \ptlder{\overline{\theta_l}}{t} \right|_{\rm{mfl}} }
\end{equation}
\begin{equation}
\label{eq_um}
\ptlder{\color{hp}\overline{u}}{t} 
= \underbrace{ - {\color{hp} \overline{w}}\ptlder{\color{hp} \overline{u}}{z} }_{ma}
  \underbrace{\color{hp} - f (v_g - \overline{v}) }_{cf/gf}
  \underbrace{ - \inverse{{\color{hp}\rho_s}}\ptlder{{\color{hp}\rho_s}{\color{cp} \overline{u'w'}}}{z} }_{ta} 
  + {\color{hp} \left. \ptlder{\overline{u}}{t} \right|_{\rm{nudging}} }
  + {\color{hp} \left. \ptlder{\overline{u}}{t} \right|_{\rm{sdmp}} }
\end{equation}
\begin{equation}
\label{eq_vm}
\ptlder{\color{hp}\overline{v}}{t} 
= \underbrace{- {\color{hp} \overline{w} }\ptlder{\color{hp}\overline{v}}{z} }_{ma}
  \underbrace{\color{hp} + f (u_g - \overline{u}) }_{cf/gf}
  \underbrace{ - \inverse{{\color{hp}\rho_s}}\ptlder{{\color{hp}\rho_s}{\color{cp} \overline{v'w'}}}{z} }_{ta}
  + {\color{hp} \left. \ptlder{\overline{v}}{t} \right|_{\rm{nudging}} }
  + {\color{hp} \left. \ptlder{\overline{v}}{t} \right|_{\rm{sdmp}} }
\end{equation}
where $\overline{RT}$ is the grid-mean radiative heating rate, $f$ the Coriolis parameter, 
and $u_g$ and $v_g$ the geostrophic winds.  The set of
equations is an anelastic set of equations, where $\rho_s$ is the dry, static,
base-state density, which only changes with respect to altitude.  
The subscript $\left.\right|_{\rm{sdmp}}$ denotes sponge layer damping, 
and the subscript $\left.\right|_{\rm{nudging}}$ denotes nudging to a prescribed state, both of which 
are available options in idealized single-column simulations.
The subscript $\left.\right|_{\rm{cl}}$ denotes the rate of change due to explicit clipping. 
Finally, the subscript $\left.\right|_{\rm{mfl}}$ denotes adjustments from CLUBB's ``monotonic" flux limiter. 
\href{https://github.com/larson-group/clubb_release/blob/da4fc00e153ee358203253ad4afde70d7ed206a5/src/CLUBB_core/mono_flux_limiter.F90#L28-L38}
{CLUBB's flux limiter}
prevents spurious extrema from forming in the mean fields due to artifacts of the numerics.
It is a ``soft" limiter that recognizes that slight extrema in the mean fields can legitimately form due to 
the presence of subgrid variability.  (To view links to CLUBB's source code browser, such as the one above,
type in your github username and password when prompted to do so.) 
Both the clipping and monotonic flux limiters are applied in a separate calculation immediately after the relevant CLUBB moment has been prognosed.  Hence the appearance of the clipping and flux limiter terms as source terms 
on the right-hand side of prognostic equations should not be taken literally.


Equations \eqref{eq_rtm}-\eqref{eq_vm} omit horizontal derivatives.  When CLUBB is run in a host model, 
the horizontal advection of first moments by the mean wind, such as $\overline{u} \partial \overline{r_t}/\partial x$, is
always calculated by the host model.  However, to date no host model has calculated the horizontal derivatives
of CLUBB's fluxes, such as $(1/\rho_s)\partial (\rho_s\overline{u' r_t'})/\partial x$.  
Such terms may become important at fine enough horizontal resolution (perhaps $\sim$2 km) and should
be considered for inclusion if CLUBB is ever used at such high horizontal resolutions. 

Next we list CLUBB's higher-order prognostic equations.  These equations satisfy the same reflectional symmetry
that the unclosed equations do (see Section \ref{sec:parity_symm}), as desired.

These equations omit related horizontal advection
terms.  They omit horizontal advection by the mean winds, which would be implemented in a host model by terms such as 
$(\overline{u}/\rho_s)\partial (\rho_s\overline{w' r_t'})/\partial x$.
They also omit \textit{turbulent} horizontal advection terms, such as $(1/\rho_s)\partial (\rho_s\overline{u' w' r_t'})/\partial x$.
Again, these terms may become important at high horizontal resolution.

\newpage

\hypertarget{url:wpxp_eqns}{}
The prognostic equations for $\overline{w'r_t'}$, $\overline{w'\theta_l'}$, and $\overline{u_h' w'}$, 
where $u_h$ denotes either horizontal component of velocity ($u$ or $v$), are all solved using 
the same LU-decomposition code that is implemented in  
\href{https://github.com/larson-group/clubb_release/blob/da4fc00e153ee358203253ad4afde70d7ed206a5/src/CLUBB_core/advance_xm_wpxp_module.F90#L44-L80}
{\texttt{subroutine advance\_xm\_wpxp}}.
\begin{equation}
\label{eq_wprtp}
\begin{split}
\ptlder{\color{cp}\overline{w'r'_t}}{t} 
= & \underbrace{ - {\color{hp} \overline{w}}\ptlder{\color{cp} \overline{w'r'_t}}{z} }_{ma}
    \underbrace{ - \inverse{{\color{hp}\rho_s}}\ptlder{{\color{hp}\rho_s} {\color{ci}\overline{w^{'2}r'_t}}}{z} }_{ta}
    \underbrace{ - {\color{cd} \overline{w^{'2}} }\ptlder{\color{hp}\overline{r_t}}{z} }_{tp}
    \underbrace{ - {\color{cp} \overline{w'r'_t} } \ptlder{\color{hp}\overline{w}}{z} }_{ac}
    \underbrace{ + \frac{g}{\theta_{vs}} {\color{ci} \overline{r'_t\theta'_v}} }_{bp} \\
  &   \underbrace{ - \frac{C_6}{\color{cd}\tau}{\color{cp} \overline{w'r'_t} } }_{pr1} 
    \underbrace{ + C_7 {\color{cp} \overline{w'r'_t}} \ptlder{\color{hp}\overline{w}}{z} }_{pr2}
    \underbrace{ - C_7 \frac{g}{\theta_{vs}} {\color{ci}\overline{r'_t\theta'_v}} }_{pr3} \\ 
 &   \underbrace{ + \ptlder{}{z} \left[ \left( {\color{cd}K_{w6}} + \nu_6 \right) 
                         \ptlder{}{z} {\color{cp} \overline{w'r'_t}} 
                   \right] }_{dp1} 
  + {\color{ci} \overline{\color{hp} \left.\ptlder{w'r'_t }{ t }\right|_{\mathrm{mc}}}}  \\
 &  
    + {\color{ut} \left. \ptlder{\overline{w'r'_t}}{t} \right|_{\rm{cl}} }
    + {\color{ut} \left. \ptlder{\overline{w'r'_t}}{t} \right|_{\rm{mfl}} }
\end{split}
\end{equation}
\begin{equation}
\label{eq_wpthlp}
\begin{split}
\ptlder{\color{cp}\overline{w'\theta'_l}}{t}
= & \underbrace{ - {\color{hp} \overline{w} }\ptlder{\color{cp} \overline{w'\theta'_l}}{z} }_{ma}	
    \underbrace{ - \inverse{{\color{hp}\rho_s}}\ptlder{{\color{hp}\rho_s}{\color{ci} \overline{w^{'2}\theta'_l}}}{z} }_{ta}
    \underbrace{ - {\color{cd}\overline{w^{'2}}}\ptlder{\color{hp}\overline{\theta_l}}{z} }_{tp}
    \underbrace{ - {\color{cp}\overline{w'\theta'_l}}\ptlder{\color{hp}\overline{w}}{z} }_{ac}
    \underbrace{ + \frac{g}{\theta_{vs}} {\color{ci}\overline{\theta'_l\theta'_v}} }_{bp} \\
 &    \underbrace{ - \frac{C_6}{\color{cd}\tau}{\color{cp}\overline{w'\theta'_l}} }_{pr1} 
  \underbrace{ + C_7 {\color{cp} \overline{w'\theta'_l} }\ptlder{\color{hp}\overline{w}}{z} }_{pr2}
    \underbrace{ - C_7 \frac{g}{\theta_{vs}} {\color{ci}\overline{\theta'_l\theta'_v}} }_{pr3} \\
 &    \underbrace{ + \ptlder{}{z} \left[ \left( {\color{cd} K_{w6}} + \nu_6 \right)
                          \ptlder{}{z} {\color{cp}\overline{w'\theta'_l}} 
                   \right] }_{dp1} 
  + {\color{ci} \overline{\color{hp} \left.\ptlder{w' \theta'_l}{ t }\right|_{\mathrm{mc}}}}  \\
 &   
    + {\color{ut} \left. \ptlder{\overline{w'\theta'_l}}{t} \right|_{\rm{cl}} }
    + {\color{ut} \left. \ptlder{\overline{w'\theta'_l}}{t} \right|_{\rm{mfl}} }
\end{split}
\end{equation}
\begin{equation}
\label{eq_upwp}
\begin{split}
\ptlder{\color{cp}\overline{u_h'w'}}{t} 
= & \underbrace{ - {\color{hp} \overline{w}}\ptlder{\color{cp} \overline{u_h'w'}}{z} }_{ma}
    \underbrace{ - \inverse{{\color{hp}\rho_s}}\ptlder{{\color{hp}\rho_s} {\color{ci}\overline{w^{'2}u_h'}}}{z} }_{ta}
    \underbrace{ - {\color{cd} \overline{w^{'2}} }\ptlder{\color{hp}\overline{u_h}}{z} }_{tp}
    \underbrace{ - {\color{cp} \overline{u_h'w'} } \ptlder{\color{hp}\overline{w}}{z} }_{ac}
    \underbrace{ + \frac{g}{\theta_{vs}} {\color{ci} \overline{u_h'\theta'_v}} }_{bp} \\
  &   \underbrace{ - \frac{C_6}{\color{cd}\tau}{\color{cp} \overline{u_h'w'} } }_{pr1} 
    \underbrace{ + C_7 {\color{cp} \overline{u_h'w'}} \ptlder{\color{hp}\overline{w}}{z} }_{pr2}
    \underbrace{ - C_7 \frac{g}{\theta_{vs}} {\color{ci}\overline{u_h'\theta'_v}} }_{pr3} 
    \underbrace{ + C^\mathrm{uu}_\mathrm{shr} {\color{cd} \overline{w^{'2}} }\ptlder{\color{hp}\overline{u_h}}{z} }_{pr4}
    \\ 
 &   \underbrace{ + \ptlder{}{z} \left[ \left( {\color{cd}K_{w6}} + \nu_6 \right) 
                         \ptlder{}{z} {\color{cp} \overline{u_h'w'}} 
                   \right] }_{dp1}  \\
 &  
    + {\color{ut} \left. \ptlder{\overline{u_h'w'}}{t} \right|_{\rm{cl}} }
    + {\color{ut} \left. \ptlder{\overline{u_h'w'}}{t} \right|_{\rm{mfl}} }
\end{split}
\end{equation}

\newpage
\hypertarget{url:xpyp_eqns}{}
The prognostic equations for $\overline{r_t'^2}$, $\overline{\theta_l'^2}$, and $\overline{r_t' \theta_l'}$
are implemented in 
\href{https://github.com/larson-group/clubb_release/blob/da4fc00e153ee358203253ad4afde70d7ed206a5/src/CLUBB_core/advance_xp2_xpyp_module.F90#L43-L71}
{\texttt{subroutine advance\_xp2\_xpyp}}.
\begin{equation}
\label{eq_rtp2}
\begin{split}
\ptlder{\color{cd} \overline{r_t^{'2}}}{t}
=& \underbrace{ - {\color{hp} \overline{w}}\ptlder{{\color{cd}\overline{r^{'2}_t}}}{z} }_{ma}
   \underbrace{ - \inverse{{\color{hp}\rho_s}}\ptlder{{\color{hp}\rho_s}{\color{ci}\overline{w'r_t^{'2}}}}{z} }_{ta}
   \underbrace{ - 2{\color{cp} \overline{w'r'_t}} \ptlder{{\color{hp}\overline{r_t}}}{z} }_{tp} \\
&   \underbrace{ - \dfrac{C_2}{\color{cd}\tau} \left(   {\color{cd} \overline{r_t^{'2}} }  
                               - r_{t,\rm{tol}}^{2} \right) }_{dp1} 
    \underbrace{ + \ptlder{}{z} \left[ \left( {\color{cd} K_{w2}} + \nu_2 \right) 
                         \ptlder{}{z} {\color{cd} \overline{r_t^{'2}}} 
                  \right] }_{dp2} 
  + {\color{ci} \overline{\color{hp} \left.\ptlder{ r_t^{'2} }{ t }\right|_{\mathrm{mc}}}}  \\
&   + {\color{ut} \left. \ptlder{\overline{r_t^{'2}}}{t} \right|_{\rm{pd}} }
   + {\color{ut} \left. \ptlder{\overline{r_t^{'2}}}{t} \right|_{\rm{cl}} }
\end{split}
\end{equation}
\begin{equation}
\label{eq_thlp2}
\begin{split}
\ptlder{\color{cd} \overline{\theta_l^{'2}}}{t}
=& \underbrace{ - {\color{hp} \overline{w}}\ptlder{{\color{cd}\overline{\theta^{'2}_l}}}{z} }_{ma}
   \underbrace{ - \inverse{{\color{hp}\rho_s}}\ptlder{{\color{hp}\rho_s}{\color{ci}\overline{w'\theta_l^{'2}}}}{z} }_{ta}
   \underbrace{ - 2{\color{cp} \overline{w'\theta'_l}} \ptlder{{\color{hp}\overline{\theta_l}}}{z} }_{tp} \\
&   \underbrace{ - \dfrac{C_2}{\color{cd}\tau} \left(   {\color{cd} \overline{\theta_l^{'2}} }  
                               - \theta_{l,\rm{tol}}^{2} \right) }_{dp1} 
    \underbrace{ + \ptlder{}{z} \left[ \left( {\color{cd} K_{w2}} + \nu_2 \right) 
                         \ptlder{}{z} {\color{cd} \overline{\theta_l^{'2}}} 
                  \right] }_{dp2} 
  + {\color{ci} \overline{\color{hp} \left.\ptlder{ \theta_l^{'2} }{ t }\right|_{\mathrm{mc}}}}  \\
&   + {\color{ut} \left. \ptlder{\overline{\theta_l^{'2}}}{t} \right|_{\rm{pd}} }
   + {\color{ut} \left. \ptlder{\overline{\theta_l^{'2}}}{t} \right|_{\rm{cl}} }
\end{split}
\end{equation}
\begin{equation}
\label{eq_rtpthlp}
\begin{split}
\ptlder{\color{cd}\overline{r'_t\theta'_l}}{t}
= & \underbrace{ - {\color{hp}\overline{w}}\ptlder{\color{cd}\overline{r'_t\theta'_l}}{z} }_{ma}
    \underbrace{ - \inverse{{\color{hp}\rho_s}}\ptlder{{\color{hp}\rho_s}{\color{ci}\overline{w'r'_t\theta'_l}}}{z} }_{ta}
    \underbrace{ - {\color{cp} \overline{w'r'_t}}\ptlder{\color{hp}\overline{\theta_l}}{z} }_{tp1}
    \underbrace{ - {\color{cp} \overline{w'\theta'_l}}\ptlder{\color{hp}\overline{r_t}}{z} }_{tp2} \\
 &    \underbrace{ - \dfrac{C_2}{\color{cd}\tau} {\color{cd}\overline{r_t' \theta_l'}} }_{dp1} 
   \underbrace{ + \ptlder{}{z} \left[ \left( {\color{cd} K_{w2}} + \nu_2 \right)
                            \ptlder{}{z} {\color{cd}\overline{r_t' \theta_l'} }
                     \right] }_{dp2} 
  + {\color{ci} \overline{\color{hp} \left.\ptlder{ r'_t \theta'_l }{ t }\right|_{\mathrm{mc}}}}  \\
&    + {\color{ut} \left. \ptlder{\overline{r'_t\theta'_l}}{t} \right|_{\rm{cl}} }
\end{split}
\end{equation}
\newpage
\hypertarget{url:wp2_wp3_eqns}{}
The prognostic equations for $\overline{w'^2}$ and $\overline{w'^3}$
are implemented in 
\href{https://github.com/larson-group/clubb_release/blob/da4fc00e153ee358203253ad4afde70d7ed206a5/src/CLUBB_core/advance_wp2_wp3_module.F90#L42-L70}
{\texttt{subroutine advance\_wp2\_wp3}}.
\begin{equation}
\label{eq_wp2}
\begin{split}
\ptlder{\color{cd}\overline{w^{'2}}}{t} 
&= \underbrace{ -{\color{hp} \overline{w}}\ptlder{\color{cd}\overline{w^{'2}}}{z} }_{ma}
   \underbrace{ - \inverse{{\color{hp}\rho_s}}\ptlder{{\color{hp}\rho_s}{\color{cd}\overline{w^{'3}}}}{z} }_{ta}
   \underbrace{ - 2{\color{cd}\overline{w^{'2}}}\ptlder{\color{hp}\overline{w}}{z} }_{ac}
   \underbrace{ + \frac{2g}{\theta_{vs}} {\color{ci} \overline{w'\theta'_v}} }_{bp} \\
&  \underbrace{ - \frac{C_4}{\color{cd} \tau} \left( {\color{cd} \overline{w^{'2}}} -\frac{2}{3}\overline{e} \right) }_{pr1} \\
&  \underbrace{  
       + C^\mathrm{uu}_\mathrm{shr} 2 {\color{cd} \overline{w^{'2}}}\ptlder{\color{hp}\overline{w}}{z}
       -C^\mathrm{uu}_\mathrm{buoy} \frac{2g}{\theta_{vs}} {\color{ci} \overline{w'\theta'_v}}
              }_{pr2}
   \underbrace{ 
   + \frac{2}{3} C^\mathrm{uu}_\mathrm{buoy} 
        \frac{g}{\theta_{vs}} {\color{ci} \overline{w'\theta'_v} } 
    + \frac{2}{3} C^\mathrm{uu}_\mathrm{shr}
     \left(
       - {\color{cp} \overline{u'w'}}\ptlder{\color{hp}\overline{u}}{z} 
       - {\color{cp} \overline{v'w'}}\ptlder{\color{hp}\overline{v}}{z} 
     \right) }_{pr3} \\
 & \underbrace{ - \dfrac{C_1}{\color{cd} \tau} \left( {\color{cd}  \overline{w^{'2}}} 
                              - w_{\rm{tol}}^{2} \right) }_{dp1}
   \underbrace{ + \ptlder{}{z} \left[ \left( {\color{cd}K_{w1}} + \nu_1 \right)
                         \ptlder{}{z} {\color{cd} \overline{w^{'2}}} 
                  \right] }_{dp2}  \\
&   + {\color{ut} \left. \ptlder{\overline{w^{'2}}}{t} \right|_{\rm{pd}} }
   + {\color{ut} \left. \ptlder{\overline{w^{'2}}}{t} \right|_{\rm{cl}} }
\end{split}
\end{equation}
\begin{equation}
\label{eq_wp3}
\begin{split}
\ptlder{\color{cd} \overline{w^{'3}}}{t}
= & \underbrace{ - {\color{hp}\overline{w}}\ptlder{\color{cd}\overline{w^{'3}}}{z} }_{ma}
    \underbrace{ - \inverse{{\color{hp}\rho_s}}\ptlder{{\color{hp}\rho_s} {\color{ci}\overline{w^{'4}}}}{z} }_{ta}
    \underbrace{ + 3\frac{\color{cd}\overline{w^{'2}}}{{\color{hp}\rho_s}}\ptlder{{\color{hp}\rho_s}\overline{\color{cd}w^{'2}}}{z} }_{tp}
    \underbrace{ - 3{\color{cd} \overline{w^{'3}}}\ptlder{\color{hp}\overline{w}}{z} }_{ac}
    \underbrace{ + \frac{3g}{\theta_{vs}} {\color{ci}\overline{w^{'2}\theta'_v}} }_{bp1} \\
  & \underbrace{ - C_{15} K_m \ptlder{}{z} \left( 
                           \frac{g}{\theta_{vs}} {\color{ci} \overline{w'\theta'_v}}  
			-  {\color{cp}\overline{u'w'}} \ptlder{\color{hp}\overline{u}}{z}
                        -  {\color{cp}\overline{v'w'}} \ptlder{\color{hp}\overline{v}}{z}
			\right)
               }_{pr\_turb} \\
  & \underbrace{ - \frac{C_8}{\color{cd}\tau} {\color{cd}\overline{w^{'3}}} }_{pr1}
    \underbrace{ - C_{11} \left(
                 - 3 {\color{cd} \overline{w^{'3}}}\ptlder{\color{hp} \overline{w}}{z}
                 + \frac{3g}{\theta_{vs}} {\color{ci} \overline{w^{'2}\theta'_v} }
             \right) }_{pr2}  \\
&    \underbrace{ + \ptlder{}{z} \left[ \left( {\color{cd} K_{w8}} + \nu_8 \right)
                          \ptlder{}{z} {\color{cd} \overline{w^{'3}} }
                   \right] }_{dp1} \\
&    + {\color{ut} \left. \ptlder{\overline{w^{'3}}}{t} \right|_{\rm{cl}} }
\end{split}
\end{equation}
\newpage
\hypertarget{url:up2_vp2_eqns}{}
The prognostic equations for $\overline{u'^2}$ and $\overline{v'^2}$
are implemented in 
\href{https://github.com/larson-group/clubb_release/blob/da4fc00e153ee358203253ad4afde70d7ed206a5/src/CLUBB_core/advance_xp2_xpyp_module.F90#L43-L71}
{\texttt{subroutine advance\_xp2\_xpyp}}.
\begin{equation}
\begin{split}
\label{eq_up2}
\ptlder{\color{cd}\overline{u^{'2}}}{t}
= & \underbrace{ - {\color{hp}\overline{w}}\ptlder{\color{cd}\overline{u^{'2}}}{z} }_{ma}
    \underbrace{ - \inverse{{\color{hp}\rho_s}}\ptlder{{\color{hp}\rho_s}{\color{ci}\overline{w'u^{'2}}}}{z} }_{ta}
    \underbrace{ - \left(1 - C^\mathrm{uu}_\mathrm{shr} \right) 2{\color{cp}\overline{u'w'}}\ptlder{\color{hp}\overline{u}}{z} }_{tp} \\
&    \underbrace{ - \frac{2}{3} 
                  C_{14} \frac{\color{cd}\overline{e}}{\color{cd}\tau}
               }_{dp1} 
    \underbrace{
    + \frac{2}{3} C^\mathrm{uu}_\mathrm{buoy}    
        \frac{g}{\theta_{vs}}  {\color{ci}\overline{w'\theta'_v}} 
    + \frac{2}{3} C^\mathrm{uu}_\mathrm{shr}
      \left(
        - {\color{cp}\overline{u'w'}}\ptlder{\color{hp}{\overline{u}}}{z} 
        -  {\color{cp} \overline{v'w'}}\ptlder{\color{hp}\overline{v}}{z} 
      \right) }_{pr2} \\
&    \underbrace{ - \dfrac{C_4}{\color{cd}\tau} \left( {\color{cd}\overline{u^{'2}}} - \frac{2}{3} {\color{cd}\overline{e}} \right) }_{pr1} 
    + \underbrace{ \ptlder{}{z} \left[ \left( {\color{cd}K_{w9}} + \nu_9 \right)
                          \ptlder{}{z} {\color{cd}\overline{u^{'2}}} 
                   \right] }_{dp2} \\
&  + {\color{ut} \left. \ptlder{\overline{u^{'2}}}{t} \right|_{\rm{pd}} }
    + {\color{ut} \left. \ptlder{\overline{u^{'2}}}{t} \right|_{\rm{cl}} }
\end{split}
\end{equation}
\begin{equation}
\begin{split}
\label{eq_vp2}
\ptlder{\color{cd}\overline{v^{'2}}}{t}
= & \underbrace{ - {\color{hp}\overline{w}}\ptlder{\color{cd}\overline{v^{'2}}}{z} }_{ma}
    \underbrace{ - \inverse{{\color{hp}\rho_s}}\ptlder{{\color{hp}\rho_s}{\color{ci}\overline{w'v^{'2}}}}{z} }_{ta}
    \underbrace{ - \left(1 - C^\mathrm{uu}_\mathrm{shr} \right) 2{\color{cp}\overline{v'w'}}\ptlder{\color{hp}\overline{v}}{z} }_{tp} \\
&    \underbrace{ - \frac{2}{3} 
                  C_{14} \frac{\color{cd}\overline{e}}{\color{cd}\tau}
               }_{dp1} 
    \underbrace{ 
    + \frac{2}{3} C^\mathrm{uu}_\mathrm{buoy}    
        \frac{g}{\theta_{vs}}  {\color{ci}\overline{w'\theta'_v}}     
    + \frac{2}{3} C^\mathrm{uu}_\mathrm{shr}
      \left(
        - {\color{cp}\overline{u'w'}}\ptlder{\color{hp}{\overline{u}}}{z} 
        -  {\color{cp} \overline{v'w'}}\ptlder{\color{hp}\overline{v}}{z} 
      \right) }_{pr2} \\
&    \underbrace{ - \dfrac{C_4}{\color{cd}\tau} \left( {\color{cd}\overline{v^{'2}}} - \frac{2}{3} {\color{cd}\overline{e}} \right) }_{pr1} 
    + \underbrace{ \ptlder{}{z} \left[ \left( {\color{cd}K_{w9}} + \nu_9 \right)
                          \ptlder{}{z} {\color{cd}\overline{v^{'2}}} 
                   \right] }_{dp2} \\
&  + {\color{ut} \left. \ptlder{\overline{v^{'2}}}{t} \right|_{\rm{pd}} }
    + {\color{ut} \left. \ptlder{\overline{v^{'2}}}{t} \right|_{\rm{cl}} }
\end{split}
\end{equation}

%
Here $g$ is acceleration due to gravity, and $\theta_{vs}$ is the dry, base-state $\theta_{v}$, 
which only varies with respect to altitude.  The following minimum threshold values of the variances 
are enforced: $\left. w \right|_{\rm{tol}}^{2}$, $\left. r_t \right|_{\rm{tol}}^{2}$, 
and $\left. \theta_l \right|_{\rm{tol}}^{2}$.  The subscript $\left.\right|_{\rm{pd}}$ stands for
the rate of change due to the positive-definite, conservative hole-filling scheme.  The
subscript $\left.\right|_{\rm{cl}}$ stands for the rate of change due to non-conservative clipping. 

\section{Numerical methods for solving CLUBB's equations}

CLUBB's equations \eqref{eq_rtm}-\eqref{eq_vp2} are partial differential equations of the
sort that is encountered in fluid dynamics.  
The system of equations is two dimensional in the vertical spatial dimension, $z$, and in time, $t$.
Hence CLUBB's equations can be solved by well-studied numerical methods.  However, CLUBB's equations are complex.  
They contain many terms.  Some sets of terms are nonlinear and interact strongly with each other.  
The numerics must be kept simple both to keep the software manageable and the algorithm comprehensible.  

\subsection{Mixed explicit and semi-implicit time stepping of CLUBB's higher-order prognostic equations} 

CLUBB is discretized in the vertical by centered differencing or else upwind differencing 
on a staggered grid (see Fig.~\ref{fig:clubb_grid}).   In particular, upwinding of the flux of vertical turbulent fluxes of scalars
is turned on by the flag \texttt{l\_upwind\_wpxp\_ta}, upwinding of the flux of variances or covariances by 
\texttt{l\_upwind\_xpyp\_ta}, and upwinding of the grid-mean vertical advection of scalars by \texttt{l\_upwind\_xm\_ma}.

Because CLUBB's equations are prognostic, they must be time stepped.  In order to save computational cost,
it is highly advantageous to use a long time step: the cost is halved if the time step is doubled.  
To increase the time step, the use of an implicit method is necessary.  However, a fully implicit method
that relies on iteration is still too expensive.  The use of several iterations is 
nearly as expensive as the use of several sub-timesteps.  Therefore, CLUBB resorts to the use of a semi-implicit
time stepper, in which certain terms in the equations are linearized and treated implicitly in a matrix on the left-hand side
of the equation and other terms are left nonlinear and treated explicitly on the right-hand side.
A single matrix solve advances the solution one time step.  The time stepping method is simple backward Euler.

The equations for $\overline{r_t'^2}$, $\overline{\theta_l'^2}$, $\overline{r_t'\theta_l'}$, 
$\overline{u'^2}$, and $\overline{v'^2}$ are each solved separately and semi-implicitly using a tridiagonal matrix equation.
For instance, for $\overline{r_t'^2}$, CLUBB solves the matrix equation:
\small
\begin{equation}
{
\underbrace{
\left[ \begin{array}{llllll}
  & &   
  \vdots     
  & & & \\ [0.5em]
  \overline{r_t'^2}^{\, \mathrm{impl.}}_{\, k-2} 
  & \overline{r_t'^2}^{\, \mathrm{impl.}}_{\, k-1}
  & \overline{r_t'^2}^{\, \mathrm{impl.}}_{\, k}
  & & &  \\ [0.5em]
  & \overline{r_t'^2}^{\, \mathrm{impl.}}_{\, k-1} 
  & \overline{r_t'^2}^{\, \mathrm{impl.}}_{\, k}
  & \overline{r_t'^2}^{\, \mathrm{impl.}}_{\, k+1}
  & &  \\ [0.5em]
  &
  & \overline{r_t'^2}^{\, \mathrm{impl.}}_{\, k}
  & \overline{r_t'^2}^{\, \mathrm{impl.}}_{\, k+1}
  & \overline{r_t'^2}^{\, \mathrm{impl.}}_{\, k+2}
  &  \\ [0.5em]
  & &
  & \overline{r_t'^2}^{\, \mathrm{impl.}}_{\, k+1}
  & \overline{r_t'^2}^{\, \mathrm{impl.}}_{\, k+2}
  & \overline{r_t'^2}^{\, \mathrm{impl.}}_{\, k+3}
  \\ [0.5em]
  & &   
  & \vdots    
  & &   
\end{array} \right] }_{\textrm{Semi-implicit terms:} -\frac{\overline{r_t'^2}}{\tau}, \textrm{etc.} }
\underbrace{
\left[ \begin{array}{l}
  \hfill \vdots \hfill \\                        [2.0em]
  \overline{r_t'^2}^{\, t+\Delta t}_{\, k-2} \\          [0.75em]
  \overline{r_t'^2}^{\, t+\Delta t}_{\, k-1} \\  [0.75em]
  \overline{r_t'^2}^{\, t+\Delta t}_{\, k} \\            [0.75em]
  \overline{r_t'^2}^{\, t+\Delta t}_{\, k+1} \\    [0.75em]
  \overline{r_t'^2}^{\, t+\Delta t}_{\, k+2} \\          
  \hfill \vdots \hfill
\end{array} \right] }_\textrm{Variables to solve for} 
=
\underbrace{
\left[ \begin{array}{l}
  \hfill \vdots \hfill \\                               [1.0em]
  \overline{r_t}^{\, \mathrm{expl.}}_{\, k-2} \\        [0.75em]
  \overline{r_t'^2}^{\,\mathrm{expl.}}_{\, k-1} \\      [0.75em]
  \overline{r_t'^2}^{\, \mathrm{expl.}}_{\, k} \\          [0.75em]
  \overline{r_t'^2}^{\,\mathrm{expl.}}_{\, k+1} \\        [0.75em]
  \hfill \vdots \hfill
\end{array} \right] }_\textrm{Explicit sources}
}
\end{equation}
\normalsize
\noindent
The matrix and the vector on the left-hand side combine with each other in order to form the terms
that are treated implicitly.  For instance, one term in $\overline{r_t'^2}^{\, \mathrm{impl.}}$ is $1/\tau$, which
gets multiplied by $\overline{r_t'^2}^{\, t+\Delta t}$ in order to form the dissipation term, $\overline{r_t'^2}/\tau$, 
where the minus sign is missing because the term has been moved to the left-hand side.  
The $\overline{r_t'^2}^{\, \mathrm{expl.}}$ terms in the right-hand side vector represent terms such as precipitation 
that cannot be linearized easily and hence are treated explicitly.  

Even if each equation is solved using a semi-implicit time stepping method, 
numerical instability may still arise from the interaction of advection terms and production terms in different equations.  
For instance, the turbulent advection term $-\partial\overline{w'r_t'}/\partial z$ 
in the $\overline{r_t}$ equation \eqref{eq_rtm} contains $\overline{w'r_t'}$,
and the turbulent production term $-\overline{w'^2} \partial \overline{r_t}/\partial z$ 
in the $\overline{w'r_t'}$ equation \eqref{eq_wprtp} contains $\overline{r_t}$.  
Over the course of a time step, these terms may interact and re-inforce each other, causing numerical instability.  
Therefore, we want to solve these terms simultaneously.  To do so, we combine the $\overline{r_t}$ 
and $\overline{w'r_t'}$ equations into a matrix system.  The left-hand side matrix has $2n \times 2n$ elements, 
where $n$ is the number of vertical grid levels.  The vector to be solved contains interleaved 
values of $\overline{r_t}$ and $\overline{w'r_t'}$ at different vertical levels $...,k-1, k, k+1,...$. 
This arrangement of the variables yields a 5-diagonal banded matrix that is solved by LU-decomposition
at each time step and in each grid column:  
\small
\begin{equation}
{
\underbrace{
\left[ \begin{array}{llllllll}
  & & &  
  \vdots &    
  & & &  \\ [0.5em]
  \overline{r_t}^{\, \mathrm{impl.}}_{\, k-1} 
  & \overline{w'r'_t}^{\, \mathrm{impl.}}_{\, k-1} 
  & \overline{r_t}^{\, \mathrm{impl.}}_{\, k}
  & \overline{w'r'_t}^{\, \mathrm{impl.}}_{\, k} 
  & \overline{r_t}^{\, \mathrm{impl.}}_{\, k+1}
  & & &  \\ [0.5em]
  & \overline{w'r'_t}^{\, \mathrm{impl.}}_{\, k-1} 
  & \overline{r_t}^{\, \mathrm{impl.}}_{\, k} 
  & \overline{w'r'_t}^{\, \mathrm{impl.}}_{\, k} 
  & \overline{r_t}^{\, \mathrm{impl.}}_{\, k+1}
  & \overline{w'r'_t}^{\, \mathrm{impl.}}_{\, k+1} 
  & &  \\ [0.5em]
  &
  & \overline{r_t}^{\, \mathrm{impl.}}_{\, k}
  & \overline{w'r'_t}^{\, \mathrm{impl.}}_{\, k} 
  & \overline{r_t}^{\, \mathrm{impl.}}_{\, k+1}
  & \overline{w'r'_t}^{\, \mathrm{impl.}}_{\, k+1}
  & \overline{r_t}^{\, \mathrm{impl.}}_{\, k+2}
  &  \\ [0.5em]
  & &
  & \overline{w'r'_t}^{\, \mathrm{impl.}}_{\, k} 
  & \overline{r_t}^{\, \mathrm{impl.}}_{\, k+1}
  & \overline{w'r'_t}^{\, \mathrm{impl.}}_{\, k+1} 
  & \overline{r_t}^{\, \mathrm{impl.}}_{\, k+2}
  & \overline{w'r'_t}^{\, \mathrm{impl.}}_{\, k+2} 
  \\ [0.5em]
  & & &  
  & \vdots    
  & & &  
\end{array} \right] }_{\textrm{Semi-implicit terms:} -\overline{w'^2}\ptlder{\overline{r_t}}{z}, -\ptlder{\overline{w'r_t'}}{z}, \textrm{etc.} }
\underbrace{
\left[ \begin{array}{l}
  \hfill \vdots \hfill \\                        [2.0em]
  \overline{r_t}^{\, t+\Delta t}_{\, k-1} \\          [0.75em]
  \overline{w'r'_t}^{\, t+\Delta t}_{\, k-1} \\  [0.75em]
  \overline{r_t}^{\, t+\Delta t}_{\, k} \\            [0.75em]
  \overline{w'r'_t}^{\, t+\Delta t}_{\, k} \\    [0.75em]
  \overline{r_t}^{\, t+\Delta t}_{\, k+1} \\          
  \hfill \vdots \hfill
\end{array} \right] }_\textrm{Variables to solve for} 
=
\underbrace{
\left[ \begin{array}{l}
  \hfill \vdots \hfill \\                               [1.0em]
  \overline{r_t}^{\, \mathrm{expl.}}_{\, k-1} \\        [0.75em]
  \overline{w'r'_t}^{\,\mathrm{expl.}}_{\, k-1} \\      [0.75em]
  \overline{r_t}^{\, \mathrm{expl.}}_{\, k} \\          [0.75em]
  \overline{w'r'_t}^{\,\mathrm{expl.}}_{\, k} \\        [0.75em]
  \hfill \vdots \hfill
\end{array} \right] }_\textrm{Explicit sources}
}
\end{equation}
\normalsize
\noindent
In the left-hand side matrix, the $\overline{r_t}^{\, \mathrm{impl.}}$ terms are the implicit terms that 
include $\overline{r_t}^{\, t+\Delta t}$, excluding $\overline{r_t}^{\, t+\Delta t}$, which comes 
from the vector on the left-hand side.  For instance, for the turbulent production term
in the $\overline{w'r_t'}$ equation, namely $\overline{w'^2} \ptlder{\overline{r_t}}{z}$, 
we have $\overline{r_t}^{\, \mathrm{impl.}} = \overline{w'^2} \ptlder{}{z}$.  

Similarly, $-\partial\overline{w'^3}/\partial z$ appears in the $\overline{w'^2}$ equation \eqref{eq_wp2},
and $-\overline{w'^2} \partial \overline{w'^2}/\partial z$ appears in the $\overline{w'^3}$ equation \eqref{eq_wp3}.    
Thus, $\overline{w'^2}$ and $\overline{w'^3}$ are solved simultaneously in a similar 5-banded matrix
equation:

\small
\begin{equation}
{
\underbrace{
\left[ \begin{array}{llllllll}
  & & &  
  \vdots &    
  & & &  \\ [0.5em]
  \overline{w'^3}^{\, \mathrm{impl.}}_{\, k-1} 
  & \overline{w'^2}^{\, \mathrm{impl.}}_{\, k-1} 
  & \overline{w'^3}^{\, \mathrm{impl.}}_{\, k}
  & \overline{w'^2}^{\, \mathrm{impl.}}_{\, k} 
  & \overline{w'^3}^{\, \mathrm{impl.}}_{\, k+1}
  & & &  \\ [0.5em]
  & \overline{w'^2}^{\, \mathrm{impl.}}_{\, k-1} 
  & \overline{w'^3}^{\, \mathrm{impl.}}_{\, k} 
  & \overline{w'^2}^{\, \mathrm{impl.}}_{\, k} 
  & \overline{w'^3}^{\, \mathrm{impl.}}_{\, k+1}
  & \overline{w'^2}^{\, \mathrm{impl.}}_{\, k+1} 
  & &  \\ [0.5em]
  &
  & \overline{w'^3}^{\, \mathrm{impl.}}_{\, k}
  & \overline{w'^2}^{\, \mathrm{impl.}}_{\, k} 
  & \overline{w'^3}^{\, \mathrm{impl.}}_{\, k+1}
  & \overline{w'^2}^{\, \mathrm{impl.}}_{\, k+1}
  & \overline{w'^3}^{\, \mathrm{impl.}}_{\, k+2}
  &  \\ [0.5em]
  & &
  & \overline{w'^2}^{\, \mathrm{impl.}}_{\, k} 
  & \overline{w'^3}^{\, \mathrm{impl.}}_{\, k+1}
  & \overline{w'^2}^{\, \mathrm{impl.}}_{\, k+1} 
  & \overline{w'^3}^{\, \mathrm{impl.}}_{\, k+2}
  & \overline{w'^2}^{\, \mathrm{impl.}}_{\, k+2} 
  \\ [0.5em]
  & & &  
  & \vdots    
  & & &  
\end{array} \right] }_{\textrm{Semi-implicit terms:} -\overline{w'^2}\ptlder{\overline{w'^2}}{z}, -\ptlder{\overline{w'^3}}{z}, \textrm{etc.} }
\underbrace{
\left[ \begin{array}{l}
  \hfill \vdots \hfill \\                        [2.0em]
  \overline{w'^3}^{\, t+\Delta t}_{\, k-1} \\          [0.75em]
  \overline{w'^2}^{\, t+\Delta t}_{\, k-1} \\  [0.75em]
  \overline{w'^3}^{\, t+\Delta t}_{\, k} \\            [0.75em]
  \overline{w'^2}^{\, t+\Delta t}_{\, k} \\    [0.75em]
  \overline{w'^3}^{\, t+\Delta t}_{\, k+1} \\          
  \hfill \vdots \hfill
\end{array} \right] }_\textrm{Variables to solve for} 
=
\underbrace{
\left[ \begin{array}{l}
  \hfill \vdots \hfill \\                               [1.0em]
  \overline{w'^3}^{\, \mathrm{expl.}}_{\, k-1} \\        [0.75em]
  \overline{w'^2}^{\,\mathrm{expl.}}_{\, k-1} \\      [0.75em]
  \overline{w'^3}^{\, \mathrm{expl.}}_{\, k} \\          [0.75em]
  \overline{w'^2}^{\,\mathrm{expl.}}_{\, k} \\        [0.75em]
  \hfill \vdots \hfill
\end{array} \right] }_\textrm{Explicit sources}
}
\end{equation}
\normalsize
\noindent

\subsection{Over-implicit time stepping of selected turbulent advection terms}


An explicit numerical treatment of turbulent advection of higher-order moments (flux-of-flux terms) 
is prone to numerical noise, and sometimes so is a semi-implicit treatment.  In order to smooth and stabilize
the solutions, CLUBB adopts an ``over-implicit" treatment of these terms.  

Consider a generic term $f(x)$.  A standard semi-implicit treatment expands $f$ in a Taylor series 
about the $n$th time step:

\begin{equation}
\begin{split}
   f(x) & \approx f^{(n)} + \left. \ptlder{f}{x} \right|_{(n)} \Delta x \\
          & \equiv  f^{(n)} + \left. \ptlder{f}{x} \right|_{(n)} \left( x^{(n+1)} - x^{(n)} \right) .
\end{split}
\end{equation}
Now substitute in
\begin{equation}
              x^{(n+1)} \approx \gamma x^{(n+1)} + (1-\gamma) x^{(n)} ,
\end{equation}
where $\gamma$ is a tunable parameter, yielding
\begin{equation}
\begin{split}
          f(x) =  f^{(n)} - \left. \ptlder{f}{x} \right|_{(n)}  x^{(n)}  
                + \left. \ptlder{f}{x} \right|_{(n)} \left( \gamma x^{(n+1)} + (1-\gamma) x^{(n)} \right) .
\end{split}
\end{equation}
When $\gamma=0$, we recover the standard explicit solution.   When $\gamma=1$, we recover a standard
semi-implicit solution.  However, CLUBB sets $\gamma=1.5$, in an attempt to provide a smoother and more robust discretization.
In the CLUBB code, $\gamma$ is named \texttt{gamma\_over\_implicit\_ts}.  CLUBB uses this discretization for the 
turbulent advection of $\overline{r_t'^2}$,  $\overline{\theta_l'^2}$, $\overline{r_t' \theta_l'}$, $\overline{u'^2}$, 
$\overline{v'^2}$, and $\overline{w'^3}$.  (The over-implicit treatment is also used for turbulent production of $\overline{w'^3}$.)

\chapter{Technical description of SILHS}

\label{chapt:silhs_tech_descr}

\section{Overview of Monte Carlo sampling of microphysics}

SILHS is used to account for the effects of subgrid variability on grid-mean time tendencies of processes such as
microphysics.  For instance, we might wish to estimate the microphysical source term, 
$\overline{w' \left.\ptlder{ r_{t} }{ t }\right|_{\mathrm{mc}}'}$, that appears
on the right-hand side of CLUBB's prognostic equation for $\overline{w'r_t'}$ \eqref{eq_wprtp_unclosed}.  
To do so, we need to average the local process rate, namely 
$ w' \left.\ptlder{ r_{t} }{ t }\right|_{\mathrm{mc}}'$, over the grid box.

More generally, let $h(\mathbf{x})$ represent a local process rate and let $\mathbf{x}$ represent 
the state variables on which $h$ depends.
Then the goal of SILHS is to numerically approximate integrals of the
following form:
\begin{equation}
{\color{ci}\overline{\color{black}h(\mathbf{x})}} =
\int h(\mathbf{x}) P(\mathbf{x}) d\mathbf{x} ,
\label{SILHS integral form}
\end{equation}
where $P(\mathbf{x})$ is the PDF that describes CLUBB's joint probability
distribution of variables, and $h(\mathbf{x})$ is a function of some or all of
these variables.  This integral yields the spatially averaged value of
$h(\mathbf{x})$ over the horizontal area spanned by a grid box.  
SILHS calculates the integral by drawing sample points 
from $P(\mathbf{x})$, feeding those sample points into $h(\mathbf{x})$,
and then averaging the resulting set of values. 

In more detail, 
to numerically approximate \eqref{SILHS integral form}, SILHS performs the following tasks 
\citep{larson_schanen_2013a}: 
\begin{enumerate}
\item \textit{Choose a set of uniform sample points at one vertical grid level.} Currently,
the vertical level is chosen to be the level where cloud water is maximized. The uniform sample
is picked using the Latin hypercube algorithm in order to reduce variance.
\item \textit{Vertically correlate these uniform sample points to the other vertical
levels in CLUBB.} Vertical correlation is the correlation in the value of a field
at one height with the value of the same field at another height.  Vertical correlation
is based on the spacing between vertical
levels, CLUBB's turbulent mixing length scale \ref{sec:mixing_length}, and an empirical constant.
\item \textit{Transform the uniformly distributed sample points to CLUBB's PDF.} At this point, SILHS
has produced a set of subcolumns drawn from the PDF $P(\mathbf{x})$ in
\eqref{SILHS integral form} (or, in the case of importance sampling, a related
PDF, as described below).
\item \textit{Evaluate the function $h(\mathbf{x})$ for each of the SILHS sample
points.} 
\item \textit{Average these sample values of $h$ in order to estimate the average
value of $h(\mathbf{x})$ over the distribution.}
\end{enumerate}

The following sections discuss some of these tasks in more depth.

\section{Sampling PDF components 1 vs 2 and precipitating vs non-precipitating regions}

SILHS draws sample points from CLUBB's PDF.  CLUBB's PDF is both multivariate and multicomponent.  
It is defined by Eqn.~(\ref{eq:CLUBB_PDF}) and described further in Section \ref{sec:clubb_pdf}.
Please look there for notation.  Because CLUBB's PDF has two components, 
a first task is to decide from which component
(i.e., from which $P_{(m)}$) to draw each sample point.  A second task is to decide
whether to draw the sample point from within a precipitating or non-precipitating region.
How often to sample from each distribution is decided by the weights $\xi_{(m)}$.  
How often to draw from each region is decided by $f_{p(m)}$. 
To decide which sample goes where, two new variates are introduced: $u_{d+1}$ 
determines whether to sample from component 1 or 2, and
$u_{d+2}$ determines whether to sample from the portion of the PDF with precipitation. 
The two variates are uniformly distributed, with $0 \le
u_{d+1},u_{d+2} \le 1$. 
The PDF in \eqref{eq:CLUBB_PDF} can then be
written as follows (see also Table 1 of \citet{raut_larson_2016_flex_sampling}):
\begin{equation}
P(\chi,\eta,w,N_{cn},\mathbf{hm},u_{d+1},u_{d+2}) =
\begin{cases}
P_{(m(u_{d+1}))}(\chi,\eta,w,N_{cn},\mathbf{hm}) & u_{d+2} < f_{p(m(u_{d+1}))}
\\ \delta(\mathbf{hm}) P_{(m(u_{d+1}))}(\chi,\eta,w,N_{cn}) & u_{d+2} \ge
f_{p(m(u_{d+1}))}
\end{cases}
\label{CLUBB PDF with uniform}
\end{equation}
where $m(u_{d+1})$ is a function that associates $u_{d+1}$ with a PDF component.
In CLUBB's two-component case, the function can simply be defined as:
\begin{equation}
m(u_{d+1}) =
\begin{cases}
1 & u_{d+1} < \xi_1 \\
2 & u_{d+1} \ge \xi_1
\end{cases}
\end{equation}

\section{Stratification of sample points in uniform space}

\hypertarget{url:lh_algorithm}{}
As a first step, sample points are picked from a uniform distribution. One
uniform variate is picked for each variate in the PDF.
To reduce sample noise, SILHS simultaneously employs both importance and stratified 
sampling strategies.  Both are done in the uniform space.

As a variance reduction technique, SILHS employs Latin hypercube sampling 
\citep{press_et_al_07a,owen_2013_monte_carlo_book}. The basic idea of 
this algorithm is to stratify each component into a checkerboard of 
rectangular regions and ensure that each row and column
of the checkerboard contains one and only one sample point.  This spreads out 
the samples in a quasi-random way that avoids undesirable clumping of sample points.  The
algorithm is implemented as follows \citep{larson_et_al_05a}.

Let $N_s$ be the number of SILHS sample points used. Including $u_{d+1}$ and $u_{d+2}$,
each sample point has $d+2$ variates, i.e., is $d+2$-dimensional.  Each of the $d+2$ uniform
variates is split into $N_s$ sections (i.e., checkerboard rows or columns): 
$\left(0,\frac{1}{N_s}\right),
\left(\frac{1}{N_s},\frac{2}{N_s}\right), \ldots,
\left(\frac{N_s-1}{N_s},1\right)$. Next, for each variate, an independent
permutation of the integers $\left(0,1,\ldots,N_s-1\right)$ is chosen,
corresponding to the $N_s$ rows/columns of the variate. These $d+2$ permutations
form a $N_s\times \left(d+2\right)$ matrix, $\bm{\Pi}$, where each column of the
matrix, $\bm{\Pi}_{\left(1\ldots N_s\right),j}$ is the permutation corresponding
to the $j$\textsuperscript{th} variate. Finally, we form another $N_s\times
\left(d+2\right)$ matrix, $\mathbf{U}$, each element of which is a random
uniform number between $0$ and $1$. The purpose of the matrix $\mathbf{U}$ is to
choose uniformly a location within each checkerboard square. Our sample matrix, $\mathbf{V}$,
is then given by: 
\begin{equation}
\mathbf{V} = \frac{1}{N_s}\left(\bm{\Pi} + \mathbf{U}\right)
\end{equation}

Each of the $N_s$ rows of $\mathbf{V}$ is a SILHS sample. In a given row of
$\mathbf{V}$, the value in each of the $\left(d+2\right)$ columns is the sample
value of the corresponding variate in the PDF.

\section{Computing importance sampling weights and scaling sample points}

\hypertarget{url:importance_sampling}{}
Importance sampling is another strategy used in SILHS for variance reduction.
The basic idea is to sample some parts of the PDF (the ``important'' regions)
more often than they would normally be sampled 
\citep{lemieux_2009a_monte_carlo,owen_2013_monte_carlo_book}.  
For instance, it may improve accuracy to place a disproportionate number of 
samples in regions of rain evaporation.
The discussion in this section follows \citet{raut_larson_2016_flex_sampling}.

First, the PDF is split into a set of disjoint categories, $C_j$. These
categories, which span the entire PDF, are currently defined as follows 
\citep[Table 1,][]{raut_larson_2016_flex_sampling}:
\begin{enumerate*}
\item In cloud, in precipitation, in mixture component 1 \\
($\chi > 0$,\quad$u_{d+2} < f_{p(1)}$,\quad$u_{d+1} < \xi_1$)
\item In cloud, in precipitation, in mixture component 2 \\
($\chi > 0$,\quad$u_{d+2} < f_{p(2)}$,\quad$u_{d+1} \ge \xi_1$)
\item Out of cloud, in precipitation, in mixture component 1 \\
($\chi \le 0$,\quad$u_{d+2} < f_{p(1)}$,\quad$u_{d+1} < \xi_1$)
\item Out of cloud, in precipitation, in mixture component 2 \\
($\chi \le 0$,\quad$u_{d+2} < f_{p(2)}$,\quad$u_{d+1} \ge \xi_1$)
\item In cloud, out of precipitation, in mixture component 1 \\
($\chi > 0$,\quad$u_{d+2} \ge f_{p(1)}$,\quad$u_{d+1} < \xi_1$)
\item In cloud, out of precipitation, in mixture component 2 \\
($\chi > 0$,\quad$u_{d+2} \ge f_{p(2)}$,\quad$u_{d+1} \ge \xi_1$)
\item Out of cloud, out of precipitation, in mixture component 1 \\
($\chi \le 0$,\quad$u_{d+2} \ge f_{p(1)}$,\quad$u_{d+1} < \xi_1$)
\item Out of cloud, out of precipitation, in mixture component 2 \\
($\chi \le 0$,\quad$u_{d+2} \ge f_{p(2)}$,\quad$u_{d+1} \ge \xi_1$)
\end{enumerate*}

Each category $C_j$ is associated with a certain amount of PDF mass, called the
category's ``PDF probability'' and denoted as:
\begin{equation}
p_j = \int \mathbf{1}_j(\mathbf{x}) P(\mathbf{x}) d\mathbf{x}
\end{equation}
where $\mathbf{1}_j(\mathbf{x})$ is the indicator function of $C_j$:
\begin{equation}
\mathbf{1}_j(\mathbf{x}) =
\begin{cases}
1 & \mathbf{x} \in C_j \\
0 & \mathbf{x} \notin C_j  .
\end{cases}
\end{equation}

Because the categories $C_j$ span the entire PDF, the PDF mass sums to unity:
\begin{equation}
\sum_{j=1}^{N_C} p_j = 1
\end{equation}
where $N_C$ is the number of categories (currently eight in CLUBB).

It can be assumed, without loss of generality, that $p_j > 0$ for all
categories, because in any category where $p_j = 0$, the corresponding portion of
the integral in \eqref{SILHS integral form} is zero, so the region of the PDF
belonging to that category can simply be left out of the integral.

It is worth noting that in general, the PDF masses $p_j$ might not always be
easy to compute. However, in the case of SILHS, because of the simple definitions of the
categories (given above) and the independence of cloud and precipitation, these
masses can be computed using information already provided to SILHS. For example,
the mass of category 1, where there is cloud and precipitation in mixture
component 1, is given simply by
\begin{equation}
p_1 = f_{c(1)} f_{p(1)} \zeta_{(1)} .
\end{equation}
Here, $f_{c(1)}$ is the fraction of (liquid) cloud in mixture component 1.
Next, we prescribe for each category another probability, $S_j$, called the
category's ``prescribed probability''. The probabilities must be prescribed such
that
\begin{equation}
\sum_{j=1}^{N_C} S_j = 1
\end{equation}
The prescribed probability $S_j$ of a given category is the probability that any
sample will fall in that category. In other words, it is the expected fraction
of sample points in the category. Therefore, intuitively, it is advantageous to
prescribe the probabilities such that the categories that are ``important'' for
a desired process are sampled more often than the unimportant categories.

We want to generate sample points such that the expected fraction of sample
points in the category $C_j$ is $S_j$ rather than $p_j$. In order to do this,
we define a set of new functions:
\[ L_j(\mathbf{x}) = 
   \begin{cases} 
      \frac{p_j}{S_j} & \mathbf{x}\in C_j \\
      0 & \text{otherwise} 
   \end{cases}
\]
We then rewrite the integral given in \eqref{SILHS integral form} as:
\begin{equation}
 \int h(\mathbf{x}) P(\mathbf{x}) d\mathbf{x} =
  \sum\limits_{j=1}^{N_C}\int h(\mathbf{x}) L_j(\mathbf{x}) Q_j(\mathbf{x})
  d\mathbf{x}
 \label{SILHS integral rewrite}
\end{equation}
where
\[Q_j(\mathbf{x}) = 
   \begin{cases} 
      \frac{P(\mathbf{x})}{L_j(\mathbf{x})} & \mathbf{x}\in C_j \\
      0 & \text{otherwise} 
   \end{cases}
\]

Here, $Q_j(\mathbf{x})$ is a set of new ``quasi-PDFs'' that has been resized to
fit our prescribed probabililties, $S_j$. Each $Q_j(\mathbf{x})$ has area $S_j$ for
$j=1,...,N_C$. 


Now, instead of drawing points from the $P(\mathbf{x})$ distribution and
evaluating the function $h(\mathbf{x})$, we draw points from the $Q(\mathbf{x})$
distribution, where
\begin{equation}
Q(\mathbf{x}) = \sum_{j=1}^{N_C} Q_j(\mathbf{x}),
\label{SILHS definition of Q}
\end{equation}
and evaluate the function $h(\mathbf{x}) L_j(\mathbf{x})$.  In the SILHS code,
drawing sample points from $Q(\mathbf{x})$ is accomplished by scaling the $\chi$ (cloud fraction),
$u_{d+1}$, and $u_{d+2}$ sample points so that the desired number of points lands in
each category, in the limit of an infinite sample.

Next, SILHS computes the weights needed to compute a weighted average
based on a finite-size sample.  
Based on the integral form given in \eqref{SILHS integral rewrite}, each sample
$h(\mathbf{x}_i)$ needs to be multiplied by $L_j(\mathbf{x}_i)$. This factor is
can be thought of as a 2D ``weight,'' $\omega_{ij}$:  
\begin{equation}
\omega_{ij} = L_j(\mathbf{x}_i) = \frac{p_j}{S_j}\ \mathbf{1}_j(\mathbf{x}_i) 
\label{SILHS importance weights def}
\end{equation}
for $i=1,...,N_s$ and $j=1,...,N_C$.  Each sample point has a 1D weight,
denoted as:
\begin{equation}
 \omega_{i} = \sum\limits_{j=1}^{N_C} \omega_{ij} .
\end{equation}
So if $\mathbf{x}_i \in C_{j'}$ for some category $j'$, then
\begin{equation}
 \omega_{i} = \frac{p_{j'}}{S_{j'}} .
\end{equation}
Thus each sample point $\mathbf{x}_i$ is associated with the weight
$\omega_i$ that corresponds to the category to which $\mathbf{x}_i$ belongs.


\section{Correlating samples in the vertical}

\hypertarget{url:vert_corr}{}
At this point, a collection of sample points in uniform space has been generated for a single
vertical grid level.  In a typical configuration, this level is the level within the grid column 
that has maximum within-cloud liquid water mixing ratio, regardless of whether there are multiple
cloud layers.  We have experimented with different methods for choosing the starting level,
such as using the \textit{grid-mean} liquid water, but the quality of sampling is relatively insensitive. 

Next, SILHS draws a sample from each other vertical grid level.  
Each vertical level in CLUBB has its own multivariate PDF, each of the form
given in \eqref{eq:CLUBB_PDF}.  This forms a vertical profile of sample points, i.e. a ``subcolumn",
that has the desired degree of vertical correlation.  Each grid level of each subcolumn in uniform space
is later transformed to the CLUBB PDF \eqref{eq:CLUBB_PDF} at that vertical level.

Each subcolumn physically represents a vertical column of space within a grid column. 
Speaking metaphorically, a subcolumn is a constant-thickness, vertical ``soda-straw" 
that is stuck into the atmosphere and pulled out.  There is some degree of vertical overlap in space. 
That is, in many cases,
the values of sample points are siFmilar between adjacent vertical levels
\citep[e.g.,][]{bergman_rasch_02a,barker_et_al_02a, pincus_et_al_03a, 
raisanen_et_al_04a,raisanen_barker_04a, raisanen_et_al_05a, pincus_et_al_06a}.

In SILHS, this is parameterized by controlling the vertical correlation between
uniform sample points. The process is as follows \citep{larson_schanen_2013a}. 

\begin{enumerate}

\item A vertical level,
$k_s$, is chosen to begin sampling.  (This variable is known as
\texttt{k\_lh\_start} in the code.)  By default, all subcolumns in a given grid column
use the same starting grid level, i.e. the same value of \texttt{k\_lh\_start}.  
E.g., it can be chosen to be the level of maximum cloud liquid water mixing ratio.

\item The vertical correlation $\rho_k$
(\texttt{vert\_corr} in the source code) is defined for each vertical level $k$ as:
\begin{equation}
\rho_k = \exp\left(\frac{-\alpha\ \Delta z_k}{L}\right) ,
\end{equation}
where $L$ is the CLUBB's turbulent mixing length scale \ref{sec:mixing_length}, 
$\Delta z_k$ is the vertical spacing between grid levels at level $k$, 
and $\alpha$ is a parameter, known in the code
as \texttt{vert\_decorr\_coef}, that controls the degree of vertical correlation.  
If $\alpha = 0$, then $\rho_k = 1$, which corresponds to maximum vertical correlation 
(or ``maximal overlap"). As
$\alpha \rightarrow \infty$, then $\rho_k \rightarrow 0$, which corresponds to
zero vertical correlation (or random overlap).

\item We form a correlated vertical profile of each of the $d+2$ uniformly 
distributed sample points.  To do so, we use a simple band-diagonal copula 
\citep{ovchinnikov_et_al_2016_overlap}.  Namely, between the ground and vertical level $k_s$, 
we set
\begin{equation}
u_{k}^{'} = u_{k+1} + u^* \left(1-\rho_k\right) ,
\end{equation}
where $u_{k+1}$ is the uniform sample at the height level immediately above $k$,
and $u^*$ is a uniform random number in the range $(-1,1)$. Between $k_s$ and the top model level,
\begin{equation}
u_{k}^{'} = u_{k-1} + u^* \left(1-\rho_k\right) .
\end{equation}

Depending on the value of $u^*$, there is a possibility that $u_{k}^{'}$ is not
in the range $(0,1)$, and so the value might need to be folded back into the
correct range. The actual uniform variate, $u_k$, is set to be a corrected
version of $u_{k}^{'}$. Specifically,
\begin{equation}
u_k =
\begin{cases}
2 - u_{k}^{'} & u_{k}^{'} > 1 \\
| u_{k}^{'} | & u_{k}^{'} < 0 \\
u_{k}^{'} & 0 \le u_{k}^{'} \le 1 .
\end{cases}
\end{equation}

\end{enumerate}

\section{Transforming the uniformly distributed sample to CLUBB's PDF}

\hypertarget{url:uniform2pdf}{}
At this stage in the calculation, each grid level contains a multivariate
sample that is uncorrelated and uniformly distributed within each category.  Each sample at each vertical level 
must now be transformed to CLUBB's PDF \eqref{eq:CLUBB_PDF} at that level. 
First, the sample value of $u_{d+1}$ is used to determine the mixture component
and $u_{d+2}$ is used to determine whether the sample is precipitating of not.
Next, the uniform values of the sample, excluding the $u_{d+1}$
and $u_{d+2}$ variates, are transformed from an uncorrelated uniform distribution 
to an uncorrelated standard normal sample using the inverse cumulative distribution function 
of the standard normal distribution:
\begin{equation}
\mathbf{Z} = \Phi^{-1} (\mathbf{u}) = 
\left(
\begin{array}{c}
\Phi^{-1} (u_1) \\
\Phi^{-1} (u_2) \\
\vdots          \\
\Phi^{-1} (u_d)
\end{array}
\right)
\end{equation}

A well-known result is that given a vector of uncorrelated standard normal
values, a sample following the desired joint normal distribution (say,
$\mathbf{x}_\mathrm{norm}$) can be obtained using the following formula 
\citep[Chapter 4,][]{johnson_87a}:
\begin{equation}
\mathbf{x}_\mathrm{norm} = \mathbf{L}\mathbf{Z} + \bm{\mu} ,
\end{equation}
where $\mathbf{L}$ is a matrix that satisfies
\begin{equation}
\mathbf{\Sigma} = \mathbf{L} \mathbf{L}^\mathbf{T} ,
\end{equation}
and $\mathbf{\Sigma}$ is the covariance matrix of the component distribution. 
The matrix $\mathbf{L}$ is calculated using the
Cholesky decomposition.


The last step is to transform from a joint normal distribution to a joint
normal-lognormal distribution. This can be done by applying the exponential
function to each log-normally distributed variate in the sample (while the
normally distributed variates remain the same).
\begin{equation}
\mathbf{x} =
\begin{array}{cc}
x_{\mathrm{norm},1} & \chi \\
x_{\mathrm{norm},2} & \eta \\
x_{\mathrm{norm},3} & w \\
\exp(x_{\mathrm{norm},4}) & N_{cn} \\
\exp(x_{\mathrm{norm},5}) & \mathrm{hm}_1 \\
\vdots & \\
\exp(x_{\mathrm{norm},d}) & \mathrm{hm}_n \\
\end{array}
\end{equation}

\section{Calculating a weighted average of the sample points}

When importance sampling is used, the sample points are picked according to the
distribution given by the PDF $Q(\mathbf{x})$, as defined in \eqref{SILHS
definition of Q}. The weight of each sample point, denoted as $\omega_i$, is
given in \eqref{SILHS importance weights def}. When importance sampling is not
used, the sample points are not weighted, and
\begin{equation}
\omega_i = 1 .
\end{equation}

Given SILHS sample points picked according to the distribution given by the
PDF $Q(\mathbf{x})$, an integral in the form of \eqref{SILHS integral form} can
be approximated using the following formulas. To integrate over the entire PDF,
use:
\begin{equation}
\int h(\mathbf{x}) P(\mathbf{x}) d\mathbf{x} \approx \frac{1}{N_s}
\sum_{i=1}^{N_s} \omega_i h(\mathbf{x}_i) ,
\label{SILHS integral estimator}
\end{equation}
where $\mathbf{x}_i$ is the $i$\textsuperscript{th} sample point and $N_s$ is
the total number of sample points.


\chapter{Comparison of CLUBB-SILHS with other parameterization methods}

\label{chapt:clubb_comparison}

CLUBB-SILHS is but one of many parameterizations, each of which has its own advantages and disadvantages.
One fundamental feature of CLUBB-SILHS that distinguishes it from other parameterizations 
is its emphasis on the green-bar integrals (see Eqns.~(\ref{eq_rtm_unclosed})-(\ref{eq_thlm_unclosed})).
The desire to perform those integrals accurately and explicitly is a chief reason that CLUBB-SILHS 
expends so much effort modeling the subgrid PDF.

\section{Comparison of CLUBB with low-order closure models}  

CLUBB prognoses 11 higher-order moments.  Doing so is complex and computationally expensive.  Low-order parameterizations
avoid some of this cost by prognosing fewer moments.  However, low-order parameterizations also omit certain terms,
which is a drawback.

\citet{mellor_yamada_82a} derive a hierarchy of equation sets, ranging from their Level 4 model, which prognoses
all second-order moments, to their Level 2 model, which diagnoses all second-order moments.  The derivation begins
by writing the equations in a tensor-invariant form.  For instance, the prognosed quantities are tensors 
(e.g., $\overline{u_i' u_j'}$) or vectors (e.g., $\overline{u_i' \theta_l'}$).  
All the equations within the tensor (e.g., $\overline{u'v'}$, etc.) are retained. 
However, the strongly anisotropic terms within each equation are dropped.    
At the simplest level, Level 2, an entirely diagnostic set of equations is obtained.  
Level 3 is attractive because it prognoses both the TKE and scalar variances, which is consistent
with scaling according to anisotropy and in principle allows upgradient scalar fluxes 
\citep{machulskaya_mironov_2013_tkesv}.
However, even at Level 4, all higher-order microphysical terms are omitted, 
and no third-order moments are prognosed.  

CLUBB's approach differs in a couple ways.  
In order to save computational expense, CLUBB predicts only the moments needed to constrain the PDF.  For instance, CLUBB 
prognoses $\overline{w'^3}$ but not $\overline{u'^3}$.  
But within each prognosed equation
that is retained, CLUBB keeps most if not all the terms.
Keeping information about third-order moments and keeping the microphysical effects on higher-order moments 
helps parameterize skewed, precipitating cumulus convection.



The Simplified Higher-Order Closure (SHOC) model uses the same PDF as CLUBB for closure, but it prognoses only one moment,
the turbulence kinetic energy \citep{bogenschutz_krueger_2013_shoc}.  SHOC diagnoses all other moments needed to specify the PDF.  
This is a considerable simplification and savings in cost.  However, SHOC's simplified diagnostic equations omit key processes.
For instance, SHOC diagnoses the scalar variances, $\overline{r_t'^2}$ and $\overline{\theta_l'^2}$, 
based on a balance of turbulent
production and scalar dissipation.  SHOC omits the turbulent advection and microphysical covariance terms.
But Fig.~\ref{fig:RICO_budgets_rtp2} shows that for the RICO case of drizzling shallow cumulus, 
these terms are larger at many altitudes than
the dissipation term that SHOC retains.


The Eddy-Diffusivity Mass-Flux (EDMF) approach parameterizes the scalar fluxes, 
$\overline{w'r_t'}$ and $\overline{w'\theta_l'}$, 
as the sum of a downgradient eddy diffusivity, representing small eddies, 
and a mass-flux contribution, representing strong organized drafts 
\citep[e.g.,][]{siebesma_et_al_07a,neggers_et_al_09a,neggers_09b,suselj_et_al_2013_stoch_edmf}.  
(For comparison, recall that SHOC includes an eddy diffusivity term but no mass-flux term.)
The approach is simpler and less expensive than CLUBB-SILHS, but it is
also more limited in scope.  Namely, although it parameterizes the scalar blue fluxes, 
it does not provide much information to perform the green-bar integrals.  
In addition, the EDMF approach inherits the assumptions and challenges of mass-flux schemes, 
noted earlier.


We see that the key trade-off between high-order and low-order closure is a trade-off between complexity 
and the number of physical processes included.

\section{Comparison of CLUBB with other third-order closure methods}  

This section compares and contrasts CLUBB with three alternative methods for closing higher-order
parameterizations. 

IPHOC is a third-order assumed PDF parameterization that forked off of the CLUBB code base 
in the early 2000s \citep{cheng_xu_06a,cheng_xu_08a}.  After the fork, IPHOC added prognostic equations
for $\overline{r_t'^3}$ and $\overline{\theta_l'^3}$.  Otherwise, IPHOC
is similar in formulation to CLUBB.  IPHOC does not include SILHS.  

Several authors have closed third-order moment equations and using 
the quasi-normal approximation \citep[e.g.,][]{andre_et_al_78a,bougeault_81b}.  
The quasi-normal approximation closes fourth-order moments,
such as $\overline{w'^4}$, in terms of second-order moments, on the assumption that the underlying subgrid PDF
is nearly normal.  This leads to inconsistency of assumptions if the parameterization 
prognoses large third-order moments. 
These parameterizations also prognose numerous third-order moments, 
unlike CLUBB, which prognoses only one.

Beside the quasi-normal approximation, other techniques for closing 
higher-order moment equations could be envisioned.
For instance, a machine learning method could be trained on LES output 
and used to close third-order moments (personal communication, Jeremy McGibbon).
Such a method would emphasize the third-order moments, in contrast to low-order
methods like SHOC, which set them to zero.
Machine learning in this context is essentially non-linear regression, 
and non-linear regression method would be expected to provide tighter fits than would integration
over an idealized subgrid PDF shape.
However, machine learning methods must be trained on a comprehensive database, 
particularly if they are intended to be used in climate simulations, 
which must handle all atmospheric conditions.  In addition, machine learning methods are opaque: 
it is difficult to understand why a machine learning method yields the answers it does.  
This is a drawback when strategizing how to modify the closures.  Finally, the same LES database 
must be used to train the machine-learning method and integrate all green-bar terms, 
including microphysical terms, 
lest there be no guarantee of internal consistency among all the terms. 

We see that a key trade-off between different strategies of closing higher-order moments is 
accuracy and convenience on the one hand versus internal consistency among closures on the other.  
Using a tailored technique to close a subset of terms is a convenient way to improve the accuracy
of that subset, but it cannot guarantee that closures within and outside the subset 
are consistent with the same underlying PDF.


\section{Comparison of CLUBB-SILHS and bulk mass flux schemes}

The overarching parameterization problem is the estimation of the blue fluxes and the green-bar integrals 
(see Chapter \ref{chapt:cloud_param_problem}).  
But what is the \textit{convective} parameterization problem?  It is the parameterization of 
the same blue fluxes and green-bar integrals, but over a multivariate 
tail of the subgrid PDF, when that tail includes large-magnitude vertical velocities.
The tail is where convection resides.  In contrast, the smaller-magnitude velocities
near the center of the subgrid PDF represent the environment.  (We focus on parameterizing PDF tails 
and neglect the problem of parameterizing 
the subgrid horizontal arrangement of convective elements, on the assumption that doing so
accurately and faithfully is simply too difficult.)   

Atmospheric PDFs often have long tails, because convective clouds are not uniform.  
For instance, cloud liquid water in a cumulus cloud exhibits
values ranging from barely supersaturated values near cloud edge to nearly adiabatic values in nearly undilute cores
(see, e.g., see Fig.~3 of \citet{larson_et_al_02a} or Fig.~11 of \citet{bogenschutz_et_al_2010a}).

Modeling tails of PDFs is notoriously difficult, and no parameterization method is likely to be fully satisfactory.

Because our goal is to point out similarities and differences between a particular assumed PDF parameterization, 
namely CLUBB-SILHS, and bulk mass-flux schemes,
we will write the mass-flux equations in a way that facilitates their comparison with assumed PDF methods.

How does a bulk mass-flux scheme approach the problem of parameterizing atmospheric tails?  Mass-flux schemes vary widely,
and we will describe only a handful of recent or operational schemes.  One commonality is that mass-flux schemes
attempt to model convective plumes directly, which is challenging.  In contrast, 
CLUBB-SILHS attempts to model only the \textit{effects} of plumes.  
We will start with the basic equations
for a prognostic \textit{spectral} (multi-plume) mass-flux scheme and then discuss the simplifications that would be needed 
in order to pass to a diagnostic \textit{bulk} scheme. 
Prototype equations for a multi-plume mass-flux scheme can be written as \citep{arakawa_schubert_1974,plant_2010_bulk_mass_flux}:
\begin{equation}
  \ptlder{\rho \sigma_i}{t} = E_i - D_i - \ptlder{M_i}{z}
  \label{eq:mass_flux_eq}
\end{equation}
\begin{equation}
  \ptlder{\rho \sigma_i s_i}{t} = E_i \overline{s} - D_i s_i - \ptlder{M_i s_i}{z} + L \rho c_i + \rho Q_{Ri}
  \label{eq:mass_flux_s}
\end{equation}
\begin{equation}
  \ptlder{\rho \sigma_i r_{ti}}{t} = E_i \overline{r_t} - D_i r_{ti} - \ptlder{M_i r_{ti}}{z} - R_i
  \label{eq:mass_flux_rt}
\end{equation}
where $i$ indexes the plumes, $\sigma_i$ is the (dimensionless) fraction occupied by the $i$th plume, 
$E_i$ and $D_i$ are the associated entrainment and detrainment rates with units of mass flux per vertical depth, 
$M_i \equiv \rho \sigma_i w_i$ is the mass flux of the $i$th plume, and $\rho$ is air density. 
In the latter two equations, the prognosed variables are the dry static energy, $s=c_pT+gz$, 
which is analogous to potential temperature $\theta$, and the total water mixing ratio (vapor plus liquid), $r_t$.   
The source terms are the $i$th radiative heating rate, $Q_{Ri}$, condensation rate $c_i$, and rate of conversion of 
liquid water to precipitation, $R_i$.

Some mass-flux schemes also add an equation with roughly the following form for vertical momentum conservation \citep[e.g.,][]{donner_1993_donner_deep,bretherton_et_al_04a}:
\begin{equation}
  \ptlder{\rho \sigma_i w_{i}}{t} = E_i \overline{w} - D_i w_{i} - \ptlder{M_i w_{i}}{z} + \mathrm{Buoyancy} - \mathrm{Pressure}   .
  \label{eq:mass_flux_w}
\end{equation}
In these equations, $\sigma_i$ is a small, discrete portion of the subgrid PDF: when we draw a sample of air at random
from the convective grid column, $\sigma_i$ indicates the probability of drawing plume $i$.  

The relationship
between equations (\ref{eq:mass_flux_eq})-(\ref{eq:mass_flux_rt}) and the underlying subgrid PDF can be clarified
by considering bulk quantities, in which the multiple plumes are represented by a single, average plume.
To do so, let us consider a continuum version of the plume equations.  Let $\lambda$ denote a continuous parameter
that indicates the presence of a convective plume when $\lambda > 0$ and the absence when $\lambda<0$.
Our parameter $\lambda$ is related to the $\lambda$ parameter in \citet{arakawa_schubert_1974}, except that
we do not assume that the value of $\lambda$ fully characterizes a plume.  Instead, we assume that the plume
depends independently on other quantities, such as vertical velocity $w$ and total water mixing ratio $r_t$.  
For instance, suppose that we choose to define the presence of a plume simply by the presence of liquid cloud.  
Then $\lambda$ would represent the super- or sub-saturation.  

Now let's write some fundamental quantities related to mass flux in terms
of continuum equations.  The resulting expressions will help elucidate 
the origin of complexity in mass-flux schemes.
The fraction of a grid box covered by convection is 
\begin{equation}
  \sigma \equiv \sum_i \sigma_i  
        \approx \int_{\lambda=-\infty}^{\lambda=\infty} 
                        H(\lambda) P(\lambda) \, d\lambda 
  \label{eq:bulk_conv_fraction}
\end{equation}
where $H$ is the Heaviside step function and $\sigma_i = P_i(\lambda) d\lambda$.
This equation states that the convective fraction is the fraction of the PDF with $\lambda>0$.
The Heaviside step function is needed because we assume that $P(\lambda)$ extends over the entire grid box,
rather than being confined to the plumes.  Now consider, for example, the calculation of the grid-mean bulk mass flux:
\begin{equation}
  M \equiv \sum_i M_i = \sum_i \rho \sigma_i w_i 
        \approx \rho \int_{w=-\infty}^{w=\infty} \int_{\lambda=-\infty}^{\lambda=\infty} 
                        H(\lambda) P(\lambda, w) \, w \, d\lambda \, dw  ,
  \label{eq:bulk_mass_flux}
\end{equation}
where here $\sigma_i = P_i(\lambda, w) \, d\lambda \, dw$.  Thus the mass flux is proportional 
to a weighted average of velocity in the convective regions.  Similarly,
the grid-mean value of $\rho r_t$ in the convective regions is
\begin{equation}
  \sum_i \rho \sigma_i r_{ti}  
        \approx \rho \int_{r_t=-\infty}^{r_t=\infty} \int_{\lambda=-\infty}^{\lambda=\infty} 
                        H(\lambda) P(\lambda, r_t) \, r_t \, d\lambda \, dr_{ti} ,
  \label{eq:bulk_rt}
\end{equation}
where here $\sigma_i = P_i(\lambda, r_t) \, d\lambda \, dr_t$.  A bulk mass-flux scheme 
writes equations similar to \eqref{eq:mass_flux_eq}-\eqref{eq:mass_flux_rt}, but with
plume quantities, denoted by subscript $i$, replaced by bulk quantities, as in 
\eqref{eq:bulk_conv_fraction}-\eqref{eq:bulk_rt}.

Working with expressions like these that are restricted to the tail of the PDF is challenging.
These expressions are conditional averages, conditioned on the presence of convection,
which is assumed to reside on the tail of the PDF.  
That is why $H(\lambda)$ appears in these expressions.
Some of the complex, unclosed terms that arise in equations for such conditional averages
are discussed in \citet{ larson_04a}.  It means, for one thing, that we must compute 
entrainment into and detrainment out of the plumes.

The tail represents a set of plumes with a complicated, irregular edge.  It is difficult to estimate 
the flux of heat and moisture into a region with such a complicated boundary.  
Moreover, mass-flux schemes must assume that $\sigma<<1$ if they wish to approximate the grid-mean blue fluxes, 
such as $\overline{w'r_t'}$, using the simple expression
\begin{equation}
  \rho \overline{w'r_t'} \approx \sum_i M_i \left( r_{ti} - \overline{r_t} \right) .
  \label{eq:mass_flux_wprtp}
\end{equation}
However, the assumption that $\sigma << 1$ restricts the usage of standard, unmodified mass flux schemes to coarse grid spacings.  
(Efforts to remove this restriction have been undertaken by, e.g., 
\citet{arakawa_2013_unified_mass_flux}, \citet{grell_freitas_2014_conv_parm}, 
and \citet{yano_2014_mass-flux}.)

To address some of these difficulties,
a bulk mass-flux approach makes simplifying assumptions.  First, simpler bulk schemes assume that 
the PDF is a double-delta function  
\citep{de_roode_et_al_00a,lappen_randall_01a,lappen_randall_01b,lappen_randall_01c,mironov_2008_2nd_order_review}.   
That is,
they assume that the convective region is internally uniform and that the environment is uniform.  This assumption
is not warranted by observations.  In fact, the tail is wide and is not well approximated by a narrow delta function
(e.g., see Fig.~3 of \citet{larson_et_al_02a}).
Even with the assumption of a double delta function, there remains the difficulty of computing entrainment and 
detrainment.  This difficulty results from the choice to work with conditional PDFs 
(see Eqns.~\eqref{eq:bulk_conv_fraction}-\eqref{eq:bulk_rt}) rather than PDFs that include 
all variability within the grid box, as does CLUBB.
Second, most bulk mass-flux schemes drop the time tendency term in the plume equation \eqref{eq:bulk_conv_fraction}.
The plume equation is usually diagnostic even in bulk schemes that prognose aspects of the convection 
at cloud base or below \citep{randall_pan_1993_mass_flux,emanuel_rothman_99a,chikira_sugiyama_2010_entrain,
park_2014_unicon_part1}.



Convective microphysics tends to be simple (aside from exceptions such as \citet{song_zhang_2011a}).  
This means that climate models that employ mass-flux schemes use the crudest microphysical assumptions in the regions
of the planet where the microphysical processes are most complex.
The use of sophisticated microphysics in convective schemes may be discouraged in part 
by the assumption that the convecting region
has uniform thermodynamic properties.  The assumption of uniformity would introduce errors 
even if a perfect microphysics scheme were employed
because cumulus clouds in nature are heterogeneous, and microphysics is nonlinear \citep{plant_2010_bulk_mass_flux}.  
In addition, the implementation of sophisticated microphysics is impeded by the assumption of a steady-state plume.
E.g., steady-state microphysics has a singularity
in the special case of a hydrometeor's growth when its fallspeed matches the plume updraft speed 
(Hugh Morrison, personal communication).   Furthermore, steady-state (diagnostic) microphysics has no memory
of previous values of hydrometeors and hence is limited in its capacity to represent the microphysical evolution
from newly activated particles to precipitation-sized particles.  
Finally, bulk schemes that omit an equation for vertical velocity, such as Eqn.~\eqref{eq:mass_flux_w}, 
have no updraft speed with which to activate aerosol or loft hydrometeors.


CLUBB-SILHS makes a different set of assumptions, with their own advantages and disadvantages.  
First, CLUBB-SILHS is not based on
\textit{conditional} moments, conditioned on the presence of convection.  
Instead, CLUBB-SILHS parameterizes the \textit{unconditional} subgrid PDF,
which encompasses all variability within a grid box.  While an unconditional PDF 
does not focus on the convective region, which
is perhaps a drawback, it is simpler to work with an unconditional PDF
because it does not have a complicated interface with the environment; rather, it includes the environment
as part of the PDF.  For instance, the width of an unconditional PDF is determined 
by the variance over the entire grid box \eqref{eq_wp2_unclosed}, 
which is easier to prognose than the variance over just the convective plumes.
Second, the use of an unconditional PDF shifts the focus from entrainment into the tail of the PDF
to a focus on turbulent mixing throughout the full width of the PDF.  Thus the entrainment
parameterization is replaced by a turbulent dissipation parameterization \citep{de_roode_et_al_00a}.
Estimating dissipation may turn out to be just as challenging as estimating entrainment, 
because information on spatial organization is lacking, but at least
dissipation has a more precise mathematical definition, which facilitates comparison with LES.
Third, instead of assuming a double-delta PDF, CLUBB-SILHS assumes a mixture of normal/lognormal components,
because such a PDF shape, while more complicated, allows within-plume variability and has a plausibly shaped tail.  
Fourth, most of CLUBB's moment equations are prognosed, which is expensive, 
but preserves a detailed memory of the variability at the prior time step.  Fifth, instead
of using a separate, simplified microphysics for convection, CLUBB-SILHS simply calls 
the same microphysics that is used for all other cloud types.



\section{Comparison of CLUBB-SILHS with suites of cloud parameterizations}

Most climate models use separate parameterizations for stratiform clouds, shallow cumulus clouds, 
and deep cumulus clouds \citep[e.g.,][]{neale_et_al_2012_cesm_descr,donner_et_al_2011_am3}.  
The cumulus parameterizations are triggered or activated only when buoyant instability is diagnosed.  
In such climate models, all three parameterizations must work together in order to span 
the variety of clouds in the atmosphere.
This ``separate schemes for separate regimes" approach embodies a different philosophy than the unified approach 
employed by CLUBB-SILHS.  A unified cloud parameterization like CLUBB-SILHS strives to create a single equation set 
that contains enough physics to parameterize all cloud types.  The difference in philosophy between the 
unified versus separate-schemes approaches leads to differences in how the two approaches are best 
modified and interpreted. 

One difference is the location of the model complexity.  In CLUBB-SILHS, 
the complexity resides in the equation set
itself, which contains nine prognostic equations with terms for 
most if not all relevant sources and sinks.  In a separate-schemes climate model, 
each scheme is simpler, but complexity resides 
in the interactions between cloud schemes.  Diagnosing and fixing errors in the interactions can be difficult 
\citep{bretherton_07a,zhang_bretherton_08a}, especially when each of the separate schemes is developed by 
a separate research group.  On the other hand, in CLUBB-SILHS, time must be spent analyzing budgets in order to
gain understanding about how the processes (that is, budget terms) interact. 

Because interactions between cloud schemes are delicate and influential, when a new parameterization is 
inserted into a separate-schemes suite, time must be spent during implementation 
to ensure that the different cloud schemes work well together.  In contrast, when CLUBB-SILHS is newly implemented 
as a unified parameterization in a climate model,
the only direct interactions are with microphysics, aerosols, and radiation. 
CLUBB-SILHS does not need to ``play well" with other cloud schemes because no other cloud schemes
are enabled when CLUBB-SILHS is running.
Hence, the new implementation's behavior is more predictable.
Namely, it can be expected to have similar errors as in prior implementations.

A third difference is the method of tuning.  Each cloud scheme in a separate-schemes climate model 
can be tuned separately, which is impossible in a climate model with a single unified parameterization.  
It is perhaps easier to tune a separate-schemes climate model, but it is still challenging to tune
transitional cloud regimes, e.g., between stratiform and shallow cumulus regimes.
Advice on tuning unified parameterizations is given in Section~\ref{sec:tuning_guide}.

Another difference is the method of comparison with observations.  Traditional parameterization suites output
familiar quantities, like stratiform and convective rain, whereas when CLUBB-SILHS is used,
there is only one source of rain.  Therefore, it may seem that CLUBB-SILHS outputs less information.  However,
it is difficult to distinguish stratiform and convective rain in the observations in the same manner that they are
distinguished in the model.  An observational analysis distinguishes the two by the \textit{properties} of the rain,
whereas the model distinguishes the two by the \textit{source} of the rain (i.e., whether it originates 
from a stratiform or convective scheme).  If a user wishes to separate convective and stratiform rain
in CLUBB output, it is perhaps best to use whatever statistical method is used for the observations.


\section{Comparison of CLUBB-SILHS with the implementations of CLUBB in CAM6 and E3SMv2}

\subsection{Use of an additional deep convective parameterization}

In the CAM6 and E3SMv2 climate models, CLUBB unifies the parameterization of 
the planetary boundary layer, shallow convection, and macrophysics.
However, the CAM6 and E3SMv2 climate models run CLUBB simultaneously with the Zhang-McFarlane (ZM) deep convective scheme
\citep{zhang_mcfarlane_1995a}.  Therefore, in these two models, the parameterization suite is only partially unified.   

In these models, CLUBB and ZM each supplies tendencies at all altitudes.
In this sense, both CLUBB and ZM's tendencies ``compete" with each other, and CLUBB has the possibility, for instance, 
of moderating ZM's strength.  If CLUBB and ZM are tuned appropriately, they appear to be quite capable of
 complementing each other's strengths and weaknesses.

What are the differences in implementation between CAM-CLUBB-SILHS on the one hand and CAM6 and E3SMv2 on the other?  
In CAM6 and E3SMv2, SILHS is shut off.  Its expense is perhaps unwarranted 
given that it would not represent variability in convective clouds.  Instead, the coupling assumptions 
between stratiform cloud and stratiform precipitation are embedded in the microphysics scheme 
(i.e., Morrison-Gettelman version 2, \citet{gettelman_morrison_2015_mg2_part1}).  In addition, 
these models have been tuned differently than CAM-CLUBB-SILHS.     
Otherwise, there are no significant differences in implementation between CAM6 and E3SMv2
versus CAM-CLUBB-SILHS.

\subsection{Diagnosis of the subgrid momentum fluxes ${\color{cp} \overline{u'w'}}$ and ${\color{cp} \overline{v'w'}}$
}

Since 14 Aug 2018, momentum fluxes have been prognosed by default in a standard clone of the CLUBB-SILHS repository.  
However, in CAM6 and E3SMv2, CLUBB diagnoses the momentum fluxes.

The momentum fluxes are closed using a down-gradient approach \citep{golaz_et_al_02a}:
\begin{subequations}
\begin{equation}
\label{eq_upwp_ed_dfsn}
{\color{cp}\overline{u'w'}} = -{\color{cd}K_m} \ptlder{\color{hp}\overline{u}}{z}
\end{equation}
\begin{equation}
\label{eq_vpwp}
{\color{cp}\overline{v'w'}} = -{\color{cd}K_m} \ptlder{\color{hp}\overline{v}}{z} .
\end{equation}
\end{subequations}
The momentum fluxes, $\overline{u'w'}$ and $\overline{v'w'}$, are also subject 
to explicit clipping.  The eddy diffusivity for momentum, $K_m$, is 
given by:
\begin{equation}
\label{eq_Km}
K_m = c_{K10} K_h,
\end{equation}
where $c_{K10}$ is a constant and $K_h$ is the eddy diffusivity for scalars:  
\begin{equation}
\label{eq_Kh}
K_h = c_K \, L \, \overline{e}^{\, 1/2} ,
\end{equation}
where $c_K$ is another constant.  

\chapter{CLUBB-SILHS FAQ}

\label{chapt:faq}





\section{Can CLUBB parameterize penetrative convection?}

\begin{itemize}


\item \textit{How can a PDF parameterization represent vertically correlated buoyant plumes (i.e., ``hot towers")?  
Where does the physics of vertically coherent plumes reside in CLUBB?}

In CLUBB, the physics of plumes resides in the equations that prognose vertical turbulent fluxes.  Unlike a mass-flux 
scheme, which contains a direct, if phenomenological, model of a plume itself, CLUBB models 
only the \textit{effects} of plumes.

What do plumes accomplish in nature?  For one thing, updrafts in plumes transport moisture upward.  
This process is represented in 
CLUBB by the vertical turbulent flux, $\overline{w'r_t'}$.  Vertical \textit{coherence} of the plume is represented by 
a positive value of the flux over the \textit{depth} of the convective layer.  

In nature, convective plumes are strengthened by buoyancy and damped by pressure perturbations.  These effects are
contained in CLUBB's prognostic equations for $\overline{w'r_t'}$ (\ref{eq_wprtp}), $\overline{w'\theta_l'}$
(\ref{eq_wpthlp}), and in other moments. 

The narrow, strong updrafts and weak, broad downdrafts that are observed in precipitating convection
are represented in CLUBB by large, positive values of skewness.

\item \textit{How can a PDF parameterization model non-local vertical transport?}


What does ``non-local" mean?  It means transporting a quantity more than one vertical grid level 
in a given time interval, such as a model physics time step of 30 min.  The duration of the time interval
influences how much non-local transport occurs.

CLUBB parameterizes non-local transport much like a LES model does.  A LES model does not contain any explicit model 
of plumes or non-local transport.  It contains only derivatives, which are local.  Nonetheless, a LES represents 
non-local transport by transporting parcels a little bit over each of many short time steps. 

In climate simulations, CLUBB typically uses a time step of 5 minutes.  A 10-m/s updraft can travel 
3 km upward in 5 minutes.  But CLUBB does not attempt to track the drafts within individual plumes;
it only attempts to model the evolution of higher-order moments.  The moments' evolution is slower and is more akin
to the rate at which cloud top rises.  This is because as a convective layer deepens, 
successive parcels must rise and moisten the environment 
before a convective layer deepens.

Some people define non-local transport as the transport of scalars between two vertically separated grid levels
without the exertion of any influence on the layers in between.  I suspect that this does not happen in the atmosphere.
Instead, clouds often undergo buoyancy sorting and hence detrain over a range of altitudes 
(see, e.g., Fig.~9 of \citet{carpenter_et_al_1998_cu_entrain}).
Additionally, within a grid-column-sized section of a field of clouds, different clouds often have 
different cloud-top altitudes; in such cases, a horizontal average will show non-zero fluxes at 
intermediate altitudes.

\item \textit{How is deep convection triggered in CLUBB?}

In nature, CAPE often builds up long before deep convection initiates.  Initially, the convection is
suppressed by a capping inversion at the top of the atmospheric boundary layer.  Only later does convective turbulence
break through the cap and initiate deep convection.

In mass-flux schemes, an explicit trigger function is often used to delay convection after CAPE has built up.  
In contrast, CLUBB contains no explicit trigger function.  Instead, CLUBB's entire equation set serves 
to model the build-up
of boundary layer turbulence and the erosion of the cap.  That is, in CLUBB, the triggering of deep convection
is an emergent property.

\item \textit{Where does CLUBB contain convective mass fluxes?}

CLUBB does not predict mass fluxes explicitly, but they are present implicitly, and they could be diagnosed
by integrating CLUBB's marginal PDF of vertical velocity, $w$, over the convective area. 

\item \textit{How can a PDF parameterization handle deep convection, given that deep convection occurs on the tail of the PDF?}

It is true that cumulus clouds often reside entirely on the tail of the subgrid PDF 
\citep[see, e.g., Fig.~3 of][]{ larson_et_al_02a}.  Therefore, it is of paramount to represent the tails accurately.
However, a basic mass flux scheme assumes that the variability is distributed according to a double delta function.
This is a poor assumption, especially in the tails, as can be seen in Fig.~3 of \citet{ larson_et_al_02a}.  
All parameterizations struggle to accurately model the tails, but a normal mixture PDF has more realistic tails 
than does the double delta PDF implied by the mass-flux approach.

\item \textit{How can a PDF parameterization handle deep convection, given that CLUBB uses pressure and dissipation parameterizations derived from the Kolmogorov inertial subrange?}


It is true that dry and moist turbulence are different.  Dry convection fills the domain with turbulence, 
whereas in a deep convecting layer, turbulence is confined mostly to the small volume 
occupied by deep cumulus clouds.  The spatial structure of turbulence intensity is not contained
in the subgrid PDF.  Nevertheless, the physics of dissipation and perturbation pressure 
in dry and moist convection has commonalities.  In both cases, turbulent dissipation 
mixes where turbulence intensity is large.  In both cases, pressure forces isotropize the turbulent components
and still oppose the acceleration of parcels under buoyant or other forces.  

Certainly improved dissipation and pressure terms would be welcome, but the existing clsoures may be adequate,
 even for deep convection.  The dissipation and pressure terms respond to sources of variability, 
and they usually damp it.
CLUBB tends to be more forgiving of errors in such responder terms.  The errors in existing closures may not have 
inordinate effects on the quantities that we care about.



\item \textit{Can a PDF parameterization model convective organization?}  

A subgrid PDF contains no direct information about the spatial arrangement of parcels.  
It does not know, for instance, 
whether cloudy elements are dispersed throughout the domain or clumped together into one mass.
The lack of information on spatial structure makes it more difficult to parameterize terms that 
cannot be closed merely by integration over the subgrid PDF, namely, the dissipation and pressure terms.
For instance, it is difficult to parameterize scalar dissipation rate generally,
because small structures dissipate more rapidly than large ones, and information
on structure size is not contained in the PDF.

Nevertheless, in many cases, it may be sufficient to parameterize some \textit{effects} of convective organization 
rather than the spatial arrangement of parcels per se.  Consider, for example, the diurnal cycle of 
surface rainfall.  Although the timing may depend on whether the rain falls from isolated cumulonimbus 
from or an organized meoscale convective system, it may be sufficient merely to parameterize 
the convective time lags and feedbacks, as done, for example, by \citet{mapes_neale_2011a} 
using their ``org" variable.  


\item \textit{Can CLUBB represent cold pools?}

Although a subgrid PDF cannot directly represent the spatial arrangement of cold pool air, 
in principle it can model some effects of cold pools.  Namely, if rain falls into cold subcloud air and evaporates,
then cold air will be further cooled, forming or strengthening cold pools.  This physics is represented by 
the $mc$ term in the prognostic equation for $\theta_l'^2$ (\ref{eq_thlp2_unclosed}).  Indeed, 
Fig.~\ref{fig:RICO_budgets_thlp2} demonstrates that CLUBB can be configured to
increase variability in temperature near the ocean surface, which is the process that creates cold pools.

\item \textit{Can CLUBB represent elevated convection?}


In principle, yes.  In practice, it has never been tested.  CLUBB's length scale represents the vertical distance a parcel
can travel under the action of buoyancy.  It is large wherever the atmosphere has buoyant instability.  Where the length
scale is large, convection can form.  Convection in CLUBB is not hard-wired to originate at the surface.


\end{itemize}

\section{CLUBB's PDF}

\begin{itemize}

\item \textit{Does CLUBB have separate univariate PDFs for $r_t$, $r_r$, $r_i$, etc., or a single, multivariate PDF?}


CLUBB has a single, multivariate PDF (see Eqn.~\ref{eq:CLUBB_PDF}).  This is important for
processes like accretion that depend on the correlation of two or more variates.

\item \textit{Is CLUBB's PDF static?}

No.  CLUBB's PDF is not a prescribed, climatological PDF.  Although its shape
is prescribed to be a normal/lognormal mixture, its mean, width, and skewness
vary with space and evolve in time as CLUBB's moments vary and evolve.
CLUBB's PDF oozes.    
E.g., CLUBB is designed to produce a more skewed PDF in cumulus layers
and a less skewed PDF in stratiform layers.

\item \textit{How can CLUBB's limited set of moments determine the PDF shape?}

A normal mixture PDF depends on more PDF parameters than the number of moments that CLUBB prognoses.
To obtain closure, assumptions about the PDF shape are made that restrict its shape.  
For more information, see Section \ref{sec:clubb_pdf}.


\item \textit{Is CLUBB's PDF general enough?}


In a unified PDF model like CLUBB, the PDF shape needs to be general enough to approximate
distributions found anywhere on the globe.  Is this achieved?  The ability of CLUBB's PDF to 
adequately diagnose basic cloud-related quantities in both shallow and deep clouds 
was tested in \citet{bogenschutz_et_al_2010a}.  The global simulations
of \citet{thayer-calder_et_al_2015_cam_clubb_silhs} suggest that the errors in CLUBB's PDF
shape are small enough to allow successful interactive simulations.

\item \textit{How does CLUBB-SILHS predict the correlation between hydrometeor species?}


CLUBB-SILHS merely prescribes the correlation matrix based on a LES of deep convection.
Clearly this assumption could be improved.  However, we have found that 
CLUBB-SILHS simulations are often more sensitive to values of the hydrometeor variances
than the hydrometeor correlations.

\item \textit{Is convection parameterizable at grey-zone resolutions with a deterministic parameterization like CLUBB?
                     Or is a stochastic scheme necessary?}


A convective flow is parameterizable if the parameterized moments capture enough of the convective physics to 
uniquely determine the short-term evolution of the flow to an acceptable approximation.



At grey-zone grid spacings (e.g.~10 km), the evolution of deep convection is not determined solely by its grid means. 
Rather, the evolution also depends on subgrid variability.  Two different convective flows that have 
identical grid means but different subgrid distributions will rapidly diverge from one another.  
This has led some researchers to introduce stochasticity in order to parameterize convection at grey-zone
grid spacings.  


However, CLUBB prognoses not only the grid means but also some higher-order moments.  
The higher-order moments specify the flow more uniquely and thereby help constrain the flow evolution.
If two simulations start with all the same means and 2nd-order moments, then those two flows will evolve 
more similarly to each other than will two flows that share only the same grid means.  
Given this, we would expect higher-order closure models to offer greater parameterizability
and have less need to introduce stochasticity.

\end{itemize}

\section{What are the effects of CLUBB's tunable parameters?}

\begin{itemize}




\item \textit{Can one predict how a change in one of CLUBB's tunable parameters will change model behavior?}


Tuning global models that use a unified parameterization like CLUBB may be more difficult than tuning global models
that use separate schemes for separate regimes.  If a model contains separate cumulus and stratocumulus
regimes, then model errors in, e.g., stratocumulus regions can be targeted in a more focussed way. 
Despite the difficulty in tuning a unified parameterization, we are gaining experience in tuning CLUBB.  
Automated tuning exercises have yielded
useful information about parameter sensitivities \citep{guo_et_al_2014_scamclubb,guo_et_al_2015_camclubb}.
Manual tuning is also helping us to develop insights.  See Section \ref{sec:tuning_guide} for more information.

\item \textit{Does CLUBB contain too many tunable parameters?}

Each parameterized term in a higher-order closure model must contain a tunable parameter (at least implicitly).
This does lead to more parameters in higher-order models than in low-order closure models.  
However, adjusting a parameter that prefixes
a minor term in a budget equation leads to only minor changes in the solution.  Because of this, the number 
of tunable parameters in a model is not a perfect metric of true model complexity.  In principle, we could omit 
such minor terms, thereby removing their parameters from CLUBB, but it costs little to compute such terms, 
and we may encounter situations in which they matter.

CLUBB does sometime allow the use of multiple parameters in front of one budget term.
For instance, one term in the $\overline{w'^3}$ equation contains three parameters: $C11$, $C11b$, $C11c$.
The use of three coefficients appeared to provide benefits in the single-column simulations of \citet{golaz_et_al_07a},
but it is not clear that there are large benefits in global simulations.  CLUBB's code keeps 
the separate parameters available,
but setting $C11b=C11$ renders $C11c$ irrelevant and effectively reduces the number of parameters from 3 to 1.

\end{itemize}

\section{How general is CLUBB?}

\begin{itemize}

\item \textit{Are CLUBB's tunable parameters universal properties of moist air turbulence?} 


As with all parameterizations, CLUBB's optimal parameter values are different for different cloud regimes 
\citep[e.g.,][]{golaz_et_al_07a,guo_et_al_2014_scamclubb}.  However, CLUBB-SILHS produces plausible results
in global simulations \citep{thayer-calder_et_al_2015_cam_clubb_silhs}, suggesting that the regime dependence is not severe.

\item \textit{ Is CLUBB applicable for paleo climate simulations?}


CLUBB has not been tested for paleo simulations.  Regardless, only the \textit{shape} of CLUBB's PDF is 
empirically prescribed using present-day data or LES.  The moments are prognosed 
and hence more universal, in principle.  
The moments are a strong constraint: e.g., the mean and standard deviation determine the PDF's position and width.  
The empirical prescription of shape affects only the tails of the PDF.

\end{itemize}

\section{What is the computational expense of CLUBB?}

The cost of CAM-CLUBB is 1.67 times that of CAM5.  The cost of CAM-CLUBB-SILHS is 2 times that of CAM5
\citep{thayer-calder_et_al_2015_cam_clubb_silhs}.  For more details, see Section \ref{sec:timing}.

\section{Level of complexity in CLUBB}


\begin{itemize}

\item \textit{Can the number of prognosed moments in CLUBB be reduced?}

Eliminating the time tendency terms in CLUBB's equations, and only those terms, would not save much computational expense,
nor would it facilitate numerical solution.  
Rather, diagnosing a moment instead of prognosing it would simplify the algorithm and reduce computational expense
only if a significant number of terms in the prognostic equation were dropped.  
But dropping terms would also discard desirable aspects of CLUBB's formulation.
For example, diagnosing scalar variances as a two-term balance between production and dissipation
would discard the turbulent advection and microphysical terms (see, e.g., Eqn.~\ref{eq_rtp2_unclosed}),
which are major terms in the budgets (Fig.~\ref{fig:RICO_budgets_rtp2}).
To cite a different example, if $\overline{w'^3}$ is diagnosed in terms of 2nd-order moments, then CLUBB would need
to distinguish skewed cumulus from less-skewed stratocumulus without the benefit of 
independent information on the skewness of $w$.  Likewise, the equations for $\overline{u'^2}$ and $\overline{v'^2}$ 
could be eliminated and the turbulence assumed to be isotropic, but in fact turbulence is often highly anistropic 
near the ground and strong inversions.  If important terms are omitted, either the results will have errors, 
or a compensating error will have to be introduced.

Balancing fidelity and cost involves difficult choices!

\item \textit{Why does CLUBB have so many lines of source code?}

CLUBB is a framework, which is to say, CLUBB contains many code options.
Many lines of code in CLUBB implement non-default options.  In addition,
CLUBB contains extensive code comments.

For these reasons, the number of lines of code is not a perfect metric of complexity.

\end{itemize}

\section{CLUBB's treatment of microphysics}

\begin{itemize}

%
%






\item \textit{Is liquid cloud fraction diagnosed by CLUBB?  If so, where is the memory in CLUBB?}    


Liquid cloud fraction is diagnosed by CLUBB.  The rationale is that small droplets
evaporate and condense quickly, and hence can be modeled by saturation adjustment.
CLUBB's saturation adjustment, however, is consistent with CLUBB's subgrid PDF.

Like prognostic parameterizations of cloud fraction \citep[e.g.,][]{tiedtke_93a},
 CLUBB's treatment of liquid clouds still has memory.  In CLUBB, the memory resides in
the moments, which are prognosed.  The advantage of representing the memory in the moments
is that, unlike cloud fraction, the moments are based on conserved variables.

\item \textit{How does CLUBB parameterize ice cloud fraction?}  


Unlike liquid clouds, ice clouds in nature sublimate and deposit slowly.  CLUBB-SILHS estimates ice cloud fraction
by the fraction of a grid box that is super-saturated with respect to ice.  This fraction is calculated as 
the area under CLUBB's PDF that exceeds ice saturation.  Clearly, this diagnostic assumption could be improved.

\item \textit{Does CLUBB's total water variable include ice?}


CLUBB's total water variable, $r_t$, includes vapor and liquid cloud water, but not ice.
However, CLUBB's PDF includes variates for cloud ice mass and number mixing ratios.  
The grid means are calculated by the microphysics, and the variances are assumed to be proportional
to the mean squared \citep{griffin_larson_2016_hydrometeor_PDF}.

\item \textit{Can CLUBB-SILHS predict the effects of microphysics on the PDF shape?}


Yes.  CLUBB-SILHS predicts the effects of microphysical processes on the second-order moments 
(see the terms with subscript ${mc}$ in Eqns.~(\ref{eq_wprtp_unclosed})-(\ref{eq_rtpthlp_unclosed})).
In this way, CLUBB-SILHS allows microphysical processes to narrow or widen the subgrid PDF.
These microphysical terms can be computed quite generally using SILHS, because the
needed moments can be estimated as sample covariances.



\end{itemize}

\section{How can CLUBB-SILHS best be compared to observations?}

The quantities in CLUBB's prognostic equations are defined precisely: they are simply horizontal averages of various sorts.
Their clear meaning facilitates comparison with LES and observations.  Although some common observables 
(e.g., updraft sizes and cloud depth) do not correspond in a one-to-one way to CLUBB variables, other quantities
that do appear explicitly in CLUBB's equations (namely, the moments) are observable, despite the lesser attention
in observational studies they have received to date.

\begin{itemize}

\item \textit{How can CLUBB be best compared to LES?}

For the purpose of either diagnosing the sources of an error in CLUBB or else calibrating a CLUBB parameter, 
the most detailed store of information is LES output.
CLUBB predicts moments based on budget terms, and most of these can be output from LES, 
providing a rather complete term-by-term
connection between the building blocks of CLUBB and ``observations." 

Because CLUBB is designed to emulate LES, CLUBB's output --- both moments and budget terms --- is directly 
comparable to LES: a horizontal average in CLUBB corresponds to a horizontal average
over the area of the LES domain.  

The capability of making detailed comparisons is a significant advantage over mass-flux schemes,
which are more phenomenological.  For instance, it is difficult to define lateral entrainment 
in a way that is both precise
and easy to diagnose in a LES.  For instance, in an environment that is inhomogeneous, 
entrainment of nearby, moist air may have
quite different consequences on cloud evolution than entrainment of more distant, drier air.  
Different definitions of entrainment can yield entrainment rates that differ 
by up to a factor of 2  \citep{romps_2010_entrain}.  

In contrast, consider the analogue of entrainment in higher-order closure models, namely, turbulent dissipation.  
Although diagnosing turbulent dissipation from a LES depends on uncertainties in the LES subgrid scheme, 
turbulent dissipation is well-defined mathematically, is easy to output from a LES, and is directly comparable to CLUBB
(see, e.g., Fig.~\ref{fig:RICO_budgets_rtp2} or \citet{griffin_larson_2016_microphys_covar}).  
In fact, CLUBB's complete budgets can be compared term by term with LES, allowing
a more detailed diagnosis of CLUBB's errors than is possible with mass-flux schemes. 

\item \textit{How can CLUBB be best compared  to aircraft observations or scanning radar?}

Either aircraft observations or scanning radar can produce horizontal averages of low-order moments
that can be compared to CLUBB, assuming that accurate initial and boundary conditions are available
to configure a CLUBB simulation.   In principle, spatial PDFs can also be observed, but in practice,
the statistics of the PDF tails (or higher-order moments) are often noisy \citep{mann_lenschow_1994_aircraft_errors}.  

Observations of conditional averages, such as cloud fraction, can be compared to existing diagnostic outputs
from CLUBB.  Other conditional averages, such as the mass flux, could be derived from CLUBB's subgrid PDF.

\item \textit{How can CLUBB-SILHS be compared to vertically pointing remote-sensing instruments?}

Both ground-based upward-staring radars and lidars, and satellite-based instruments, can provide useful information
that can be compared to CLUBB-SILHS output.  In particular, the subcolumns produced by SILHS are comparable to 
vertically pointing observations, and many useful statistics, such as the variance of liquid water path, 
can be easily computed from a collection of subcolumns.  

\item \textit{How can convective and stratiform rain best be distinguished in diagnoses of global simulations using CLUBB?}  


It is difficult to distinguish, say, stratiform versus convective rain 
in the same way in observations as in a model.  E.g., an observational analysis might distinguish the two
based on a characteristic of the observed rain itself, whereas a model might distinguish them based on a 
parameterized function that triggers a shallow convection scheme. 

Rather than creating arbitrary definitions of cumulus vs.~stratiform clouds, it may be preferable
to think in terms of conditional averages, e.g., rain rate conditioned on cloud fraction.  In this way,
observations and CLUBB can be compared in an apples-to-apples way.

\end{itemize}



\section{How does CLUBB's behavior change as the grid spacing is changed?}

\begin{itemize}

\item \textit{How does CLUBB adjust to changes in the horizontal grid spacing?}

At large horizontal grid spacings, we want the turbulent fluxes to be produced by the subgrid parameterization (CLUBB).  
At small enough grid spacings, the convective motions will be resolved, and we want CLUBB to shut itself down.  

To achieve this, CLUBB limits its turbulent mixing length to a fraction of the horizontal grid spacing 
\citep{larson_2012_2km_16km}.  The fraction is chosen to be one-fourth, based on empirical experimentation.  
As the grid spacing is reduced,
the mixing length and hence turbulent dissipation time scale are reduced.  This reduces the magnitude of the turbulent moments.  
Since CLUBB's length scale in convective flows might have a length scale of several hundred meters, 
the reduction does not kick in until the grid spacing falls below approximately 2 km.  At such fine grid spacings,
however, the host model may wish to include horizontal advection terms, 
which are discussed in Chapter \ref{chapt:clubb_closed_eqns}.


\item \textit{Can higher-order closure work at coarse vertical grid spacing?}

In single-column simulations, CLUBB produces more accurate results at high vertical resolution than low vertical resolution.  
Parameterizing clouds at coarse vertical grid spacing
is inherently difficult.  When a host model contains only, say, 30 vertical grid levels, 
then there is little information available to distinguish boundary-layer cumulus and stratocumulus.  
For instance, at such a coarse grid spacing, the model cannot resolve a stratocumulus cloud-top inversion.  
With only 30 grid levels, even interpolating profiles in the vertical is difficult: 
a single interpolation of a thin stratocumulus layer can match either 
peak liquid water content within the profile or vertical integrated liquid water path, but not both.  

To combat these problems, a parameterization could, for example, attempt to build in assumptions 
about the strength of inversions.  But it would be difficult to find a general formula 
that works for both cumulus and stratocumulus layers. 

\end{itemize}

\section{CLUBB's treatment of momentum}

\begin{itemize}




\item \textit{Does CLUBB represent the effects of vertical wind shear?}


In CLUBB, mean wind shear generates turbulence kinetic energy 
(TKE $= 1/2 (\overline{u'^2}+\overline{v'^2}+\overline{w'^2}$)).  A wind shear generation term 
appears on the right-hand side of the prognostic equations for $\overline{u'^2}$ (Eqn.~\ref{eq_up2_unclosed}) 
and $\overline{v'^2}$ (Eqn.~\ref{eq_vp2_unclosed}).  
The generation of $\overline{u'^2}$ and $\overline{v'^2}$ 
is redistributed to $\overline{w'^2}$ by two parameterized pressure terms, the return-to-isotropy
 and ``slow" pressure terms.    The values of $\overline{u'^2}$, $\overline{v'^2}$, 
and $\overline{w'^2}$ appear in CLUBB in several places, including in CLUBB's eddy diffusivity.  

However, CLUBB does not parameterize the effect of shear on the \textit{spatial organization} of deep convective systems,
 which in nature may take the form of, say, squall lines or supercells.  No basic PDF method, without additional assumptions, 
outputs explicit information on spatial structure. 


\item \textit{Are CLUBB's momentum equations tensorially self-consistent?}


Unfortunately, no.  Ideally, the filtered equations would be derived in a coordinate-free, tensorially invariant form.
E.g., a prognostic equation for the 2nd-order moment tensor, $\overline{u_i' u_j'}$, and another for the 
third-order moment tensor, $\overline{u_i' u_j' u_k'}$, would be derived.  
Then the two closed equations would be projected
onto the coordinate system, e.g., Cartesian coordinates with a horizontal lower boundary.  
In practice, prognosing so many moments would be prohibitively expensive.
CLUBB's philosophy, in contrast, is to prognose the minimum number of moments needed to determine the PDF shape.  
For this reason, CLUBB retains an equation for $\overline{w'^3}$ but none for $\overline{u'^3}$ and $\overline{v'^3}$.  

The practical consequence of CLUBB's approach is that in CLUBB, the acceleration due to gravity, $g$, is assumed to point downwards.  This is satisfied in a host model that uses a sigma coordinate, so that the lower boundary is effectively horizontal.
 If CLUBB were used in a model with non-sigma, sloped boundaries, then, e.g., the $\overline{u'^2}$ equation would have no dependence on $g$, as it should.  CLUBB opts to sacrifice generality for speed.


\end{itemize}

\chapter{Where one can find simulation results using CLUBB}

\label{chapt:results}


\section{Peer-reviewed publications that document simulations that include CLUBB}

CLUBB has been implemented as a parameterization in models with a range of grid spacings,
ranging from a cloud-resolving model to climate models.  Results from moderately recent simulations
can be found in the papers listed in Table \ref{tab:clubb_results_refs}.  This is not a complete
list of publications, but it is a starting point.

\begin{table}
\caption{``Recent" papers that display results of simulations that use CLUBB.}
\begin{tabular}{p{2in}p{3.5in}}
\hline
\textbf{Simulation type}     &  \textbf{Reference}        \\
\hline
{Single-column model}   & \citet{guo_et_al_10a,bogenschutz_et_al_2012a,storer_et_al_2015a}         \\
{Cloud-resolving model (SAM)}   & \citet{larson_2012_2km_16km}         \\
{Regional forecast model (WRF)}   &  \citet{larson_et_al_2013b}         \\
{Multi-scale modeling framework (SP-CAM)}   &  \citet{wang_et_al_2015_mmf_clubb}         \\
{Climate model with a separate deep convective parameterization (CAM-CLUBB-ZM,AM3-CLUBB)}   & \citet{bogenschutz_et_al_2013a,guo_et_al_2014_am3clubb}         \\
{Unified climate model (CAM-CLUBB-SILHS, AM3-CLUBB)} & \citet{thayer-calder_et_al_2015_cam_clubb_silhs,guo_et_al_2015_am3_clubb_plus}         \\
\hline
\end{tabular}
\label{tab:clubb_results_refs}
\end{table}

\section{Timing}

\label{sec:timing}

As compared to CAM5, the use of CLUBB and 8 subcolumns in CAM approximately doubles the model cost 
\citep{thayer-calder_et_al_2015_cam_clubb_silhs}.  Slightly more than half the added cost comes from
CLUBB, rather than SILHS.  The additional cost of SILHS is about 6\% per subcolumn.

In cloud resolving simulations, the time step is short, and CLUBB need not be called every time step.
Using the SAM cloud-resolving model \citep{khairoutdinov_randall_03a} 
with the Morrison microphysics scheme \citep{morrison_et_al_09b}, if CLUBB is called
every 12th time step, then the total additional cost of CLUBB is about 20\% \citep{larson_2012_2km_16km}.
These timing tests did not include SILHS.

If CLUBB-SILHS doubles the model cost, then using CLUBB-SILHS is equivalent in cost to increasing the grid spacing
in both horizontal directions by about 40\%, assuming that the vertical grid spacing and time step remain unchanged.

\chapter{Code documentation}

\label{chapt:code_doc}

\section{CLUBB user license agreement}

The CLUBB license appears in files \texttt{advance\_clubb\_core\_module.F90} 
and \texttt{silhs\_api\_module.F90} of the source code:

\sffamily
Cloud Layers Unified By Binormals (CLUBB) user license agreement.

Thank you for your interest in CLUBB. We work hard to create a code that implements the best software engineering practices, is supported to the extent allowed by our limited resources, and is available without cost to non-commercial users. You may use CLUBB if, in return, you abide by these conditions:

1. Please cite CLUBB in presentations and publications that contain results obtained using CLUBB.

2. You may not use any part of CLUBB to create or modify another single-column (1D) model that is not called CLUBB. However, you may modify or augment CLUBB or parts of CLUBB if you include "CLUBB" in the name of the resulting single-column model. For example, a user at MIT might modify CLUBB and call the modified version "CLUBB-MIT." Or, for example, a user of the CLM land-surface model might interface CLM to CLUBB and call it "CLM-CLUBB." This naming convention recognizes the contributions of both sets of developers.

3. You may implement CLUBB as a parameterization in a large-scale host model that has 2 or 3 spatial dimensions without including "CLUBB" in the combined model name, but please acknowledge in presentations and publications that CLUBB has been included as a parameterization.

4. You may not provide all or part of CLUBB to anyone without prior permission from Vincent Larson (vlarson@uwm.edu). 

5. You may not use CLUBB for commercial purposes unless you receive permission from Vincent Larson.

6. You may not re-license all or any part of CLUBB.

7. CLUBB is provided "as is" and without warranty.

We hope that CLUBB will develop into a community resource. We encourage users to contribute their 
CLUBB modifications or extensions to the CLUBB development group. We will then consider them for inclusion in CLUBB. 
We would be pleased to acknowledge contributors and list their CLUBB-related papers 
on our \href{https://carson.math.uwm.edu/larson-group/clubb_site/about.html}{"About CLUBB"} webpage  
for those contributors who so desire.

Thanks so much and best wishes for your research!

The CLUBB Development Group 
\rmfamily

\section{Where to download CLUBB}

\hypertarget{url:download_clubb}{}

CLUBB is stored in a git repo on github.  In order to download and run CLUBB:

\begin{enumerate}


\item Download CLUBB to your computer by going to a Linux or Mac prompt and typing: 

\begin{verbatim}
git clone https://github.com/larson-group/clubb_release.git
\end{verbatim}

Use your github username and password to log in.

\item  To compile and run CLUBB, follow the instructions in the README file in CLUBB's top directory.  

\end{enumerate}

The latest updates to CLUBB and SILHS are made available immediately after they are committed to CLUBB's git repository.
In other words, CLUBB is ``open source in real time." 
 New commits are typically made every week, and each commit contains a comment documenting the intent and content of the code change. 

\section{Supported operating systems and Fortran compilers}

UWM tests CLUBB on computers running the Centos Linux and Mac OS X operating systems.

CLUBB is written in Fortran 2003.  UWM tests the compilation of CLUBB using the following compilers:
GNU Fortran (gfortran), Intel (ifort), Portland Group or Nvidia (pgfortran), and Oracle (sunf95).

\section{The CLUBB standalone single-column model}

A checkout of CLUBB-SILHS includes its own self-contained single-column model (SCM) that is independent 
of CAM, E3SM, or any other host model.  We call this SCM ``CLUBB-SCM" or ``CLUBB standalone."
It is called a "standalone" model because it runs separately from CAM and E3SM.  However, it contains
more than a parcel model of the ``core" of CLUBB; it also includes its own vertical grid, microphysics, and radiation scheme
taken from codes other than CAM and E3SM. 

CLUBB-SCM can be compiled and run by following the instructions in the \href{https://github.com/larson-group/clubb_release/blob/master/README}{README} file
that is located in the top-level directory of a CLUBB-SILHS checkout.  CLUBB-SCM
is tested regularly to ensure that it compiles using a free compiler (gfortran) and
two popular commercial compilers (PGI/Nvidia and Intel). 

Included with a checkout of CLUBB-SILHS is a python script (\href{https://github.com/larson-group/clubb_release/tree/085f7215ca3e99bd2e5521f74b5750d68c9069e1/postprocessing/pyplotgen}{``pyplotgen"}) that creates standard plots of output.
CLUBB-SCM also includes a number of unit tests related to CLUBB, SILHS,
and the analytic integration of microphysics.  If you change the code, 
it may be useful to run the unit tests in order to check for regression bugs.
The driver script for the tests is \href{https://github.com/larson-group/clubb_release/blob/085f7215ca3e99bd2e5521f74b5750d68c9069e1/run_scripts/run_G_unit_tests.bash}{run\_scripts/run\_G\_unit\_tests.bash}.

A checkout of CLUBB-SILHS comes with a number of pre-configured 
\href{https://github.com/larson-group/clubb_release/tree/085f7215ca3e99bd2e5521f74b5750d68c9069e1/input/case_setups}
{benchmark boundary-layer cases}.
The suite of cases includes ones that 
simulate shallow cumulus, marine stratocumulus, mixed-phase Arctic stratus, 
a clear convective boundary layer, and stable boundary layers.  
Each of these cases runs in a few minutes or less.
Running these cases can help verify that a user's download and compilation
were successful.  Results can be compared with previous results in the literature.
The cases also provide useful code examples for setting up new cases.
But in addition to their usefulness for mundane testing and debugging, the cases are realistic enough 
that they can be used to address scientific questions of practical interest.

Running CLUBB-SCM is much easier than running a global model that contains CLUBB. 
For instance, CLUBB standalone can be run on a laptop without waiting in a supercomputer queue.
Because only a single column is simulated, the output is less voluminous and debugging is facilitated.
But CLUBB standalone also has limitations.
For instance, CLUBB standalone's benchmark cases do not span the full range of atmospheric
conditions.  A problem that appears in a global simulation may not be reproducible in 
a single-column simulation.

CLUBB-SCM also has advantages and disadvantages as compared to  
the single-column version of CAM-CLUBB-SILHS or E3SM-CLUBB-SILHS.  
CLUBB-SCM compiles faster and doesn't require
the download of large global forcing datasets.  CLUBB-SCM also
has options not included in the CAM or E3SM versions 
(e.g., analytic integration over the microphysics).  However, CLUBB-SCM does 
not include CAM or E3SM's microphysics or radiation parameterizations, and it would not 
be useful for troubleshooting the interface between CLUBB and the host models.

\section{Code outline}

This section lists excerpts of CLUBB code, with details removed for the sake of clarity.

\subsection{Call order of CLUBB standalone's single-column driver}
 
The source code is in
\href{https://github.com/larson-group/clubb_release/blob/da4fc00e153ee358203253ad4afde70d7ed206a5/src/clubb_driver.F90#L51-L55}{\texttt{subroutine run\_clubb}}.

\begin{lstlisting}[escapechar=|]
!------------------------------------------------------------------
!            Main Time Stepping Loop
!------------------------------------------------------------------

do itime = iinit, ifinal, 1

  ! Advance CLUBB's prognosed moments, and
  !   diagnose terms involving buoyancy, cloud, 
  !   and turbulent advection.
  call advance_clubb_core

  ! Diagnose CLUBB's PDF parameters, including those related 
  !   to hydrometeors.
  call setup_pdf_parameters
/*  These lines are commented out. 
  ! Calculate moments used to compute liquid/ice water loading,
  !   i.e., < rt'hm' >, < thl'hm' >, and < w'^2 hm' >.
  call hydrometeor_mixed_moments
*/
  ! Use SILHS to draw subcolumns from CLUBB's PDF.
  call generate_silhs_sample_api
/*  These lines are commented out.
  ! Convert subcolumns from ``extended" variables chi and Ncn 
  !   to non-negative variables that can be input 
  !   into microphysics (rt, thl, rc, rv, Nc).
  call clip_transform_silhs_output_api
*/
  ! Call a microphysics scheme in order to calculate 
  !   microphysical tendencies.
  call calc_microphys_scheme_tendcies

  ! Advance predictive microphysics fields one model time step.
  call advance_microphys

end do ! itime=1, ifinal

!------------------------------------------------------------------
!            End Main Time Stepping Loop
!------------------------------------------------------------------
\end{lstlisting}

\subsection{Inputs and outputs of CLUBB's core}

The main CLUBB subroutine, \href{https://github.com/larson-group/clubb_release/blob/da4fc00e153ee358203253ad4afde70d7ed206a5/src/CLUBB_core/advance_clubb_core_module.F90#L127}{\texttt{advance\_clubb\_core}},
advances CLUBB one time step.

\begin{lstlisting}

subroutine advance_clubb_core &
           ( l_implemented, dt, fcor, sfc_elevation, hydromet_dim, & ! intent(in)
             thlm_forcing, rtm_forcing, um_forcing, vm_forcing, & ! intent(in)
             sclrm_forcing, edsclrm_forcing, wprtp_forcing, &     ! intent(in)
             wpthlp_forcing, rtp2_forcing, thlp2_forcing, &       ! intent(in)
             rtpthlp_forcing, wm_zm, wm_zt, &                     ! intent(in)
             wpthlp_sfc, wprtp_sfc, upwp_sfc, vpwp_sfc, &         ! intent(in)
             wpsclrp_sfc, wpedsclrp_sfc, &                        ! intent(in)
             p_in_Pa, rho_zm, rho, exner, &                       ! intent(in)
             rho_ds_zm, rho_ds_zt, invrs_rho_ds_zm, &             ! intent(in)
             invrs_rho_ds_zt, thv_ds_zm, thv_ds_zt, hydromet, &   ! intent(in)
             rfrzm, radf, &                                       ! intent(in)
             wphydrometp, wp2hmp, rtphmp_zt, thlphmp_zt, &        ! intent(in)
             host_dx, host_dy, &                                  ! intent(in) 
             um, vm, upwp, vpwp, up2, vp2, &                      ! intent(inout)
             thlm, rtm, wprtp, wpthlp, &                          ! intent(inout)
             wp2, wp3, rtp2, rtp3, thlp2, thlp3, rtpthlp, &       ! intent(inout)
             sclrm,   &                                           ! intent(inout)
             sclrp2, sclrprtp, sclrpthlp, &                       ! intent(inout)
             wpsclrp, edsclrm, &                                  ! intent(inout)
             rcm, cloud_frac, &                                   ! intent(inout)
             wpthvp, wp2thvp, rtpthvp, thlpthvp, &                ! intent(inout)
             sclrpthvp, &                                         ! intent(inout)
             pdf_params, pdf_params_zm, &                         ! intent(inout)
#if defined(CLUBB_CAM) || defined(GFDL)
             khzm, khzt, &                                        ! intent(out)
#endif
             wprcp, ice_supersat_frac, &                          ! intent(out)
             rcm_in_layer, cloud_cover )                          ! intent(out)

!!! Input Variables
logical, intent(in) ::  & 
  l_implemented ! True if CLUBB is being run within a large-scale host model, 
                !   rather than a standalone single-column model.

real( kind = core_rknd ), intent(in) ::  & 
  dt  ! Current timestep duration    [s]

real( kind = core_rknd ), intent(in) ::  & 
  fcor,  &          ! Coriolis forcing             [s^-1]
  sfc_elevation     ! Elevation of ground level    [m above MSL]

integer, intent(in) :: &
  hydromet_dim      ! Total number of hydrometeor species        [#]

! Input Variables
real( kind = core_rknd ), intent(in), dimension(gr%nz) ::  & 
  thlm_forcing,    & ! liquid potential temp forcing (thermo levs) [K/s]
  rtm_forcing,     & ! total water forcing (thermo levs)        [(kg/kg)/s]
  um_forcing,      & ! eastward wind forcing (thermo levs)     [m/s/s]
  vm_forcing,      & ! northward wind forcing (thermo levs)     [m/s/s]
  wprtp_forcing,   & ! total water turbulent flux forcing (mom levs) [m*K/s^2]
  wpthlp_forcing,  & ! liq pot temp turb flux forcing (mom levs) [m*(kg/kg)/s^2]
  rtp2_forcing,    & ! total water variance forcing (mom levs)   [(kg/kg)^2/s]
  thlp2_forcing,   & ! liq pot temp variance forcing (mom levs)  [K^2/s]
  rtpthlp_forcing, & ! <r_t'th_l'> covariance forcing (mom levs) [K*(kg/kg)/s]
  wm_zm,           & ! vertical mean wind component on mom levs  [m/s]
  wm_zt,           & ! vertical mean wind component on thermo levs [m/s]
  p_in_Pa,         & ! Air pressure (thermodynamic levels)       [Pa]
  rho_zm,          & ! Air density on momentum levels            [kg/m^3]
  rho,             & ! Air density on thermodynamic levels       [kg/m^3]
  exner,           & ! Exner function (thermodynamic levels)     [-]
  rho_ds_zm,       & ! Dry, static density on momentum levels    [kg/m^3]
  rho_ds_zt,       & ! Dry, static density on thermo. levels     [kg/m^3]
  invrs_rho_ds_zm, & ! Inverse dry, static density on momentum levs. [m^3/kg]
  invrs_rho_ds_zt, & ! Inverse dry, static density on thermo levs.  [m^3/kg]
  thv_ds_zm,       & ! Dry, base-state theta_v on momentum levs. [K]
  thv_ds_zt,       & ! Dry, base-state theta_v on thermo levs.   [K]
  rfrzm              ! Total ice-phase water mixing ratio        [kg/kg]

real( kind = core_rknd ), dimension(gr%nz,hydromet_dim), intent(in) :: &
  hydromet           ! Array of hydrometeors                [units vary]

real( kind = core_rknd ), dimension(gr%nz), intent(in) :: &
  radf ! Buoyancy production at cloud top due to longwave radiative cooling [m^2/s^3]

real( kind = core_rknd ), dimension(gr%nz, hydromet_dim), intent(in) :: &
  wphydrometp, & ! Covariance of w and a hydrometeor      [(m/s) <hm units>]
  wp2hmp,      & ! < w'^2 hm' > (hm = hydrometeor) [(m/s)^2 <hm units>]
  rtphmp_zt,   & ! Covariance of rt and hm (on thermo levs.) [(kg/kg) <hm units>]
  thlphmp_zt     ! Covariance of thl and hm (on thermo levs.)      [K <hm units>]

real( kind = core_rknd ), intent(in) ::  &
  wpthlp_sfc,   & ! w'theta_l' at surface   [(m K)/s]
  wprtp_sfc,    & ! w'r_t' at surface       [(kg m)/( kg s)]
  upwp_sfc,     & ! u'w' at surface         [m^2/s^2]
  vpwp_sfc        ! v'w' at surface         [m^2/s^2]

! Passive scalar variables
real( kind = core_rknd ), intent(in), dimension(gr%nz,sclr_dim) :: &
  sclrm_forcing    ! Passive scalar forcing         [{units vary}/s]

real( kind = core_rknd ), intent(in),  dimension(sclr_dim) ::  &
  wpsclrp_sfc      ! Passive scalar flux at surface         [{units vary} m/s]

! Eddy passive scalar variables
real( kind = core_rknd ), intent(in), dimension(gr%nz,edsclr_dim) :: &
  edsclrm_forcing  ! Eddy-diffusion passive scalar forcing    [{units vary}/s]

real( kind = core_rknd ), intent(in),  dimension(edsclr_dim) ::  &
  wpedsclrp_sfc    ! Eddy-diffusion passive scalar flux at surface    [{units vary} m/s]

! Host model horizontal grid spacing, if part of host model.
real( kind = core_rknd ), intent(in) :: & 
  host_dx,  & ! East-west horizontal grid spacing     [m]
  host_dy     ! North-south horizontal grid spacing   [m]

!!! Input/Output Variables
! These are prognostic or are planned to be in the future
real( kind = core_rknd ), intent(inout), dimension(gr%nz) ::  &
  um,      & ! eastward grid-mean wind component (thermo levs)   [m/s]
  upwp,    & ! u'w' (momentum levels)                         [m^2/s^2]
  vm,      & ! northward grid-mean wind component (thermo levs)   [m/s]
  vpwp,    & ! v'w' (momentum levels)                         [m^2/s^2]
  up2,     & ! u'^2 (momentum levels)                         [m^2/s^2]
  vp2,     & ! v'^2 (momentum levels)                         [m^2/s^2]
  rtm,     & ! total water mixing ratio, r_t (thermo. levels) [kg/kg]
  wprtp,   & ! w' r_t' (momentum levels)                      [(kg/kg) m/s]
  thlm,    & ! liq. water pot. temp., th_l (thermo. levels)   [K]
  wpthlp,  & ! w'th_l' (momentum levels)                      [(m/s) K]
  rtp2,    & ! r_t'^2 (momentum levels)                       [(kg/kg)^2]
  rtp3,    & ! r_t'^3 (thermodynamic levels)                  [(kg/kg)^3]
  thlp2,   & ! th_l'^2 (momentum levels)                      [K^2]
  thlp3,   & ! th_l'^3 (thermodynamic levels)                 [K^3]
  rtpthlp, & ! r_t'th_l' (momentum levels)                    [(kg/kg) K]
  wp2,     & ! w'^2 (momentum levels)                         [m^2/s^2]
  wp3        ! w'^3 (thermodynamic levels)                    [m^3/s^3]

! Passive scalar variables
real( kind = core_rknd ), intent(inout), dimension(gr%nz,sclr_dim) :: &
  sclrm,     & ! Passive scalar mean (thermo. levels) [units vary]
  wpsclrp,   & ! w'sclr' (momentum levels)            [{units vary} m/s]
  sclrp2,    & ! sclr'^2 (momentum levels)            [{units vary}^2]
  sclrprtp,  & ! sclr'rt' (momentum levels)           [{units vary} (kg/kg)]
  sclrpthlp    ! sclr'thl' (momentum levels)          [{units vary} K]

real( kind = core_rknd ), intent(inout), dimension(gr%nz) ::  & 
  rcm,        & ! cloud water mixing ratio, r_c (thermo. levels) [kg/kg]
  cloud_frac, & ! cloud fraction (thermodynamic levels)          [-]
  wpthvp,     & ! < w' th_v' > (momentum levels)                 [kg/kg K]
  wp2thvp,    & ! < w'^2 th_v' > (thermodynamic levels)          [m^2/s^2 K]
  rtpthvp,    & ! < r_t' th_v' > (momentum levels)               [kg/kg K]
  thlpthvp      ! < th_l' th_v' > (momentum levels)              [K^2]

real( kind = core_rknd ), intent(inout), dimension(gr%nz,sclr_dim) :: &
  sclrpthvp    ! < sclr' th_v' > (momentum levels)   [units vary]

type(pdf_parameter), dimension(gr%nz), intent(inout) :: &
  pdf_params,    & ! Fortran structure of PDF parameters on thermo levs    [units vary]
  pdf_params_zm    ! Fortran structure of PDF parameters on mom levs        [units vary]

! Eddy passive scalar variable
real( kind = core_rknd ), intent(inout), dimension(gr%nz,edsclr_dim) :: & 
  edsclrm   ! Eddy passive scalar grid-mean (thermo. levels)   [units vary]

! Variables that need to be output for use in other parts of the CLUBB
! code, such as microphysics (rcm, pdf_params), forcings (rcm), and/or
! BUGSrad (cloud_cover).
real( kind = core_rknd ), intent(out), dimension(gr%nz) ::  & 
  rcm_in_layer, & ! rcm within cloud layer                          [kg/kg]
  cloud_cover     ! cloud cover                                     [-]

! Variables that need to be output for use in host models
real( kind = core_rknd ), intent(out), dimension(gr%nz) ::  &
  wprcp,            & ! w'r_c' (momentum levels)                  [(kg/kg) m/s]
  ice_supersat_frac   ! ice cloud fraction (thermodynamic levels) [-]

#if defined(CLUBB_CAM) || defined(GFDL)
real( kind = core_rknd ), intent(out), dimension(gr%nz) :: &
  khzt, &       ! eddy diffusivity on thermo levels
  khzm          ! eddy diffusivity on momentum levels
#endif


\end{lstlisting}

\subsection{Internal call order of CLUBB's core}

CLUBB's core is contained in 
\href{https://github.com/larson-group/clubb_release/blob/da4fc00e153ee358203253ad4afde70d7ed206a5/src/CLUBB_core/advance_clubb_core_module.F90#L127}{\texttt{subroutine advance\_clubb\_core}}

\begin{lstlisting}
! Diagnose CLUBB's turbulent mixing length scale.
call compute_length

! Calculate CLUBB's turbulent eddy-turnover time scale as 
!   CLUBB's length scale divided by a velocity scale.
tau_zt = MIN( Lscale / sqrt_em_zt, taumax )

! Compute CLUBB's eddy diffusivity as  
!   CLUBB's length scale times a velocity scale.
Kh_zt = c_K * Lscale * sqrt_em_zt

! Diagnose surface variances based on surface fluxes.
call surface_varnce

! Advance the prognostic equations for 
!   the scalar grid means (rtm, thlm, sclrm) and 
!   scalar turbulent fluxes (wprtp, wpthlp, and wpsclrp) 
!   by one time step. 
call advance_xm_wpxp

! Advance the prognostic equations 
!   for scalar variances and covariances,
!   plus the horizontal wind variances, by one time step.
call advance_xp2_xpyp

! Advance the 2nd- and 3rd-order moments 
!   of vertical velocity (wp2, wp3) by one timestep.
call advance_wp2_wp3

! Advance the horizontal mean winds (um, vm),
!   the mean of the eddy-diffusivity scalars (i.e. edsclrm),
!   and their fluxes (upwp, vpwp, wpedsclrp) by one time step.
call advance_windm_edsclrm

! Given CLUBB's prognosed moments, diagnose CLUBB's PDF parameters
!   and quantities integrated over that PDF, including
!   quantities related to clouds, buoyancy, and turbulent advection. 
call generate_pdf_params
\end{lstlisting}

\subsection{Inputs and outputs to SILHS}

SILHS' driver is
\href{https://github.com/larson-group/clubb_release/blob/da4fc00e153ee358203253ad4afde70d7ed206a5/src/SILHS/latin_hypercube_driver_module.F90#L27-L45}{\texttt{subroutine generate\_silhs\_sample}}.

\begin{lstlisting}

subroutine generate_silhs_sample &
           ( iter, pdf_dim, num_samples, sequence_length, nz, & ! intent(in)
             l_calc_weights_all_levs_itime, &                   ! intent(in)
             pdf_params, delta_zm, rcm, Lscale, &               ! intent(in)
             rho_ds_zt, mu1, mu2, sigma1, sigma2, &             ! intent(in)
             corr_cholesky_mtx_1, corr_cholesky_mtx_2, &        ! intent(in)
             hydromet_pdf_params, &                             ! intent(in)
             X_nl_all_levs, X_mixt_comp_all_levs, &             ! intent(out)
             lh_sample_point_weights )                          ! intent(out)

    ! Input Variables
    integer, intent(in) :: &
      iter,            & ! Model iteration (time step) number
      pdf_dim,         & ! Number of variables to sample
      num_samples,     & ! Number of samples per variable
      sequence_length, & ! nt_repeat/num_samples; 
                         !   number of timesteps before sequence repeats
      nz                 ! Number of vertical model levels

    type(pdf_parameter), dimension(nz), intent(in) :: &
      pdf_params ! PDF parameters       [units vary]

    real( kind = core_rknd ), dimension(nz), intent(in) :: &
      delta_zm, &  ! Difference in momentum altitudes    [m]
      rcm          ! Liquid water mixing ratio          [kg/kg]

    real( kind = core_rknd ), dimension(nz), intent(in) :: &
      Lscale       ! Turbulent mixing length            [m]

    real( kind = core_rknd ), dimension(nz), intent(in) :: &
      rho_ds_zt    ! Dry, static density on thermo. levels    [kg/m^3]

    logical, intent(in) :: &
      l_calc_weights_all_levs_itime ! determines if vertically correlated 
                                    !   sample points are needed
    
    
    ! Output Variables
    real( kind = core_rknd ), intent(out), dimension(nz,num_samples,pdf_dim) :: &
      X_nl_all_levs ! Sample that is transformed ultimately to normal-lognormal

    integer, intent(out), dimension(nz,num_samples) :: &
      X_mixt_comp_all_levs ! Which mixture component we're in

    real( kind = core_rknd ), intent(out), dimension(nz,num_samples) :: &
      lh_sample_point_weights ! Weight of each sample point

    ! More Input Variables!
    real( kind = core_rknd ), dimension(pdf_dim,pdf_dim,nz), intent(in) :: &
      corr_cholesky_mtx_1, & ! Correlation Cholesky matrix (1st comp) [-]
      corr_cholesky_mtx_2    ! Correlation Cholesky matrix (2nd comp) [-]

    real( kind = core_rknd ), dimension(pdf_dim,nz), intent(in) :: &
      mu1,    & ! 1st-comp hydrometeor means (chi, eta, w, <hydrometeors>)  
      mu2,    & ! 2nd-comp hydrometeor means (chi, eta, w, <hydrometeors>)  
      sigma1, & ! 1st-comp stdevs of hydrometeors (chi, eta, w, <hydrometeors>) 
      sigma2    ! 2nd-comp stdevs of hydrometeors (chi, eta, w, <hydrometeors>) 

    type(hydromet_pdf_parameter), dimension(nz), intent(in) :: &
      hydromet_pdf_params ! Hydrometeor PDF parameters  [units vary]


\end{lstlisting}

\subsection{Internal call order of SILHS}

SILHS' driver,
\href{https://github.com/larson-group/clubb_release/blob/da4fc00e153ee358203253ad4afde70d7ed206a5/src/SILHS/latin_hypercube_driver_module.F90#L27-L45}{\texttt{subroutine generate\_silhs\_sample}},
carries out the following main steps.

\begin{lstlisting}

! Compute k_lh_start, the starting vertical grid level 
!   for SILHS sampling
k_lh_start = compute_k_lh_start( . . . )

! Generate a uniformly distributed sample at k_lh_start
call generate_all_uniform_samples 

! Transform the uniformly distributed samples to
!   ones distributed according to CLUBB's PDF.
call transform_uniform_samples_to_pdf

\end{lstlisting}

\subsection{CLUBB-SILHS' variable naming convention}

Certain abbreviations are used repeatedly in CLUBB variables names.  Some of the more common ones
are listed in Table \ref{tab:variable_letter_meanings}.

\begin{table}
\caption{Meanings of certain letters in CLUBB's variable names.}
\begin{tabular}{p{0.75in}p{2.25in}p{2.5in}}
\hline
\textbf{Letters}  &  \textbf{Meaning}     &  \textbf{Example}        \\
\hline
\texttt{m}   &  horizontal grid mean             &   \texttt{wm} is the mean vertical velocity.  \\
\texttt{p}   &  prime (i.e.~perturbation)     &   \texttt{wp2} is the variance of vertical velocity.  \\  
\texttt{r}   &  mixing ratio                     &   \texttt{rtm} is the mean total water mixing ratio. \\
\texttt{th}  & theta (i.e.~potential temp)  &   \texttt{thlm} is the liquid water pot. temp.            \\   
\texttt{hydromet} &  hydrometeor array       &    \texttt{wphydrometp} is the vertical turbulent flux of a hydrometeor. \\
\texttt{hm} &  abbreviation for \texttt{hydromet}   &    \texttt{hm\_1} is the mean of a hydrometeor in PDF component 1. \\
\texttt{\_zm} & interpolated to the momentum grid (interface levels)    &   \texttt{rho\_zm} is the air density on the momentum grid.  \\
\texttt{\_zt} &  interpolated to the thermodynamic grid (full levels)    &   \texttt{wp2\_zt} is \texttt{wp2} on the thermodynamic grid.  \\
\texttt{\_sfc} & ground or ocean surface  &   \texttt{T\_sfc} is the temperature of the ground or ocean.  \texttt{upwp\_sfc} is the east-west momentum flux just above the ground or ocean. \\
\texttt{\_1} &  PDF component 1     &   \texttt{cloud\_frac\_1} is the fraction of the first Gaussian component that is occupied by cloud.  \\
\texttt{\_n} &  normal-space version     &   \texttt{mu\_x\_1\_n} is mean of variable \texttt{x} in component 1 after being transformed from lognormal to normal space.  \\
\texttt{microphys} &  microphysics      &    \texttt{advance\_microphys} is a subroutine that computes microphysics. \\
\texttt{l\_} &  a logical variable     &   \texttt{l\_implemented} is true if CLUBB is being used in a host model instead of in a single-column simulation.  \\
\hline
\end{tabular}
\label{tab:variable_letter_meanings}
\end{table}

\subsection{Key variables and Fortran structures in CLUBB}

Key CLUBB variables are listed in Table \ref{tab:key_variables_in_clubb}. Key CLUBB arrays
are listed in Table \ref{tab:fortran_structures_in_clubb}.  Key SILHS arrays are listed in Table \ref{tab:fortran_structures_in_silhs}.

\begin{table}
\caption{Selected variables in CLUBB}
\begin{tabular} {p{0.3in}p{0.6in}p{4in}}
   & \textbf{Variable Name in CLUBB} & \textbf{Description} \\ \hline
$\overline{w'r'_t}$ & \texttt{wprtp} & Vertical turbulent flux of total water mixing ratio (vapor+cloud liquid) \\
$\overline{w'\theta'_l}$ & \texttt{wpthlp} & Vertical turbulent flux of liquid water potential temperature \\
$\overline{u'w'}$ & \texttt{upwp} & Covariance of east-west and vertical velocity \\
$\overline{v'w'}$ & \texttt{vpwp} & Covariance of north-south and vertical velocity \\
$\overline{r^{'2}_t}$ & \texttt{rtp2} & Variance of $r_t$ \\
$\overline{\theta^{'2}_l}$ & \texttt{thlp2} & Variance of $\theta_l$ \\
$\overline{r'_t\theta'_l}$ & \texttt{rtpthlp} & Covariance of $r_t$ and $\theta_l$ \\
$\overline{w^{'2}}$ & \texttt{wp2} & Variance of $w$\\
$\overline{u'^2}$ & \texttt{up2} & Variance of $u$ \\
$\overline{v'^2}$ & \texttt{vp2} & Variance of $v$ \\
$\overline{w^{'3}}$ & \texttt{wp3} & Third-order Moment of $w$ \\
$K_{h}$ & \texttt{Kh\_zt} & Eddy diffusivity of scalars \\
$\overline{r_c}$ & \texttt{rcm} & Liquid cloud water mixing ratio \\
$C$ & \texttt{cloud\_frac} & Liquid cloud fraction  \\
   & \texttt{edsclrm} &  Grid means of scalars computed using eddy diffusivity  \\
   & \texttt{sclrm}     & Grid means of scalars computed using higher-order closure  \\
\label{tab:key_variables_in_clubb}
\end{tabular}
\end{table}

\begin{table}
\caption{Selected Fortran structures and arrays in CLUBB.}
\begin{tabular}{p{1in}p{3.5in}}
\hline
\textbf{Fortran structure or array name}     &  \textbf{Description}        \\
\hline
\texttt{hydromet}         &   Array of grid means of hydrometeors              \\
\texttt{wphydrometp}  &   Array of vertical turbulent fluxes of hydrometeors              \\
\texttt{pdf\_params}   &  PDF parameters of all variates except hydrometeors               \\
\hline
\end{tabular}
\label{tab:fortran_structures_in_clubb}
\end{table}

\begin{table}
\caption{Selected Fortran structures and arrays in SILHS.}
\begin{tabular}{p{1.5in}p{3.5in}}
\hline
\textbf{Fortran structure or array name}     &  \textbf{Description}        \\
\hline
\texttt{mu\_x\_1\_n}                      &  Multivariate mean of PDF component 1 in normal space               \\
\texttt{sigma\_x\_1\_n}                  &  Multivariate standard deviation of PDF component 1 in normal space                \\
\texttt{corr\_cholesky\_mtx\_1}     &  Cholesky decomposition of the hydrometeor correlation matrix 
                                                             for PDF component 1 in normal space               \\
\texttt{hydromet\_pdf\_params}    &  PDF parameters of all variates, including hydrometeors               \\
\texttt{X\_nl\_all\_levs}                   &  Array of multivariate subcolumns                 \\
\hline
\end{tabular}
\label{tab:fortran_structures_in_silhs}
\end{table}

\subsection{CLUBB's vertical grid}

\begin{figure}[htp]
\hypertarget{url:clubb_grid}{}
\centering
\begin{tikzpicture}

\draw [line width=2pt] (-4,17) node [left,fill=white] {CAM/E3SM interface level 1} -- (4,17) node [midway,fill=white] {zm(nzmax)};
\draw[line width=1pt,dashed] (-4,16) node [left,fill=white] {CAM/E3SM full level 1} -- (4,16)  
node [midway,fill=white] {zt(nzmax)};

\draw [line width=3pt,line cap=round,dash pattern=on 0pt off 3\pgflinewidth] (0,15) -- (0,14);

\draw [line width=2pt] (-4,13) -- (4,13) node [midway,fill=white] {zm(k+1)};

\draw[line width=1pt,dashed] (-4,12) -- (4,12)  node [midway,fill=white] {zt(k+1)};
\draw[|-|] (5,13) -- (5,11) node [midway,fill=white] {dzt(k+1)};

\draw[line width=2pt] (-4,11) -- (4,11) node [midway,fill=white]  {zm(k)};
\draw[|-|] (-5,10) -- (-5,12) node [midway,fill=white] {dzm(k)};

\draw[line width=1pt,dashed] (-4,10) -- (4,10) node [midway,fill=white]  {zt(k)};
\draw[|-|] (5,9) -- (5,11) node [midway,fill=white] {dzt(k)};

\draw [line width=2pt] (-4,9) -- (4,9) node [midway,fill=white] {zm(k-1)};
\draw[|-|] (-5,10) -- (-5,8) node [midway,fill=white] {dzm(k-1)};

\draw[line width=1pt,dashed] (-4,8) -- (4,8)  node [midway,fill=white] {zt(k-1)};
\draw[|-|] (5,9) -- (5,7) node [midway,fill=white] {dzt(k-1)};

\draw[line width=2pt] (-4,7) -- (4,7) node [midway,fill=white]  {zm(k-2)};

\draw [line width=3pt,line cap=round,dash pattern=on 0pt off 3\pgflinewidth] (0,6) -- (0,5);

\draw[line width=2pt] (-4,4) node [left,fill=white] {CAM/E3SM interface level nzmax-1} -- (4,4) node [midway,fill=white]  {zm(2)} node [right,fill=white] {$\overline{w'r_t'}$, $\overline{w'\theta_l'}$, $\overline{w^{'2}}$, $\overline{w^{'4}}$, $\ldots$};

\draw[line width=1pt,dashed] (-4,3) node [left,fill=white] {CAM/E3SM full level nzmax} -- (4,3) node [midway,fill=white]  {zt(2)}  node [right,fill=white] {$\overline{r_t}$, $\overline{\theta_l}$, $\overline{u}$, $\overline{v}$, $\overline{w^{'3}}$, $\overline{w^{'2}\theta_v'}$, $\ldots$};
\draw[pattern=north west lines] (-4,1.5) rectangle (4,2);

\draw [line width=2pt] (-4,2) -- (4,2) node [midway,fill=white] {zm(1)} node [right,fill=white] {Surface} ;
\draw[line width=1pt,dashed] (-4,1)  node [left,fill=white] {No level here in CAM} -- (4,1)  node [midway,fill=white] {zt(1)} node [right,fill=white,align=left] {Below-ground  ghost point};

     \node[align=center,font=\bfseries,  yshift=3em] (title) 
         at (current bounding box.north)
         {\Large CLUBB's vertical grid};

\end{tikzpicture}
\caption{CLUBB's \texttt{zm} grid levels are the ``momentum" levels, 
which correspond to CAM and E3SM's interface levels.
At the \texttt{zm} levels reside the even-order moments.  CLUBB's \texttt{zt} (``thermodynamic", or CAM full) 
levels are where the odd-order moments reside.  CLUBB has a below-ground level (\texttt{zt(1)}) 
that is absent from CAM.  A truncated schematic of CLUBB's grid appears in Fig.~2 of \citet{golaz_et_al_02a} 
(\href{https://www.ametsoc.org/ams/index.cfm/publications/authors/journal-and-bams-authors/author-resources/copyright-information/copyright-policy/}{\textcircled{c} Copyright 2002 AMS}).}
\label{fig:clubb_grid}
\end{figure}
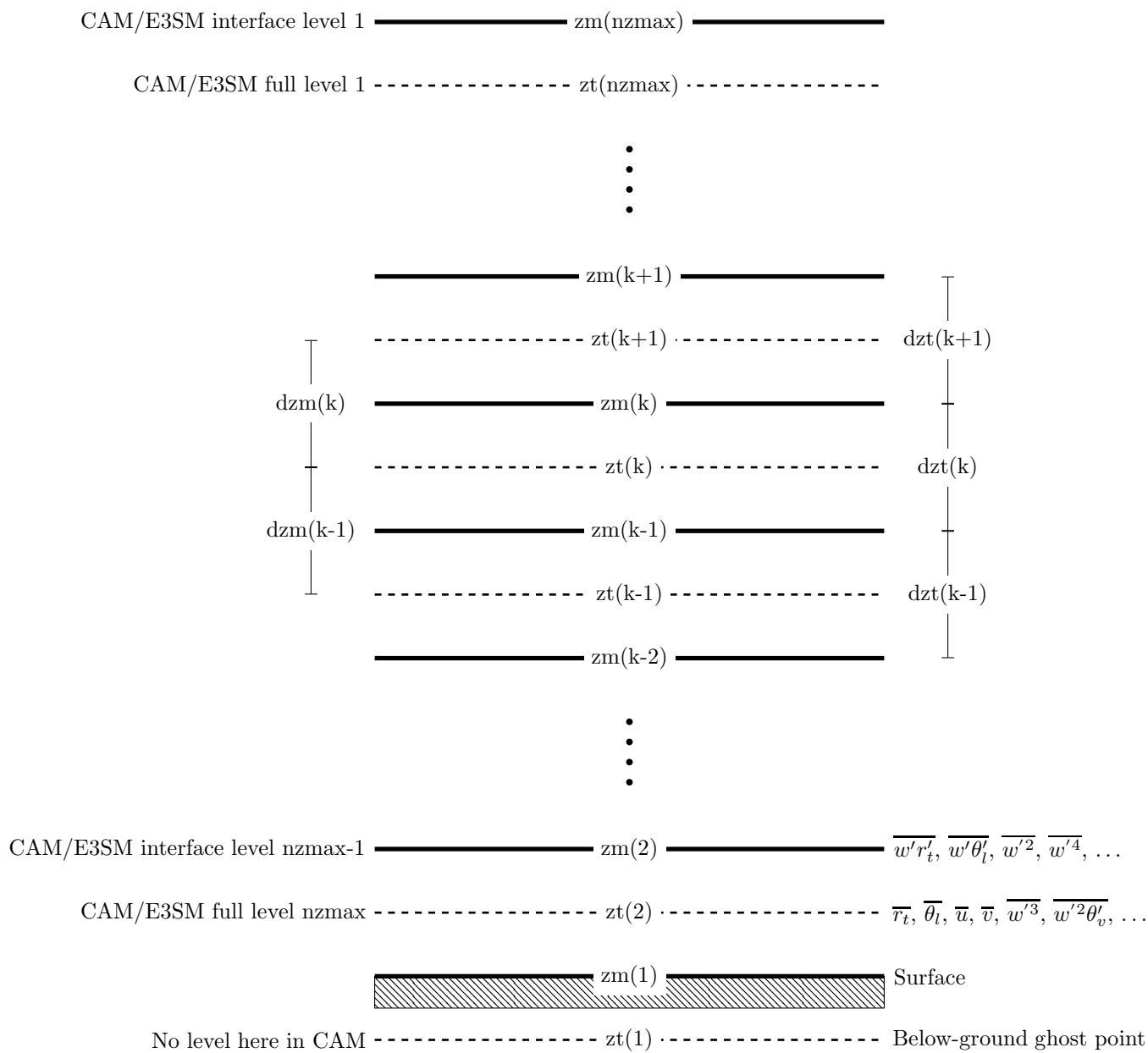

CLUBB uses a staggered grid that places the even-order and odd-order moments on interleaving grid levels.
The grid levels are numbered from $k=1$ at the ground upward to $k=nzmax$ at the top of the model domain.
The grid structure is displayed in Fig.~\ref{fig:clubb_grid}.

\subsection{Optional configurations in CLUBB-SILHS and the model flags that control them}

CLUBB is not merely one specific parameterization.  Instead, it is a framework or platform
for the community to build on.  
CLUBB's software is designed in a modular way that allows old components to be replaced with new ones.
Several configuration options already exist in CLUBB-SILHS.  Different users may choose different 
options depending on the desired balance of features, accuracy, and cost.  As examples, we discuss two of these
options below.

\subsubsection{Analytic versus Monte Carlo integration of microphysics}

Although the green-bar integrals ought to be estimated in some way, a variety of
quadrature methods are competitive \citep[e.g.,][]{chowdhary_et_al_2015_quad}.  CLUBB-SILHS implements two.

First, the integrals may be estimated by Monte Carlo integration using SILHS.
To choose this option in CLUBB, set 
\texttt{lh\_microphys\_type = "interactive"} and \texttt{l\_local\_kk = .true.}.

Second, the integrals may be calculated analytically, if the Khairoutdinov-Kogan 
microphysics is used \citep{khairoutdinov_kogan_00a}.  To choose this option, set 
\texttt{microphys\_scheme = "khairoutdinov\_kogan"} and \texttt{l\_local\_kk = .false.}.

\subsubsection{Omission of horizontal wind variances, $\overline{u'^2}$ and $\overline{v'^2}$}

In its default configuration, CLUBB prognoses $\overline{u'^2}$, $\overline{v'^2}$, and $\overline{w'^2}$ 
separately.  This allows CLUBB to predict anistropic turbulence near the ground,
where the horizontal component of TKE may be significantly larger than the vertical component.
However, to avoid the cost of prognosing $\overline{u'^2}$ and $\overline{v'^2}$,
CLUBB allows the user to omit these equations and instead assume that the turbulence is 
isotropic.  In this case, CLUBB assumes that the turbulence kinetic energy, $\overline{e}$,
is proportional to the vertical velocity variance $\overline{w^{'2}}$:
\begin{equation}
\label{eq_tke_iso}
\overline{e} = \frac{3}{2} \overline{w^{'2}}.
\end{equation}
To select this configuration, set \texttt{l\_tke\_aniso = .false.}.




\section{How to implement CLUBB-SILHS in a host model}

In order to implement CLUBB, a host model needs to call, every time step,
\href{https://github.com/larson-group/clubb_release/blob/da4fc00e153ee358203253ad4afde70d7ed206a5/src/CLUBB_core/clubb_api_module.F90#L1279-L1280}
{\texttt{subroutine advance\_clubb\_core\_api}}.

To implement SILHS, the host model 
needs to call 
\href{https://github.com/larson-group/clubb_release/blob/da4fc00e153ee358203253ad4afde70d7ed206a5/src/CLUBB_core/clubb_api_module.F90#L3093}
{\texttt{subroutine setup\_pdf\_parameters\_api}}
and 

\href{https://github.com/larson-group/clubb_release/blob/da4fc00e153ee358203253ad4afde70d7ed206a5/src/SILHS/silhs_api_module.F90#L322}
{\texttt{generate\_silhs\_sample\_api}}
every time step.  In addition, various variables need to be initialized once at the beginning of the simulation.

\subsection{Special consideration when implementing in a fine-scale host model}

CLUBB-SILHS is a single-column model, and as such, its equation set does not include horizontal advection terms.  
However, if CLUBB-SILHS is implemented in a host model with horizontal grid spacing finer than 2 km, 
the horizontal advection of (at least some of) CLUBB's higher-order moments ought to be implemented in the host model.    
Aside from that, an implementation of CLUBB-SILHS in a cloud-resolving model is the same as an implementation of CLUBB-SILHS in a coarse-resolution global climate model.

\subsection{CLUBB API and SILHS API}

CLUBB's native habitat is its own self-contained single-column model (``CLUBB standalone").
CLUBB standalone does not contain any code from CAM or any other global host model.

When a researcher wishes to implement CLUBB into a new host model, it may not be obvious
which pieces of CLUBB standalone code ought to be called from the host model.  
Re-implementing CLUBB's pre-existing functionality into a host model is 
time-consuming, duplicative, error-prone, and confusing to users. 
To avoid this, CLUBB standalone contains an Application Programming Interface called ``CLUBB API."
It is a Fortran module, 
\href{https://github.com/larson-group/clubb_release/blob/da4fc00e153ee358203253ad4afde70d7ed206a5/src/CLUBB_core/clubb_api_module.F90}
{\texttt{clubb\_api\_module}}, 
that aims to contain all the CLUBB
subroutines that might be useful to call in a host model.  For instance, it contains
\texttt{subroutine advance\_clubb\_core\_api}, which simply calls \texttt{subroutine advance\_clubb\_core}.
It also contains \texttt{subroutine setup\_pdf\_parameters\_api} and various initialization and utility subroutines.  

One advantage of CLUBB API is that it serves as documentation.  It tells host modelers what CLUBB provides.
Researchers who wish to implement CLUBB in a host model
can find all subroutines that need to be implemented in one module.  Moreover, any CLUBB subroutines 
that are called from the host model should end with \texttt{\_api}.  This provides a useful check.

Likewise, a similar API exists for SILHS: 
\href{https://github.com/larson-group/clubb_release/blob/da4fc00e153ee358203253ad4afde70d7ed206a5/src/SILHS/silhs_api_module.F90}
{\texttt{silhs\_api\_module}}.  
It contains
\texttt{generate\_silhs\_sample\_api}, plus some utilities.  






\section{Tuning guide}

\label{sec:tuning_guide}

\subsection{Background on tuning CLUBB}

Each higher-order moment equation contains multiple unclosed terms.  Each of these unclosed terms, 
in turn, must include a tunable parameter, either explicitly or implicitly by, for instance, 
setting its value to 1 (Chapter \ref{chapt:clubb_closed_eqns}).  This leads to a large number of tunable parameters.

However, some parameters are more important than others.  The higher-order equations are necessarily sensitive 
to the large terms, and hence the parameters in those terms are important.  They are the ones to focus on
if a significant change in model behavior is desired.  In addition, the tunable parameters in the PDF
($\gamma$ and $\beta$) are important.  On the other hand, the parameters in the 
smaller terms are less important.  However, the only way to remove the parameters for the smaller terms is to delete those terms entirely, and CLUBB has chosen not to do this.

CLUBB appears to contain more tunable parameters than it does in practice.  Historically, we have introduced some tuning parameters in CLUBB, found them to be ineffective, and have not gone back and deleted them.  For instance, at one point in the past, we experimented with tuning $r_t$ differently than $\theta_l$, 
but we abandoned this approach, and so now, for instance, I set \texttt{C2rt=C2thl}.  Likewise, \citet{golaz_et_al_07a} introduced skewness dependence into some parameters in an attempt to better distinguish stratocumulus from cumulus:
\begin{equation}
        \mathrm{C\_Skw\_fnc} 
                = \mathrm{Cb} + (\mathrm{C}-\mathrm{Cb}) 
                                                   \exp\left( -0.5 \left( \frac{\mathrm{Sk}_w}{Cc} \right)^2 \right) .
\end{equation}
However, for several (but not all) such parameters, I set $\mathrm{C}=\mathrm{C_b}$, thereby reducing the number of parameters from 3 ($\mathrm{C}$, $\mathrm{C_b}$, and $\mathrm{C_c}$) to 1.

\subsection{List of selected tuning parameters}

Selected tuning parameters are listed in Table \ref{tab:tuning_parameters_in_clubb}.  The effects of changes to these parameters is documented in \citet{guo_et_al_2014_scamclubb} for single-column simulations and in \citet{guo_et_al_2015_camclubb} for global simulations.

\begin{table}
\caption{Selected tuning parameters in CLUBB}
\begin{tabular} {p{1in}p{0.6in}p{2in}p{2in}}
\textbf{Name}   & \textbf{Range} & \textbf{Meaning} & \textbf{Notes} \\ \hline
\texttt{gamma\_coef} & $0.25-0.36$ &   Larger $\gamma$ means larger width of $w$ components (Eqn.~\ref{eq_sc}) 
                  &  Increasing $\gamma$ strongly brightens clouds \\
\texttt{beta} & $ 1.2-2.6 $ &   $\beta$ controls scalar skewnesses (Eqn.~\ref{eq:scalar_Sk}) 
                  &  Increasing $\beta$ dims clouds \\
\texttt{C11,C11b} & $ 0.2-0.8 $ &   Buoyancy pressure damping of $\overline{w'^3}$ (Eqn.~\ref{eq_wp3}) 
                       &   Increases in \texttt{C11} and \texttt{C11b} reduce $\mathrm{Sk}_w$ and brighten clouds              \\
\texttt{C8} & $ 3-5 $ &   Pressure damping of $\overline{w'^3}$ (Eqn.~\ref{eq_wp3}) 
                       &   Increases in \texttt{C8} reduce $\mathrm{Sk}_w$ and brighten clouds              \\
\texttt{C1} & $ 0.5-2.5 $ &    Dissipation of $\overline{w'^2}$ (Eqn.~\ref{eq_wp2}) 
                   &   Increasing \texttt{C1} tends to increase $\mathrm{Sk}_w$, which favors the presence of Cu rather than Sc \\
\texttt{C14} & $ 0.3-2.0 $ &    Dissipation of $\overline{u'^2}$ and $\overline{v'^2}$ (Eqns.~\ref{eq_up2},\ref{eq_vp2}) 
                   &   Increasing \texttt{C14} damps $\overline{u'^2}$ and $\overline{v'^2}$ \\
\texttt{C6rt=C6thl} & $ 3-7 $ 
                  &   Low-skewness value of pressure damping of $\overline{w'r_t'}$ and $\overline{w'\theta_l'}$ (Eqn.~\ref{eq_wprtp},\ref{eq_wpthlp})  
                   &    Decreasing \texttt{C6rt=C6thl} increases vertical scalar fluxes, especially in Sc      \\
\texttt{C6rtb=C6thlb} & $ 3-7 $ 
                   &   High-skewness value of pressure damping of $\overline{w'r_t'}$ and $\overline{w'\theta_l'}$ (Eqns.~\ref{eq_wprtp},\ref{eq_wpthlp}) 
                   &    Decreasing \texttt{C6rtb=C6thlb} increases vertical scalar fluxes, especially in Cu       \\
\texttt{C7=C7b} & $ 0.3-0.8 $ &   Buoyancy portion of pressure damping of $\overline{w'r_t'}$ and $\overline{w'\theta_l'}$ (Eqns.~\ref{eq_wprtp},\ref{eq_wpthlp}) &   Decreasing \texttt{C7=C7b} increases vertical scalar fluxes \\
\texttt{C2rt=C2thl} & $ 0.2-2 $  &   Dissipation of $\overline{r_t'^2}$ and $\overline{\theta_l'^2}$  (Eqns.~\ref{eq_rtp2},\ref{eq_thlp2}) 
                                 &  Increasing \texttt{C2rt=C2thl} can dim clouds, or if rain matters, brighten clouds \citep{guo_et_al_2015_camclubb}    \\
\texttt{c\_K10} & $ 0.2-0.6 $  &    \texttt{c\_K10} pre-multiplies the eddy diffusivity for momentum (Eqn.~\ref{eq_Km}) 
                   &  Increasing \texttt{c\_K10} increases vertical transport of momentum    \\
                                        &     \\
\label{tab:tuning_parameters_in_clubb}
\end{tabular}
\end{table}


\subsection{Tuning trade-offs}


Tuning of certain parameters can be used to improve aspects of global CAM or E3SM simulations,
but the tuning changes also leads to side effects.  Therefore, there exist tuning trade-offs.

\textit{Brightening or dimming low clouds}

To achieve global radiative balance, one can adjust \texttt{C11}, \texttt{C11b}, and/or \texttt{C8}.  
For instance, increasing these parameters damps $\overline{w'^3}$, thereby reducing skewness 
and increasing the stratiform nature of the clouds, which brightens them.  A similar effect can be achieved by decreasing \texttt{gamma\_coef}, which is a very sensitive parameter.  

If Sc and Cu needs to be better distinguished, one can try decreasing \texttt{gamma\_coef} while increasing
\texttt{C14} (personal communication, P.~Bogenschutz).  

\textit{Increasing or decreasing the magnitude of sea-level pressure (SLP)}

In CAM-CLUBB's surface pressure field, often the highs are too high and the lows too low.  To improve SLP, 
one can increase \texttt{c\_K10}.  However, increasing it too much increases surface wind stresses on the ocean too much,
causing excessive cooling in coupled climate simulations (personal communication, P.~Bogenschutz).  

\textit{Smoothing out noisy surface precipitation fields}

In CAM-CLUBB-SILHS, precipitation is often too spotty, with intermittent but intense rainfall.  To smooth the rainfall
fields, one can reduce the values of \texttt{C2rt=C2thl}.  However, this may degrade the spatial pattern of precipitation
in the Indian ocean, and it may lead to too much rain evaporation and precipitable water.



\chapter{Annotated bibliography}

\label{chapt:annot_bib}

This bibliography does not aspire to be a complete list of CLUBB papers.
It merely describes starting points for more information on CLUBB.  Much of this information
appears also at \href{http://www.uwm.edu/\~vlarson}{http://www.uwm.edu/$\sim$vlarson}.

\section{Formulation of CLUBB}

{\color{gray} 
The following papers discuss the formulation of the core of CLUBB. 
}

\textbf{\bibentry{golaz_et_al_02a} }

\textbf{\bibentry{golaz_et_al_02b} }

{\color{gray} 
Traditionally, cloud parameterization has been viewed as a multiplicity of tasks. Such tasks include the prediction of heat flux, moisture flux, cloud fraction, and liquid water. In contrast, the papers above adopt the alternative viewpoint that the goal of parameterization consists largely of a single task: the prediction of the joint PDF of vertical velocity, heat, and moisture. Once the PDF is given, the fluxes, cloud fraction, and liquid water can be diagnosed.

The above papers present a parameterization that can model both stratocumulus and cumulus clouds without case-specific adjustments. This avoids the difficulty of having to construct a “trigger function” that determines which cloud type should be modeled under which meteorological conditions. 
}

\textbf{\bibentry{larson_et_al_02a} }

{\color{gray} 
This paper discusses joint PDFs that include the vertical velocity. Joint PDFs allow us to diagnose the buoyancy flux, which is the means by which convection generates turbulence. Joint PDFs also allow us
to diagnose fluxes of heat and moisture. Therefore, joint PDFs can serve as the foundation of cloud and turbulence parameterizations in numerical models, as proposed and explored in the two following papers.
}

\textbf{\bibentry{larson_golaz_05a} }

{\color{gray} 
The aforementioned papers show that if we choose an accurate PDF family, then we can solve for many of the unknowns in our one-dimensional cloud parameterization. For some of these unknown terms, the present paper lists simple, analytic approximations. All approximated formulas are based on the same PDF and hence are consistent with each other.

A PDF may be constructed from a set of means, variances, and other moments of velocity, moisture, and temperature. It is possible that a particular set of moments does not correspond to any real PDF
in the family. We call such a set of moments ``specifically unrealizable." For instance, a set that includes asymmetric moments is specifically unrealizable with respect a PDF family of symmetric, bell-shaped curves. This is because the bell shape family is too restrictive to include asymmetric moments. We show that a broad class of moments is 
specifically realizable with respect to CLUBB's PDF family. That is, CLUBB's PDF family is not restrictive.
}

\textbf{\bibentry{golaz_et_al_07a} }

{\color{gray} 
It is often easy to see when an atmospheric model disagrees with data. It is usually much harder to locate the ultimate sources of model error.

It is particularly difficult to diagnose errors in a model's structure, that is, errors in the functional form of the model equations. One technique that may help is parameter estimation, that is, the optimization of model parameter values. Typically, parameter estimation is used solely to improve the fit between a model and observational data. In the process, however, parameter estimation may cover up structural model errors.

In a quite opposite application, parameter estimation may be used to uncover the ways in which a model is wrong. The basic idea is to separately optimize model parameters to two different data sets, and then identify parameter values that differ between the two optimizations. When no single value of a particular parameter fits both datasets, then there must exist a related structural error.
}

\section{Coupling CLUBB to microphysics using analytic integration}

\textbf{\bibentry{larson_et_al_2011b}}

{\color{gray} 
In order to drive microphysics using subgrid variability, we need to know the correlations between hydrometeor species. For instance, the correlation between cloud water and rain water influences the rate of accretion of cloud droplets by rain drops. If cloud and rain are correlated, then cloud and rain co-exist, and accretion occurs rapidly. This paper proposes a method to diagnose correlations based on information that is typically available in cloud models.
}

\textbf{\bibentry{larson_griffin_2013a}}

\textbf{\bibentry{griffin_larson_2013a}}

{\color{gray} 
One reason to predict the subgrid PDF is to drive microphysical parameterizations more accurately. For instance, once we know the subgrid PDF, then we know what percentage of a grid box is precipitating strongly, and so forth. 
In these papers, we integrate a microphysics scheme analytically over CLUBB's PDF. We are able 
to do this exactly for the drizzle parameterization of Khairoutdinov and Kogan, 
which is relatively simple in formulation. We find that, for a marine stratocumulus case, 
accounting for subgrid variability leads to considerably more simulated drizzle at the ocean surface.
}

\textbf{\bibentry{chowdhary_et_al_2015_quad}}

{\color{gray} 
Analytic integration over microphysics is restricted in applicability, and Monte Carlo sampling introduces sampling noise. Here, the integration is performed using a third alternative: deterministic quadrature. This method is more general than analytic integration and more accurate than Monte Carlo integration.
}

\section{Coupling CLUBB to microphysics using SILHS}

\textbf{\bibentry{larson_et_al_05a}}

\textbf{\bibentry{larson_schanen_2013a}}

{\color{gray} 
The most accurate way to drive microphysics using a PDF is to integrate the relevant microphysical formulas analytically over the PDF. However, this may be intractable for some microphysics schemes or may require rewriting the microphysics code. To avoid this, one may draw sample points from the PDF and input them into the microphysics code one at a time. This allows the use of existing microphysics codes, but it also introduces statistical noise due to imperfect sampling. To reduce the noise, sample points may be spread out in a quasi-random fashion using “Latin hypercube sampling,” and the sample points may be clustered in important regions, such as cloud.
}

\textbf{\bibentry{ovchinnikov_et_al_2016_overlap}}

{\color{gray} 
This paper analyzes the vertical overlap of hydrometeors.  CLUBB vertical overlap assumption is embodied
in a copula.  This copula is compared to one obtained from a LES of deep convection.
}

\section{Simulations that use CLUBB-SILHS as a deep convective parameterization}

{\color{gray} 
In these papers, no mass-flux scheme is turned on, and deep convection is represented by 
CLUBB-SILHS alone.}

\textbf{\bibentry{storer_et_al_2015a}}

{\color{gray} 
This paper tests CLUBB-SILHS' capabilities as a deep convection parameterization
in a single-column setting, where none of the convection can be handled by the dynamical core.
CLUBB-SILHS's single equation set 
is used to simulate stratocumulus, shallow cumulus, and deep cumulus layers.
The deep convective simulations exhibit profiles of rain, snow, and graupel 
that are comparable to LES benchmark simulations.
}


\textbf{\bibentry{thayer-calder_et_al_2015_cam_clubb_silhs}}

{\color{gray}
This paper tests CLUBB-SILHS in global CAM simulations with no other deep convective parameterization.
The model climatology is competitive with CAM5 except that the precipitation fields are too noisy.
The horizontal resolutions tested are $1^\circ$ and $2^\circ$.
This demonstrates that CLUBB can go deep in a realistic global configuration.
}

\section{CAM-CLUBB}

{\color{gray} 
In these papers, the implementation of CLUBB in CAM is tested using a variety of configurations and observational datasets.
}

\textbf{\bibentry{bogenschutz_et_al_2013a}}

{\color{gray} 
In this paper, CLUBB is implemented in CAM and tested in global simulations. CLUBB is used in these simulations to parameterize all shallow (stratocumulus and cumulus) clouds, but not deep cumulus.
}

\textbf{\bibentry{kubar_et_al_2015_camclubb}}

{\color{gray} 
Here, CAM-CLUBB's depiction of low clouds is evaluated using satellite data. CAM-CLUBB simulates a smoother transition between marine stratocumulus and shallow cumulus clouds.
}

\textbf{\bibentry{guo_et_al_2014_scamclubb}}

\textbf{\bibentry{guo_et_al_2015_camclubb}}

{\color{gray} 
In these two papers, the sensitivity of CAM-CLUBB to changes in parameter values is tested using single-column and global simulations. These papers provide valuable guidance not only on the practical issue of tuning CAM-CLUBB, but also on the issue of understanding how changes in the strength of various small-scale processes affects the emergent cloud behavior.
}

\textbf{\bibentry{wang_et_al_2015_mmf_clubb}}

{\color{gray} 
Here CLUBB is implemented in a cloud-resolving model with 4-km horizontal grid spacing, which in turn is implemented in each grid column of CAM5. The model behavior is similar to CAM-CLUBB at 100-km horizontal grid spacing. This indicates that CLUBB behaves similarly over a range of horizontal grid spacings. 
}


\section{Implementation of CLUBB in cloud-resolving and regional models}

{\color{gray} 
The fact that CLUBB works in host models with a wide range of grid spacings (4 to 100 km) suggests 
that CLUBB is relatively insensitive to horizontal grid spacing.
}

\textbf{\bibentry{larson_2012_2km_16km}}

{\color{gray} 
This paper implements CLUBB in a convection-permitting model, SAM. The use of CLUBB in SAM is tested for various boundary-layer cloud cases. We introduce a simple method for damping CLUBB’s effects at high resolution, thereby reducing undesirable sensitivities to horizontal grid spacing. We find that the use of CLUBB can improve the simulations for grid spacings of 4 km or greater.
}

\textbf{\bibentry{larson_et_al_2013b}}

\textbf{\bibentry{larson_et_al_2012b}}

{\color{gray} 
These conference papers show simulations of a marine stratocumulus case using CLUBB 
implemented in a weather-forecast model, WRF, at moderate resolution.
}

\section{Participation by CLUBB in single-column model intercomparisons}

{\color{gray} 
In single-column intercomparisons, CLUBB has been tested in a wide variety of cloud regimes.
}

\textbf{\bibentry{svensson_et_al_11a}}

\textbf{\bibentry{bosveld_et_al_2014_gabls3}}

{\color{gray} 
These two intercomparisons demonstrate that CLUBB can simulate stable boundary layers, including those that form at night after the occurrence of daytime boundary-layer turbulence.
}

\textbf{\bibentry{klein_et_al_09a}}

\textbf{\bibentry{morrison_et_al_09a}}

{\color{gray} 
Clouds in the Arctic are often mixed-phase: that is, they often contain both liquid and ice. Long-lived mixed-phase clouds are difficult to simulate because ice naturally tends to grow at the expense of liquid. Models may overdeplete liquid unless the ice particles are limited in number and sediment out of cloud base rapidly enough. Our cloud parameterization, CLUBB, was used to simulate mixed-phase clouds during the M-PACE experiment. CLUBB was able to maintain liquid water in these clouds, as was observed.
}

\textbf{\bibentry{davies_et_al_2013a}}

{\color{gray} 
This paper compares several internationally recognized parameterizations of deep convection. The simulated observations were obtained during the Tropical Warm Pool International Cloud Experiment (TWP-ICE) near Darwin, Australia. CLUBB simulated this deep convective case using the same configuration that is used to simulate boundary-layer clouds. CLUBB's results for TWP-ICE are competitive with those of the other participating parameterizations. The results suggest that CLUBB contains enough physics to serve as a unified parameterization of both shallow and deep clouds.
}

\textbf{\bibentry{zhang_et_al_2013_cgils}}

{\color{gray} 
This intercomparison demonstrates that CLUBB can simulate marine shallow clouds that are driven to equilibrium in month-long simulations.
}

\textbf{\bibentry{wyant_et_al_07a}}

{\color{gray} 
This paper compared the output from numerous single-column models that were set up identically to simulate a cloud layer observed during the DYCOMS-II field experiment. Part of the challenge was simulating drizzle. In order to couple drizzle to the cloud fields, instead of drawing sample points from the PDF using the Latin hypercube method discussed above, we analytically integrated over the PDF.
}

\clearpage
\bibliography{mybibabbr.bib}

\end{document}